\documentclass[11pt]{article}

\usepackage{authblk}
\usepackage{fancyhdr}
\usepackage{float}

\usepackage[round, sort]{natbib}

\usepackage{amssymb}
\usepackage{amsmath,amsthm}

\usepackage{hyperref}
\usepackage{xcolor}
\hypersetup{
    colorlinks,
    linkcolor={red!50!black},
    citecolor={blue!50!black},
    urlcolor={blue!80!black}
}

\usepackage{multirow}
\usepackage{pgfplots}
\usepackage{arydshln}
\usepackage{paralist}
\usepackage{booktabs}
\usepackage{pbox}

\newcounter{rqnum} %research question number

\newcommand{\rqref}[1]{RQ\ref{#1}}

\newcounter{pnum} %pain point number

\newcommand{\ppref}[1]{P\ref{#1}}

\newcounter{qnum} %quality number

\newcommand{\qref}[1]{Q\ref{#1}}

\newcommand{\CC}{C\nolinebreak\hspace{-.05em}\raisebox{.4ex}{\small\bf
+}\nolinebreak\hspace{-.10em}\raisebox{.4ex}{\small\bf +}}

\pgfplotsset{compat=1.13}

\begin{document}

\title{State of the Practice for Medical Imaging Software}

\author[1,*]{W.\ Spencer Smith}
\author[1]{Ao Dong}
\author[1]{Jacques Carette}
\author[2]{Michael D.\ Noseworthy}

\affil[1]{McMaster University, Computing and Software Department, {Canada}}
\affil[2]{McMaster University, Electrical \& Computer Engineering Department,
{Canada}}
\affil[*]{Corresponding Author}

\maketitle

\begin{abstract}

We present the state of the practice for Medical Imaging (MI) software. We
selected 29 medical imaging projects from 48 candidates, assessed 10 software
qualities (installability, correctness/ verifiability, reliability, robustness,
usability, maintainability, reusability, understandability,
visibility/transparency and reproducibility) by answering 108 questions for each
software project, and interviewed 8 of the 29 development teams. Based on the
quantitative data for the first 9 qualities, we ranked the MI software with the
Analytic Hierarchy Process (AHP). The four top ranked software products are:
\textit{3D Slicer}, \textit{ImageJ}, \textit{Fiji}, and \textit{OHIF Viewer}.
Our ranking is mostly consistent with the community's ranking, with four of our
top five projects also appearing in the top five of a list ordered by
stars-per-year. Generally, MI software is in a healthy state as shown by the
following: in the repositories we observed 88\% of the documentation artifacts
recommended by research software development guidelines, 100\% of MI projects
use version control tools, and developers appear to use the common quasi-agile
research software development process. However, the current state of the
practice deviates from the existing guidelines because of the rarity of some
recommended artifacts (like test plans, requirements specification, code of
conduct, code style guidelines, product roadmaps, and Application Program
Interface (API) documentation), low usage of continuous integration (17\% of the
projects), low use of unit testing (about 50\% of projects), and room for
improvement with documentation (six of nine developers felt their documentation
wasn't clear enough). From interviewing the developers, we identified five pain
points and two qualities of potential concern: lack of development time, lack of
funding, technology hurdles, ensuring correctness, usability, maintainability,
and reproducibility. The interviewees proposed strategies to improve the state
of the practice, to address the identified pain points, and to improve software
quality. Combining their ideas with ours, we have the following list of
recommendations: increase documentation, increase testing by enriching datasets,
increase continuous integration usage, move to web applications, employ linters,
use peer reviews, design for change, add assurance cases, and incorporate a
``Generate All Things'' approach.

\end{abstract}


\section{Introduction} \label{ch_intro}

\begin{figure}[!ht]
    \begin{center}
        \includegraphics[scale=0.25]{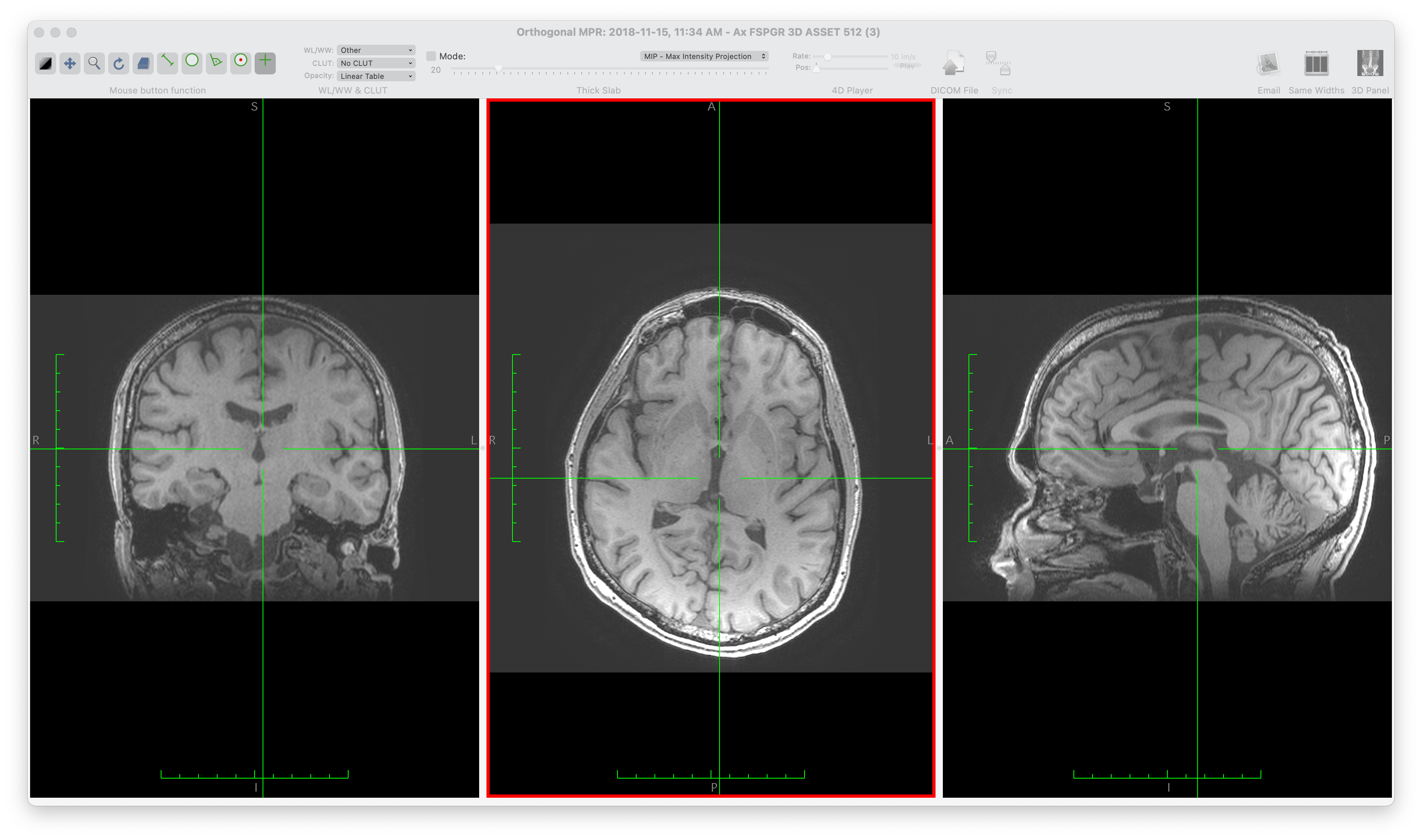}        
    \end{center}
    \caption{Example brain image showing a multi-planar reformat using Horos
	(free open-source medical imaging/DICOM viewer for OSX, based on OsiriX)}
    \label{Fig_Example}
\end{figure}

We aim to study the state of software development practice for Medical Imaging
(MI) software.  MI tools use images of the interior of the body (from sources
such as Magnetic Resonance Imaging (MRI), Computed Tomography (CT), Positron
Emission Tomography (PET) and Ultrasound) to provide information for diagnostic,
analytic, and medical applications \citep{FDA2021, enwiki:1034887445,
Zhang2008}.  Figure~\ref{Fig_Example}, which shows an image of the brain,
highlights the importance and value of MI. Through MI medical practitioners and
researchers can noninvasively gain insights into the human body, including
information on injuries and illnesses. Given the importance of MI software and
the high number of competing software projects, we wish to understand the merits
and drawbacks of the current development processes, tools, and methodologies. We
aim to assess through a software engineering lens the quality of the existing
software with the hope of highlighting standout examples, understanding current
pain points and providing guidelines and recommendations for future development.

\subsection{Research Questions} \label{sec_motivation}

Not only do we wish to gain insight into the state of the practice for MI
software, we also wish to understand the development of research software in
general. We wish to understand the impact of the often cited gap, or chasm,
between software engineering and research software \citep{Kelly2007,
Storer2017}. Although scientists spend a substantial proportion of their working
hours on software development \citep{Hannay2009, Prabhu2011}, many developers
learn software engineering skills by themselves or from their peers, instead of
from proper training \citep{Hannay2009}. \citet{Hannay2009} observe that many
scientists showed ignorance and indifference to standard software engineering
concepts. For instance, according to a survey by \citet{Prabhu2011}, more than
half of their 114 subjects did not use a proper debugger when coding.

To gain insights, we devised 10 research questions, which can be applied to MI,
as well as to other domains, of research software \citep{SmithEtAl2021,
SmithAndMichalski2022}.  We designed the questions to learn about the
community's interest in, and experience with, software artifacts, tools,
principles, processes, methodologies, and qualities.  When we mention artifacts
we mean the documents, scripts and code that constitutes a software development
project. Example artifacts include requirements, specifications, user manuals,
unit tests, system tests, usability tests, build scripts, API (Application
Programming Interface) documentation, READMEs, license documents, process
documents, and code.  Once we have learned what MI developers do, we then put
this information in context by contrasting MI software against the trends shown
by developers in other research software communities.  Our aim is to collect
enough information to understand the current pain points experienced by the MI
software development community so that we can make some preliminary
recommendations for future improvements. 

We based the structure of the paper on the research questions, so for each
research question below we point to the section that contains our answer.  We
start with identifying the relevant examples of MI software for the assessment
exercise:

\begin{enumerate}
	\item[RQ\refstepcounter{rqnum}\therqnum \label{RQ_WhatProjects}:] What MI
	software projects exist, with the constraint that the source code must be
	available for all identified projects? (Section~\ref{ch_results})
	\item [RQ\refstepcounter{rqnum}\therqnum \label{RQ_HighestQuality}:] Which
	of the projects identified in \rqref{RQ_WhatProjects} follow current best
	practices, based on evidence found by experimenting with the software and
	searching the artifacts available in each project's repository?
	(Section~\ref{ch_results})
	\item [RQ\refstepcounter{rqnum}\therqnum \label{RQ_CompareHQ2Popular}:] How
	similar is the list of top projects identified in \rqref{RQ_HighestQuality}
	to the most popular projects, as viewed by the scientific community?
	(Section~\ref{Sec_VsCommunityRanking})
    \item [RQ\refstepcounter{rqnum}\therqnum \label{RQ_CompareArtifacts}:] How
	do MI projects compare to research software in general with respect to the
	artifacts present in their repositories?
	(Section~\ref{Sec_CompareArtifacts})
	\item [RQ\refstepcounter{rqnum}\therqnum \label{RQ_CompareToolsProjMngmnt}:]
	How do MI projects compare to research software in general with respect to
	the use of tools (Section~\ref{Sec_CompareTools}) for:
	\begin{enumerate} 
		\item [\rqref{RQ_CompareToolsProjMngmnt}.a] development; and,
		\item [\rqref{RQ_CompareToolsProjMngmnt}.b] project management?
	\end{enumerate}
	\item [RQ\refstepcounter{rqnum}\therqnum \label{RQ_CompareMethodologies}:]
	How do MI projects compare to research software in general with respect to
	principles, processes, and methodologies used?
	(Section~\ref{Sec_CompareMethodologies})
	\item [RQ\refstepcounter{rqnum}\therqnum \label{RQ_PainPoints}:] What are
	the pain points for developers working on MI software projects?
	(Section~\ref{painpoints})
	\item [RQ\refstepcounter{rqnum}\therqnum \label{RQ_ComparePainPoints}:] How
	do the pain points of developers from MI compare to the pain points
	for research software in general? (Section~\ref{painpoints})
	\item [RQ\refstepcounter{rqnum}\therqnum \label{RQ_Concerns}:] For MI
	developers what specific best practices are taken to address the pain points
	and software quality concerns? (Section~\ref{painpoints})
	\item [RQ\refstepcounter{rqnum}\therqnum \label{RQ_Recommend}:]
	What research software development practice could potentially address the
	pain point concerns identified in \rqref{RQ_PainPoints}?
	(Section~\ref{ch_recommendations})

\end{enumerate}

\subsection{Scope} \label{sec_scope}

To make the project feasible, we only cover MI visualization software.  As a
consequence we are excluding many other categories of MI software, including
Segmentation, Registration, Visualization, Enhancement, Quantification,
Simulation, plus MI archiving and telemedicine systems (Compression, Storage,
and Communication) (as summarized by \citet{Bankman2000} and
\citet{Angenent2006}).  We also exclude Statistical Analysis and Image-based
Physiological Modelling \citep{enwiki:1034877594} and Feature Extraction,
Classification, and Interpretation \citep{Kim2011}. Software that provides MI
support functions is also out of scope; therefore, we have not assessed the
toolkit libraries VTK \citep{SchroederEtAl2006} and ITK \citep{McCormick2014}.
Finally, Picture Archiving and Communication System (PACS), which helps users to
economically store and conveniently access images \citep{Choplin1992}, are
considered out of scope. 

\subsection{Methodology Overview}

We designed a general methodology to assess the state of the practice for
research software \citep{SmithEtAl2021, SmithAndMichalski2022}. Details can be
found in Section~\ref{ch_methods}.  Our methodology has been applied to MI
software \citep{Dong2021} and Lattice Boltzmann Solvers \citep{Michalski2021,
SmithEtAl2024}.  This methodology builds off prior work to assess the state of
the practice for such domains as Geographic Information Systems
\citep{smith2018state}, Mesh Generators \citep{smith2016state}, Seismology
software \citep{Smith2018Seismology}, and Statistical software for psychology
\citep{smith2018statistical}.  In keeping with the previous methodology, we have
maintained the constraint that the work load for measuring a given domain should
be feasible for a team as small as one person, and for a short time, ideally
around a person month of effort. We consider a person month as $20$ working days
($4$ weeks in a month, with $5$ days of work per week) at $8$ person hours per
day, or $20 \times 8 = 160$ person hours.

With our methodology, we first choose a research software domain (in the current
case MI) and identify a list of about 30 software packages. (For measuring MI we
used 29 software packages.)  We approximately measure the qualities of each
package by filling in a grading template. Compared with our previous
methodology, the new methodology also includes repository based metrics, such as
the number of files, number of lines of code, percentage of issues that are
closed, etc.  With the quantitative data in the grading template, we rank the
software with the Analytic Hierarchy Process (AHP) (Section~\ref{ch_background}
provides details). After this, as another addition to our previous methodology,
we interview some development teams to further understand the status of their 
development process.

\section{Background} \label{ch_background}

To measure the existing MI software we need two sets of definitions: i) the
definitions of relevant software license models (Section
\ref{sec_software_categories}); and, ii) the definitions of the software
qualities that we will be assessing (Section \ref{sec_software_quality}). In our
assessment we rank the software packages for each quality; therefore, this
section also provides the background on our ranking process --- the Analytic
Hierarchy Process (Section \ref{sec_AHP}).

\subsection{Software Categories} \label{sec_software_categories}

When assessing software packages, we need to know the software's license.  In
particular, we need to know whether the source code will be available to us or
not.  We define three common software categories.  We will only assess software
that fits under the Open Source Software license.

\begin{itemize}

\item \textbf{Open Source Software (OSS)} For OSS, the source code is openly
accessible. Users have the right to study, change and distribute it under a
license granted by the copyright holder. For many OSS projects, the development
process relies on the collaboration of different contributors worldwide
\citep{Corbly2014}. Accessible source code usually exposes more ``secrets'' of a
software project, such as the underlying logic of software functions, how
developers achieve their works, and the flaws and potential risks in the final
product. Thus, OSS is suitable for researchers analyzing the qualities of a
project.

\item \textbf{Freeware} Freeware is software that can be used free of charge.
Unlike OSS, the authors of do not allow access or modify the source code
\citep{LINFO2006}. To many end-users, the differences between freeware and OSS
may not be relevant. However, software developers who wish to modify the source
code, and researchers looking for insight into software development process may
find the inaccessible source code a problem. 

\item \textbf{Commercial Software} ``Commercial software is software developed
by a business as part of its business'' \citep{GNU2019}. Typically speaking,
commercial software requires users to pay to access all of its features,
excluding access to the source code. However, some commercial software is also
free of charge \citep{GNU2019}. Based on our experience, most commercial
software products are not OSS.

\end{itemize}

\subsection{Software Quality Definitions} \label{sec_software_quality}

Quality is defined as a measure of the excellence or worth of an entity.  As is
common practice, we do not think of quality as a single measure, but rather as a
set of measures.  That is, quality is a collection of different qualities, often
called ``ilities.''  Below we list the 10 qualities of interest for this study.
The order of the qualities follows the order used in \citet{GhezziEtAl2003},
which puts related qualities (like correctness and reliability) together.
Moreover, the order is roughly the same as the order developers consider
qualities in practice.

\begin{itemize}
	\item \textbf{Installability} The effort required for the installation
    and/or uninstallation of software in a specified environment
    \citep{ISO/IEC25010, lenhard2013measuring}.

	\item \textbf{Correctness \& Verifiability} A program is correct if it
    matches its specification \citep[p.\ 17]{GhezziEtAl2003}.  The specification
    can either be explicitly or implicitly stated.  The related quality of
    verifiability is the ease with which the software components or the
    integrated product can be checked to demonstrate its correctness. 

	\item \textbf{Reliability} The probability of failure-free operation of a
	computer program in a specified environment for a specified time
	\citep{musa1987software}, \citep[p.\ 357]{GhezziEtAl2003}.

	\item \textbf{Robustness} Software possesses the characteristic of
	robustness if it behaves ``reasonably'' in two situations: i) when it
	encounters circumstances not anticipated in the requirements specification,
	and ii) when users violate the assumptions in its requirements specification 
	\citep[p.\ 19]{GhezziEtAl2003}, \citep{boehm2007software}.

	\item \textbf{Usability} ``The extent to which a product can be used by
	specified users to achieve specified goals with effectiveness, efficiency,
	and satisfaction in a specified context of use'' \citep{ISO/TR16982:2002,
	ISO9241-11:2018}.

	\item \textbf{Maintainability} The effort with which a software system or
	component can be modified to i) correct faults; ii) improve performance or
	other attributes; iii) satisfy new requirements
	\citep{IEEEStdGlossarySET1990, boehm2007software}.

	\item \textbf{Reusability} ``The extent to which a software component can be
	used with or without adaptation in a problem solution other than the one for
	which it was originally developed'' \citep{kalagiakos2003non}.

	\item \textbf{Understandability} ``The capability of the software product to
	enable the user to understand whether the software is suitable, and how it
	can be used for particular tasks and conditions of use'' \citep{iso2001iec}.

	\item \textbf{Visibility/Transparency} The extent to which all the steps
	of a software development process and the current status of it are conveyed
	clearly \citep[p.\ 32]{GhezziEtAl2003}.

	\item \textbf{Reproducibility} ``A result is said to be reproducible if
	another researcher can take the original code and input data, execute it,
	and re-obtain the same result'' \citep{BenureauAndRougier2017}.
\end{itemize}

\subsection{Analytic Hierarchy Process (AHP)} \label{sec_AHP}

Saaty developed AHP in the 1970s, and people have widely used it since to make
and analyze multiple criteria decisions \citep{VaidyaEtAl2006}. AHP organizes
multiple criteria in a hierarchical structure and uses pairwise comparisons
between alternatives to calculate relative ratios \citep{Saaty1990}. AHP works
with sets of $n$ \textit{options} and $m$ \textit{criteria}.  In our project
$n=29$ and $m=9$ since there are 29 options (software products) and 9 criteria
(qualities). We rank the software for each of the qualities, and then we combine
the quality rankings into an overall ranking based on the relative priorities
between qualities.

The first step for ranking the software choices for a given quality involves a
pairwise comparison between each of the $n$ software options for that quality.
AHP expresses the comparison through an $n \times n$ matrix $A$. When comparing
option $i$ and option $j$, the value of $A_{ij}$ is decided as follows, with the
value of $A_{ji}$ generally equal to $1/A_{ij}$ \citep{Saaty1990}: $A_{ij} = 1$
if criterion $i$ and criterion $j$ are equally important, while $A_{ij} = 9$ if
criterion $i$ is extremely more important than criterion $j$.  The natural
numbers between 1 and 9 are used to show the different levels of relative
importance between these two extremes. The above assumes that option $i$ is of
equal, or more, importance compared to option $j$ ($i \geq j$).  If that is not
the case, we reverse $i$ and $j$ and determine $A_{ji}$ first, then $A_{ij} =
1/A_{ji}$.

Section~\ref{sec_grading_software} shows how we measure the software via a
grading template.  For the AHP process, the relevant measure is the subjective
score from $1$ to $10$ for each quality for each package. To turn these
subjective measures $x_{\text{sub}}$ and $y_{\text{sub}}$ into Saaty's
pair-wise scores for option $x$ versus option $y$, respectively, we use the
following calculation:
\[
\begin{cases}
\min\{9, x_{\text{sub}} - y_{\text{sub}} + 1\} & x_{\text{sub}} \geq y_{\text{sub}} \\
1 / \min\{9, y_{\text{sub}} - x_{\text{sub}} + 1\} & x_{\text{sub}} < y_{\text{sub}}
\end{cases}
\]

\noindent For example, we measured the usability for 3D Slicer and Ginkgo CADx
as $8$ and $7$, respectively; therefore, on the 9-point scale, 3D Slicer compared
to Ginkgo CADx is 2 and Ginkgo CADx versus 3D Slicer is 1/2, as shown in the
sample AHP calculations (Table~\ref{Tbl_SampleAHP}).

The second step is to calculate the priority vector $w$ from $A$.  The
vector $w$ ranks the software options by how well they achieve the given
quality.  The priority vector can be calculated by solving the equation
\citep{Saaty1990}:
\begin{equation} 
    A w = \lambda_{\text{max}} w,
\end{equation}
where $\lambda_{\text{max}}$ is the maximal eigenvalue of $A$.  In this project,
$w$ is approximated with the classic \textit{mean of normalized values} approach
\citep{AlessioEtAl2006}:

\begin{equation}
w_i = \frac{1}{n}\sum_{j=1}^{n}\frac{A_{ij}}{\sum_{k=1}^{n}A_{kj}}
\end{equation}

Table~\ref{Tbl_SampleAHP} summarizes the above two steps for the quality of
installability.  The matrix $A$ is shown in the first set of columns, then the
normalized version of $A$ and finally the average of the normalized values to
form the vector $w$ in the last column.

\begin{table}[h!]
\begin{center}
\begin{tabular}{ l c c c c c | c c c c c | c }
 \toprule
 ~ & \multicolumn{5}{c|}{$A_{ij}$} & \multicolumn{5}{c|}{${A_{ij}}/{\sum_{k=1}^{n}A_{kj}}$} & ~\\
 \midrule
 ~ & \rotatebox{90}{3D Slicer} & \rotatebox{90}{Ginkgo} & \rotatebox{90}{XMedCon} & $\cdots$ & \rotatebox{90}{Gwyddion} & \rotatebox{90}{3D Slicer} & \rotatebox{90}{Ginkgo} & \rotatebox{90}{XMedCon} & $\cdots$ & \rotatebox{90}{Gwyddion} & AVG \\
 \midrule
 3D Slicer & 1 & 2 & 4 & $\cdots$ & 2 & 0.071 & 0.078 & 0.060 & $\cdots$ & 0.078 & 0.068\\
 Ginkgo & 1/2 & 1 & 3 & $\cdots$ & 1 & 0.036 & 0.039 & 0.045 & $\cdots$ & 0.039 & 0.041\\
 XMedCon & 1/4 & 1/3 & 1 & $\cdots$ & 1/3 & 0.018 & 0.013 & 0.015 & $\cdots$ & 0.013 & 0.015\\
 $\vdots$ & $\vdots$ & $\vdots$ & $\vdots$ & $\ddots$ & $\vdots$ & $\vdots$ & $\vdots$ & $\vdots$ & $\ddots$ & $\vdots$ & $\vdots$\\
 Gwyddion & 1/2 & 1 & 3 & $\cdots$ & 1 & 0.036 & 0.039 & 0.045 & $\cdots$ & 0.039 & 0.041\\  
 \midrule
 SUM = & 14.01 & 25.58 & 66.75 & $\cdots$ & 25.58 & 1.000 & 1.000 & 1.000 & $\cdots$ & 1.000 & 1.000\\
 \bottomrule
\end{tabular}
\end{center}
\caption{Sample AHP Calculations for the Quality of Usability} \label{Tbl_SampleAHP}
\end{table}

We repeat the first and second steps for each of the qualities.  The third step
combines the quality rankings into an overall ranking.  Following AHP, we need
to first prioritize the qualities.  The AHP method finds the priority of quality
$i$ ($p_i$) in the same way that the score ($w_j$) was found for software
package $j$ evaluated for a given quality (as shown above).  That is, we
conduct a pairwise comparison between the priority of different qualities to
construct the $m \times m$ matrix $A$, and then we take the mean of normalized
values for row $i$ to find the priority value $p_i$ for quality $i$.  If we
introduce the notation that $w^i_j$ is the score for quality $i$ for package
$j$, then the overall score $S_j$ for package $j$ is found via:

$$S_j = \sum_{i=1}^m w^i_j p_i$$ 

\section{Methodology} \label{ch_methods}

We developed a methodology for evaluating the state of the practice of research
software \citep{SmithEtAl2021, SmithAndMichalski2022}.  The methodology can be
instantiated for a specific domain of scientific software, which in the current
case is medical imaging software for visualization.  Our methodology involves
and engages a domain expert partner throughout, as discussed in
Section~\ref{sec_vet_software_list}.  The four main steps of the methodology
are:

\begin{enumerate}
\item Identify list of representative software packages
(Section~\ref{sec_software_selection});
\item Measure (or grade) the selected software
(Section~\ref{sec_grading_software});
\item Interview developers (Section~\ref{sec_interview_methods});
\item Answer the research questions (as given in Section~\ref{sec_motivation}).
\end{enumerate}

In the sections below we provide additional detail on the above steps, while
concurrently giving examples of how we applied the methodology to the MI domain.

\subsection{Interaction With Domain Expert} \label{sec_vet_software_list}

The Domain Expert is an important member of the state of the practice assessment
team. Pitfalls exist if non-experts attempt to acquire an authoritative list of
software, or try to definitively rank the software. Non-experts have the problem
that they can only rely on information available on-line, which has the
following drawbacks:
\begin{inparaenum}[i)]
  \item the on-line resources could have false or inaccurate information; and,
  \item the on-line resources could leave out relevant information that is so
in-grained with experts that nobody thinks to explicitly record it.
\end{inparaenum}

Domain experts may be recruited from academia or industry.  The only
requirements are knowledge of the domain and a willingness to be engaged in the
assessment process.  The Domain Expert does not have to be a software developer,
but they should be a user of domain software.  Given that the domain experts are
likely to be busy people, the measurement process cannot put too much of a burden
on their time.  For the current assessment, our Domain Expert (and paper
co-author) is Dr.\ Michael Noseworthy, Professor of Electrical and Computer
Engineering at McMaster University, Co-Director of the McMaster School of
Biomedical Engineering, and Director of Medical Imaging Physics and Engineering
at St.\ Joseph's Healthcare, Hamilton, Ontario, Canada.  

In advance of the first meeting with the Domain Expert, they are asked to
create a list of top software packages in the domain.  This is done to help
the expert get in the right mind set in advance of the meeting.  Moreover,
by doing the exercise in advance, we avoid the potential pitfall of the expert
approving the discovered list of software without giving it adequate thought.

The Domain Experts are asked to vet the collected data and analysis.  In
particular, they are asked to vet the proposed list of software packages and the
AHP ranking.  These interactions can be done either electronically or with
in-person (or virtual) meetings.

\subsection{List of Representative Software} \label{sec_software_selection}

We have a two-step process for selecting software packages: i) identify software
candidates in the chosen domain; and, ii) filter the list to remove less
relevant members \citep{SmithEtAl2021}.

We initially identified 48 MI candidate software projects from the literature
\citep{Bjorn2017, Bruhschwein2019, Haak2015}, on-line articles \citep{Emms2019,
Hasan2020, Mu2019}, and forum discussions \citep{Samala2014}.  The full list of
48 packages is available in \citet{Dong2021}.  To reduce the length of the list
to a manageable number (29 in this case, as given in Section~\ref{ch_results}),
we filtered the original list as follows:

\begin{enumerate}

\item We removed the packages that did not have source code available, such as
\textit{MicroDicom}, \textit{Aliza}, and \textit{jivex}.

\item We focused on the MI software that provides visualization functions, as
described in Section~\ref{sec_scope}. Furthermore, we removed seven packages that were
toolkits or libraries, such as \textit{VTK}, \textit{ITK}, and \textit{dcm4che}.
We removed another three that were for PACS.

\item We removed \textit{Open Dicom Viewer}, since it has not received any
updates in a long time (since 2011).

\end{enumerate}

The Domain Expert provided a list of his top 12 software packages.  We compared
his list to our list of 29.  We found 6 packages were on both lists: \textit{3D
Slicer}, \textit{Horos}, \textit{ImageJ}, \textit{Fiji}, \textit{MRIcron} (we
actually use the update version \textit{MRIcroGL}) and \textit{Mango} (we
actually use the web version \textit{Papaya}).  Six software packages
(\textit{AFNI}, \textit{FSL}, \textit{Freesurfer}, \textit{Tarquin},
\textit{Diffusion Toolkit}, and \textit{MRItrix}) were on the Domain Expert
list, but not on our filtered list.  However, when we examined those packages,
we found they were out of scope, since their primary function was not
visualization.  The Domain Expert agreed with our final choice of 29 packages.

\subsection{Grading Software} \label{sec_grading_software}

\begin{figure}[!ht]
\includegraphics[scale=0.66]{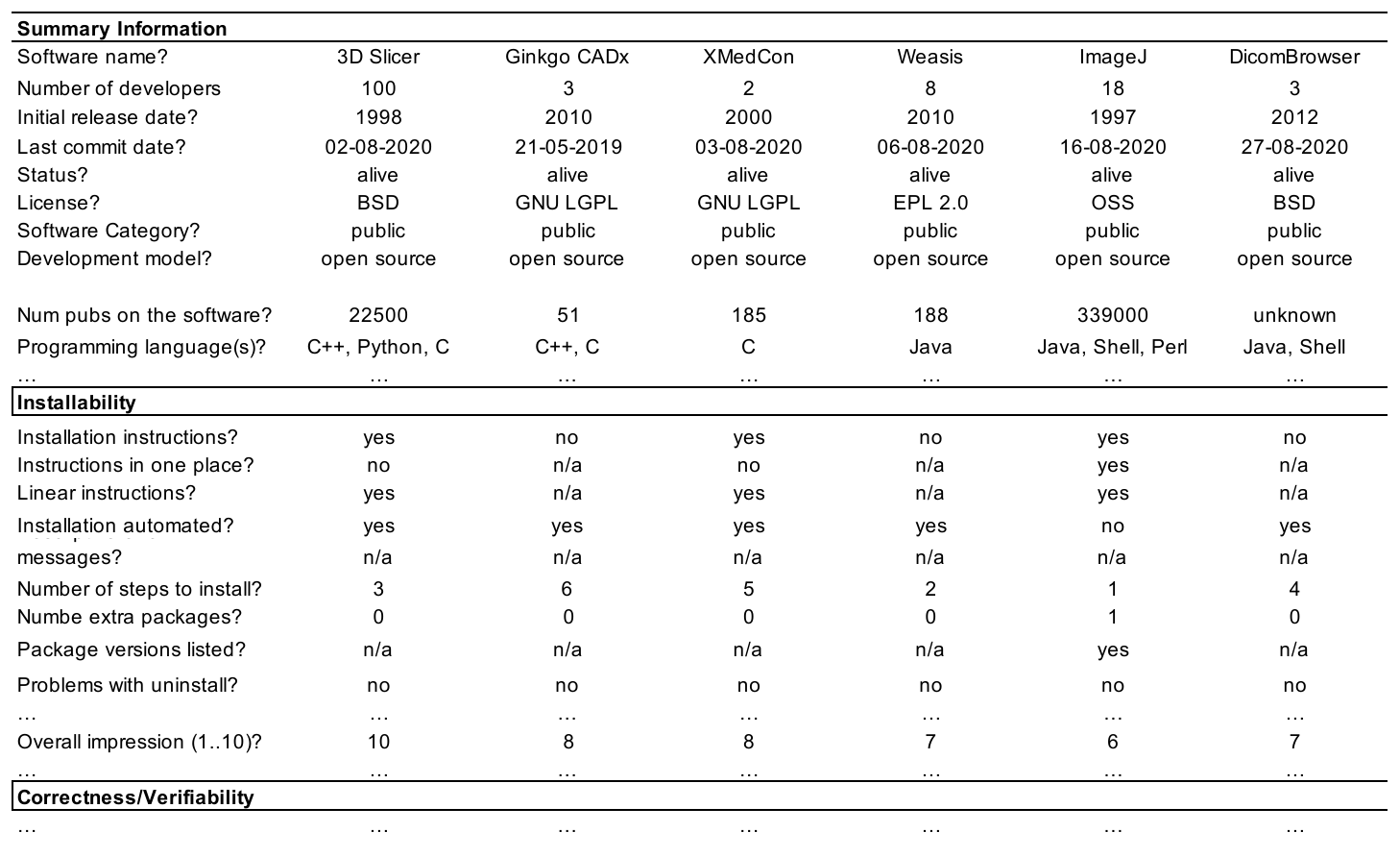}
\caption{Grading template example}
\label{fg_grading_template_example}
\end{figure}

We grade the selected software using the measurement template summarized in
\citet{SmithEtAl2021}.  The template provides measures of the qualities listed
in Section~\ref{sec_software_quality}, except for reproducibility, which is
assessed through the developer interviews (Section~\ref{sec_interview_methods}).
For each software package, we fill in the template questions. To stay within the
target of 160 person hours to measure the domain, we allocated between one and
four hours for each package. Project developers can be contacted for help
regarding installation, if necessary, but we impose a cap of about two hours on
the installation process, to keep the overall measurement time feasible.
Figure~\ref{fg_grading_template_example} shows an excerpt of the spreadsheet.
The spreadsheet includes a column for each measured software package. 

The full template consists of 108 questions categorized under 9 qualities.  We
designed the questions to be unambiguous, quantifiable, and measurable with
limited time and domain knowledge. We group the measures under headings for each
quality, and one for summary information. The summary information (shown in
Figure~\ref{fg_grading_template_example}) is the first section of the template.
This section summarizes general information, such as the software name, purpose,
platform, programming language, publications about the software, the first
release and the most recent change date, website, source code repository of the
product, number of developers, etc.  We follow the definitions given by
\citet{GewaltigAndCannon2012} for the software categories.  Public means
software intended for public use.  Private means software aimed only at a
specific group, while the concept category is for software written simply to
demonstrate algorithms or concepts. The three categories of development models
are (open source, free-ware and commercial) are discussed in
Section~\ref{sec_software_categories}.  Information in the summary section sets
the context for the project, but it does not directly affect the grading scores.

For measuring each quality, we ask several questions and the typical answers are
among the collection of ``yes'', ``no'', ``n/a'', ``unclear'', a number, a
string, a date, a set of strings, etc. The grader assigns each quality an
overall score, between 1 and 10, based on all the previous questions.  Several
of the qualities use the word ``surface''.  This is to highlight that, for these
qualities in particular, the best that we can do is a shallow measure.  For
instance, we are not currently doing any experiments to measure usability.
Instead, we are looking for an indication that the developers considered
usability.  We do this by looking for cues in the documentation, like a getting
started manual, a user manual and a statement of expected user characteristics.
Below is a summary of how we assess adoption of best practices by measuring each
quality.

\begin{itemize}

\item \textbf{Installability} We assess the following: 
\begin{inparaenum}[i)]
    \item existence and quality of installation instructions;
    \item the quality of the user experience via the ease of following
    instructions, number of steps, automation tools; and,
    \item whether there is a means to verify the installation.
\end{inparaenum}
If any problem interrupts the process of installation or uninstallation, we give
a lower score. We also record the Operating System (OS) used for the
installation test.

\item \textbf{Correctness \& Verifiability} We check each project to identify
any techniques used to ensure this quality, such as literate programming,
automated testing, symbolic execution, model checking, unit tests, etc. We also
examine whether the projects use Continuous Integration and Continuous Delivery
(CI/CD). For verifiability, we go through the documents of the projects to check
for the presence of requirements specifications, theory manuals, and getting
started tutorials. If a getting started tutorial exists and provides expected
results, we follow it to check the correctness of the output.

\item \textbf{Surface Reliability} We check the following: 
\begin{inparaenum}[i)]
    \item whether the software breaks during installation;
    \item the operation of the software following the getting started tutorial
    (if present);
    \item whether the error messages are descriptive; and,
    \item whether we can recover the process after an error.
\end{inparaenum}

\item \textbf{Surface Robustness} We check how the software handles
unexpected/unanticipated input. For example, we prepare broken image files for
MI software packages that load image files. We use a text file (.txt) with a
modified extension name (.dcm) as an unexpected/unanticipated input. We load a
few correct input files to ensure the function is working correctly before
testing the unexpected/unanticipated ones.

\item \textbf{Surface Usability} We examine the project's documentation,
checking for the presence of getting started tutorials and/or a user manual. We
also check whether users have channels to request support, such as an e-mail
address, or issue tracker. Our impressions of usability are based on our
interaction with the software during testing.  In general, an easy-to-use
graphical user interface will score high.

\item \textbf{Maintainability} We believe that the artifacts of a project,
including source code, documents, and building scripts, significantly influence
its maintainability. Thus, we check each project for the presence of such
artifacts as API documentation, bug tracker information, release notes, test
cases, and build scripts. We also check for the use of tools supporting issue
tracking and version control, the percentages of closed issues, and the
proportion of comment lines in the code.

\item \textbf{Reusability} We count the total number of code files for each
project. Projects with numerous components potentially provide more choices for
reuse. Furthermore, well-modularized code, which tends to have smaller parts in
separate files, is typically easier to reuse. Thus, we assume that projects with
more code files and fewer Lines of Code (LOC) per file are more reusable. We also
consider projects with API documentation as delivering better reusability.

\item \textbf{Surface Understandability} Given that time is a constraint, we
cannot look at all code files for each project; therefore, we randomly examine
10 code files for their understandability. We check the code's style within each
file, such as whether the identifiers, parameters, indentation, and formatting
are consistent, whether the constants (other than 0 and 1) are not hardcoded, and
whether the code is modularized. We also check the descriptive information for
the code, such as documents mentioning the coding standard, the comments in the
code, and the descriptions or links for details on algorithms in the code. 

\item \textbf{Visibility/Transparency} To measure this quality, we check the
existing documents to find whether the software development process and
current status of a project are visible and transparent. We examine the
development process, current status, development environment, and release notes
for each project.
\end{itemize}

As part of filling in the measurement template, we use freeware tools to collect
repository related data. \href{https://github.com/tomgi/git_stats}{GitStats}
\citep{Gieniusz2019} is used to measure the number of binary files as well as
the number of added and deleted lines in a repository. We also use this tool to
measure the number of commits over different intervals of time.
\href{https://github.com/boyter/scc}{Sloc Cloc and Code (scc)}
\citep{Boyter2021} is used to measure the number of text based files as well as
the number of total, code, comment, and blank lines in a repository.

Both tools measure the number of text-based files in a git repository and lines
of text in these files. Based on our experience, most text-based files in a
repository contain programming source code, and developers use them to compile
and build software products. A minority of these files are instructions and
other documents. So we roughly regard the lines of text in text-based files as
lines of programming code. The two tools usually generate similar but not
identical results. From our understanding, this minor difference is due to the
different techniques to detect if a file is text-based or binary.

For projects on GitHub we manually collect additional information, such as the
numbers of stars, forks, people watching this repository, open pull requests,
closed pull requests, and the number of months a repository has been on GitHub.
We need to take care with the project creation date, since a repository can have
a creation date much earlier than the first day on GitHub.  For example, the
developers created the git repository for \textit{3D Slicer} in 2002, but did
not upload a copy of it to GitHub until 2020. Some GitHub data can be found
using its GitHub Application Program Interface (API) via the following url:
\textit{https://api.github.com/repos/[owner]/[repository]} where [owner] and
[repository] are replaced by the repo specific values. The number of months a
repository has been on GitHub helps us understand the average change of metrics
over time, like the average new stars per month. 

The repository measures help us in many ways. Firstly, they help us get a fast
and accurate project overview. For example, the number of commits over the last
12 months shows how active a project has been, and the number of stars and forks
may reveal its popularity (used to assess \rqref{RQ_CompareHQ2Popular}).
Secondly, the results may affect our decisions regarding the grading scores for
some software qualities. For example, if the percentage of comment lines is low,
we double-check the understandability of the code; if the ratio of open versus
closed pull requests is high, we pay more attention to maintainability.

As in \citet{SmithEtAl2016}, Virtual machines (VMs) were used to provide an
optimal testing environment for each package. We used VMs because it is easier
to start with a fresh environment, without having to worry about existing
libraries and conflicts. Moreover, when the tests are complete the VM can be
deleted, without any impact on the host operating system. The most significant
advantage of using VMs is to level the playing field. Every software install
starts from a clean slate, which removes ``works-on-my-computer'' errors. When
filling in the measurement template, the grader notes the details for each VM,
including hypervisor and operating system version.

When grading the software, we found 27 out of the 29 packages are compatible
with two or three different OSes, such as Windows, macOS, and Linux, and 5 of
them are browser-based, making them platform-independent. However, in the
interest of time, we only performed the measurements for each project by
installing it on one of the platforms.  When it was an option, we selected
Windows as the host OS.

\subsection{Interview Methods} \label{sec_interview_methods}

The repository-based measurements summarize the information we can collect from
on-line resources. This information is incomplete because it doesn't generally
capture the development process, the developer pain points, the perceived
threats to software quality, and the developers' strategies to address these
threats.  Therefore, part of our methodology involves interviewing developers.

We based our interviews on a list of 20 questions, which can be found in
\citet{SmithEtAl2021}. Some questions are about the background of the software,
the development teams, the interviewees, and how they organize their projects.
We also ask about the developer's understanding of the users. Some questions
focus on the current and past difficulties, and the solutions the team has
found, or plan to try. We also discuss documentation, both with respect to how
it is currently done, and how it is perceived. A few questions are about
specific software qualities, such as maintainability, understandability,
usability, and reproducibility. The interviews are semi-structured based on the
question list; we ask follow-up questions when necessary.  The interview process
presented here was approved by the McMaster University Research Ethics Board
under the application number
\href{https://github.com/smiths/AIMSS/blob/master/StateOfPractice/MACREM/Application.pdf}
{MREB\#: 5219}.

We sent interview requests to all 29 projects using contact information from
projects websites, code repository, publications, and from biographic pages at
the teams' institutions.  In the end nine developers from eight of the projects
agreed to participate: \textit{3D Slicer}, \textit{INVESALIUS 3}, \textit{dwv},
\textit{BioImage Suite Web}, \textit{ITK-SNAP}, \textit{MRIcroGL},
\textit{Weasis}, and \textit{OHIF}. We spent about 90 minutes for each
interview. One participant was too busy to have an interview, so they wrote down
their answers. In one case two developers from the same project agreed to be
interviewed. We held the meetings on-line using either Zoom or Teams, which
facilitated recording and automatic transcription. The full interview answers
can be found in \citet{Dong2021}.

\section{Measurement Results} \label{ch_results}

Table~\ref{tab_final_list} shows the 29 software packages that we measured,
along with summary data collected in the year 2020. We arrange the items in
descending order of LOC. We found the initial release dates (Rlsd) for most
projects and marked the two unknown dates with ``?''. The date of the last
update is the date of the latest update, at the time of measurement. We found
funding information (Fnd) for only eight projects.  For the Number Of
Contributors (NOC) we considered anyone who made at least one accepted commit as
a contributor. The NOC is not usually the same as the number of long-term
project members, since many projects received change requests and code from the
community.  With respect to the OS, 25 packages work on all three OSs: Windows
(W), macOS (M), and Linux (L). Although the usual approach to cross-platform
compatibility was to work natively on multiple OSes, five projects achieved
platform-independence via web applications. The full measurement data for all
packages is available in \citet{Dong2021-Data}.

\begin{table}[!ht]
\centering
\begin{tabular}{p{6cm}lllllllll}
\toprule
\multirow{2}{*}{Software} & \multirow{2}{*}{Rlsd} & \multirow{2}{*}{Updated} & \multirow{2}{*}{Fnd} & \multirow{2}{*}{NOC} & \multirow{2}{*}{LOC} & \multicolumn{3}{c}{OS} & \multirow{2}{*}{Web} \\ \cline{7-9}
 &  &  &  &  &  & W & M & L &  \\ \midrule
ParaView \citep{Ahrens2005} & 2002 & 2020-10 & \checkmark & 100 & 886326 & \checkmark & \checkmark & \checkmark & \checkmark \\
Gwyddion \citep{Nevcas2012} & 2004 & 2020-11 &  & 38 & 643427 & \checkmark & \checkmark & \checkmark &  \\
Horos \citep{horosproject2020} & ? & 2020-04 &  & 21 & 561617 &  & \checkmark &  &  \\
OsiriX Lite \citep{PixmeoSARL2019} & 2004 & 2019-11 &  & 9 & 544304 &  & \checkmark &  &  \\
3D Slicer \citep{Kikinis2014} & 1998 & 2020-08 & \checkmark & 100 & 501451 & \checkmark & \checkmark & \checkmark &  \\
Drishti \citep{Limaye2012} & 2012 & 2020-08 &  & 1 & 268168 & \checkmark & \checkmark & \checkmark &  \\
Ginkgo CADx \citep{Wollny2020} & 2010 & 2019-05 &  & 3 & 257144 & \checkmark & \checkmark & \checkmark &  \\
GATE \citep{Jan2004} & 2011 & 2020-10 &  & 45 & 207122 &  & \checkmark & \checkmark &  \\
3DimViewer \citep{TESCAN2020} & ? & 2020-03 & \checkmark & 3 & 178065 & \checkmark & \checkmark &  &  \\
medInria \citep{Fillard2012} & 2009 & 2020-11 &  & 21 & 148924 & \checkmark & \checkmark & \checkmark &  \\
BioImage Suite Web \citep{Papademetris2005} & 2018 & 2020-10 & \checkmark & 13 & 139699 &
\checkmark & \checkmark & \checkmark & \checkmark \\
Weasis \citep{Roduit2021} & 2010 & 2020-08 &  & 8 & 123272 & \checkmark & \checkmark & \checkmark &  \\
AMIDE \citep{Loening2017} & 2006 & 2017-01 &  & 4 & 102827 & \checkmark & \checkmark & \checkmark &  \\
XMedCon \citep{Nolf2003} & 2000 & 2020-08 &  & 2 & 96767 & \checkmark & \checkmark & \checkmark &  \\
ITK-SNAP \citep{Yushkevich2006} & 2006 & 2020-06 & \checkmark & 13 & 88530 & \checkmark & \checkmark & \checkmark &  \\
Papaya \citep{UTHSCSA2019} & 2012 & 2019-05 &  & 9 & 71831 & \checkmark & \checkmark & \checkmark &  \\
OHIF Viewer \citep{Ziegler2020} & 2015 & 2020-10 &  & 76 & 63951 & \checkmark & \checkmark & \checkmark & \checkmark \\
SMILI \citep{Chandra2018} & 2014 & 2020-06 &  & 9 & 62626 & \checkmark & \checkmark & \checkmark &  \\
INVESALIUS 3 \citep{Amorim2015} & 2009 & 2020-09 &  & 10 & 48605 & \checkmark & \checkmark & \checkmark &  \\
dwv \citep{Martelli2021} & 2012 & 2020-09 &  & 22 & 47815 & \checkmark & \checkmark & \checkmark & \checkmark \\
DICOM Viewer \citep{Afsar2021} & 2018 & 2020-04 & \checkmark & 5 & 30761 & \checkmark & \checkmark & \checkmark &  \\
MicroView \citep{ParallaxInnovations2020} & 2015 & 2020-08 &  & 2 & 27470 & \checkmark & \checkmark & \checkmark &  \\
MatrixUser \citep{Liu2016} & 2013 & 2018-07 &  & 1 & 23121 & \checkmark & \checkmark & \checkmark &  \\
Slice:Drop \citep{Haehn2013} & 2012 & 2020-04 &  & 3 & 19020 & \checkmark & \checkmark & \checkmark & \checkmark \\
dicompyler \citep{Panchal2010} & 2009 & 2020-01 &  & 2 & 15941 & \checkmark & \checkmark &  &  \\
Fiji \citep{Schindelin2012} & 2011 & 2020-08 & \checkmark & 55 & 10833 & \checkmark & \checkmark & \checkmark &  \\
ImageJ \citep{Rueden2017} & 1997 & 2020-08 & \checkmark & 18 & 9681 & \checkmark & \checkmark & \checkmark &  \\
MRIcroGL \citep{Rorden2021} & 2015 & 2020-08 &  & 2 & 8493 & \checkmark & \checkmark & \checkmark &  \\
DicomBrowser \citep{Archie2012} & 2012 & 2020-08 &  & 3 & 5505 & \checkmark & \checkmark & \checkmark &  \\ \bottomrule
\end{tabular}
\caption{Final software list (sorted in descending order of the number of Lines
Of Code (LOC))}
\label{tab_final_list}
\end{table}

\begin{figure}[!ht]
\centering
\includegraphics[scale=0.5]{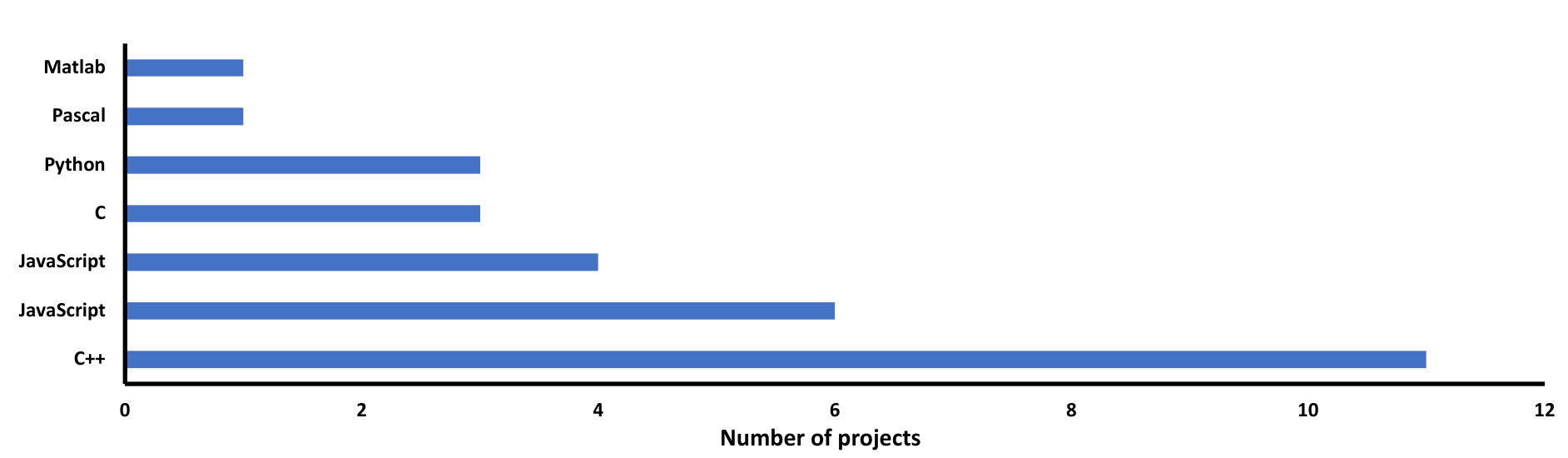}
\centering
\caption{\label{fig_language}Primary languages versus number of projects}
\end{figure}

Figure \ref{fig_language} shows the primary languages versus the number of
projects using them.  The primary language is the language used for the majority
of the project's code; in most cases projects also use other languages.  The
most popular language is \CC, with almost 40\% of projects (11 of 29).  The two
least popular choices are Pascal and Matlab, with around 3\% of projects each
(1 of 29).

\subsection{Installability} \label{sec_result_installability}

\begin{figure}[!ht]
\includegraphics[scale=0.47]{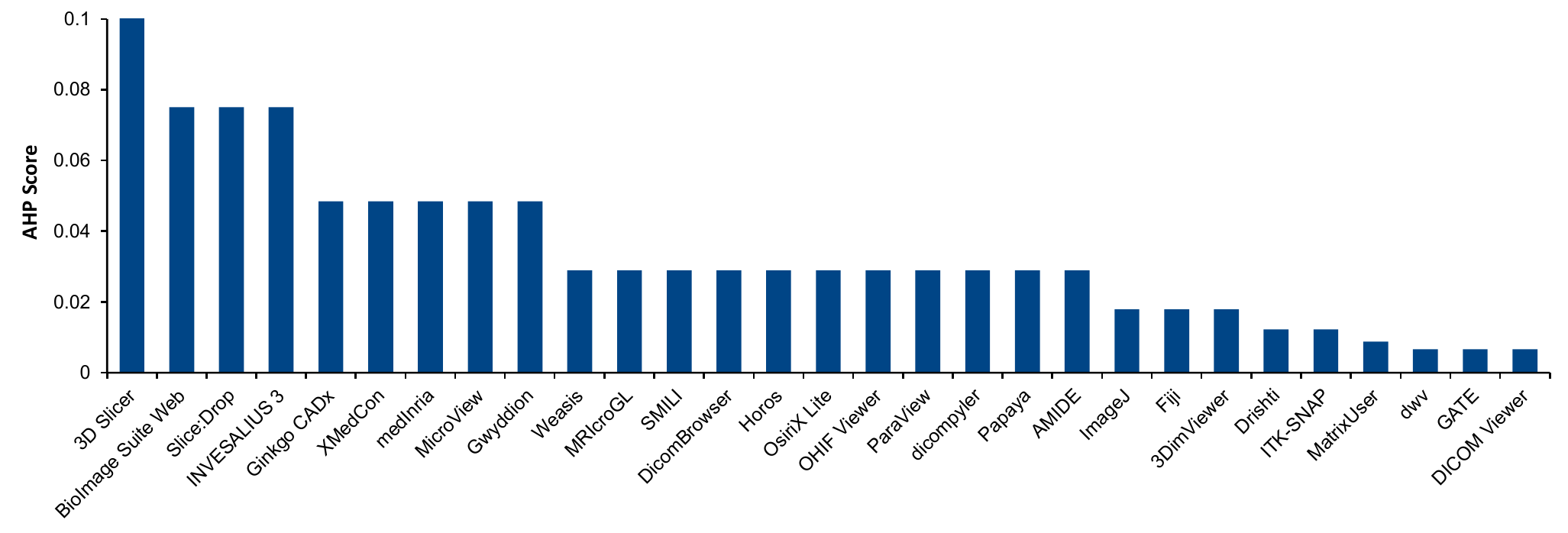}
\caption{AHP installability scores}
\label{fg_installability_scores}
\end{figure}

Figure \ref{fg_installability_scores} lists the installability scores.  We found
installation instructions for 16 projects. Among the ones without instructions,
\textit{BioImage Suite Web} and \textit{Slice:Drop} do not need installation,
since they are web applications. Installing 10 of the projects required extra
dependencies. Five of them are web applications (as shown in
Table~\ref{tab_final_list}) and depend on a browser; \textit{dwv}, \textit{OHIF
Viewer}, and \textit{GATE} needs extra dependencies to build; \textit{ImageJ}
and	\textit{Fiji} need an unzip tool; \textit{MatrixUser} is based on Matlab;
\textit{DICOM Viewer} needs to work on a Nextcloud platform.

\textit{3D Slicer} has the highest score because it had easy to follow
installation instructions, and an automated, fast, frustration-free installation
process. The installer added all dependencies automatically and no errors
occurred during the installation and uninstallation steps. Many other software
packages also had installation instructions and automated installers.  We had no
trouble installing the following packages: \textit{INVESALIUS 3},
\textit{Gwyddion}, \textit{XMedCon}, and \textit{MicroView}. We determined their
scores based on the understandability of the instructions, installation steps,
and user experience. Since \textit{BioImage Suite Web} and \textit{Slice:Drop}
needed no installation, we gave them high scores. \textit{BioImage Suite Web}
also provided an option to download cache for offline usage, which was easy to
apply.

\textit{GATE}, \textit{dwv}, and \textit{DICOM Viewer} showed severe
installation problems. We were not able to install them, even after a reasonable
amount of time (2 hours).  For \textit{dwv} and \textit{GATE} we failed to build
from the source code, but we were able to proceed with measuring other qualities
using a deployed on-line version for \textit{dwv}, and a VM version for
\textit{GATE}. For \textit{DICOM Viewer} we could not install the NextCloud
dependency, and we did not have another option for running the software.
Therefore, for \textit{DICOM Viewer} we could not measure reliability or
robustness.  The other seven qualities could be measured, since they do not
require installation.

\textit{MatrixUser} has a lower score because it depends on Matlab. We assessed
the score from the point of view of a user that would have to install Matlab and
acquire a license.  Of course, for users that already work within Matlab, the
installability score should be higher.

\subsection{Correctness \& Verifiability} \label{sec_result_correctness_verifiability}

\begin{figure}[!ht]
\includegraphics[scale=0.47]{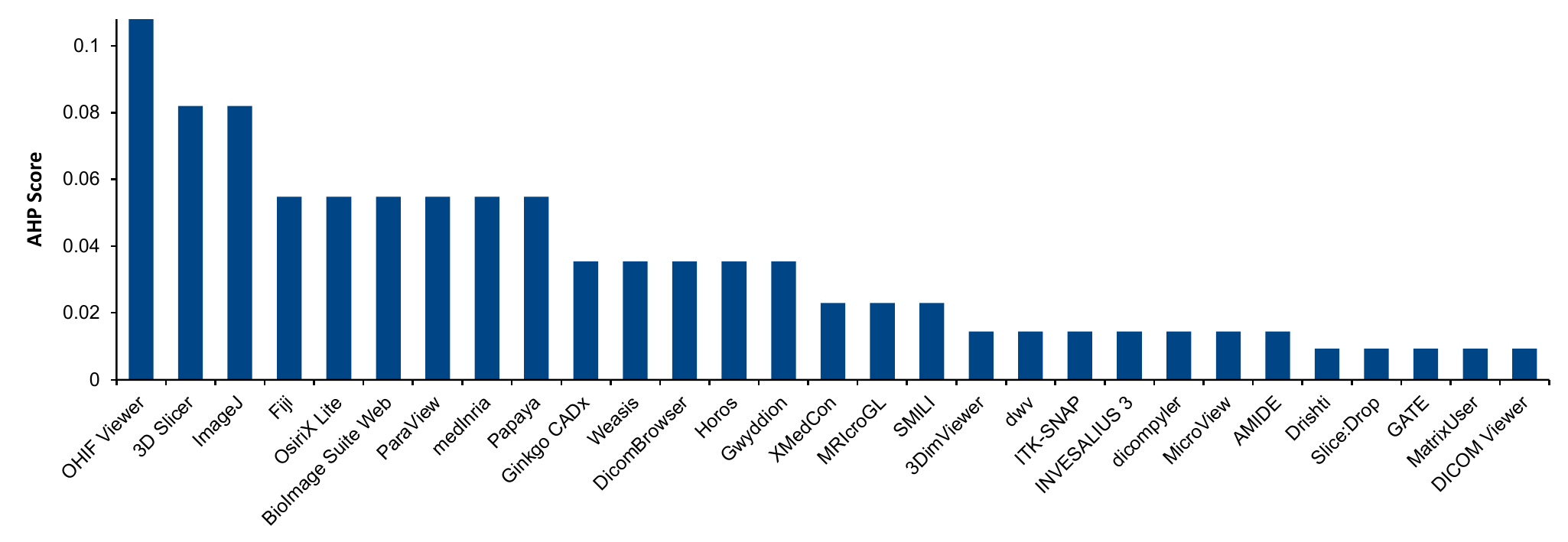}
\caption{AHP correctness \& verifiability scores}
\label{fg_correctness_verifiability_scores}
\end{figure}

Figure~\ref{fg_correctness_verifiability_scores} shows the scores of correctness
and verifiability. Generally speaking, the packages with higher scores adopted
more techniques to improve correctness, and had better documentation for us to
verify against.  For instance, we looked for evidence of unit testing, since it
benefits most parts of the software's life cycle, such as designing, coding,
debugging, and optimization \citep{Hamill2004}.  We only found evidence of unit
testing for about half of the projects. We identified five projects using CI/CD
tools: \textit{3D Slicer}, \textit{ImageJ}, \textit{Fiji}, \textit{dwv}, and
\textit{OHIF Viewer}.

Even for some projects with well-organized documentation, requirements
specifications and theory manuals were still missing.  We could not identify
theory manuals for all projects, and we did not find requirements specifications
for most projects. The only requirements-related document we found was a road
map of \textit{3D Slicer}, which contained design requirements for upcoming
changes.

\subsection{Surface Reliability} \label{sec_result_reliability}

\begin{figure}[!ht]
\includegraphics[scale=0.47]{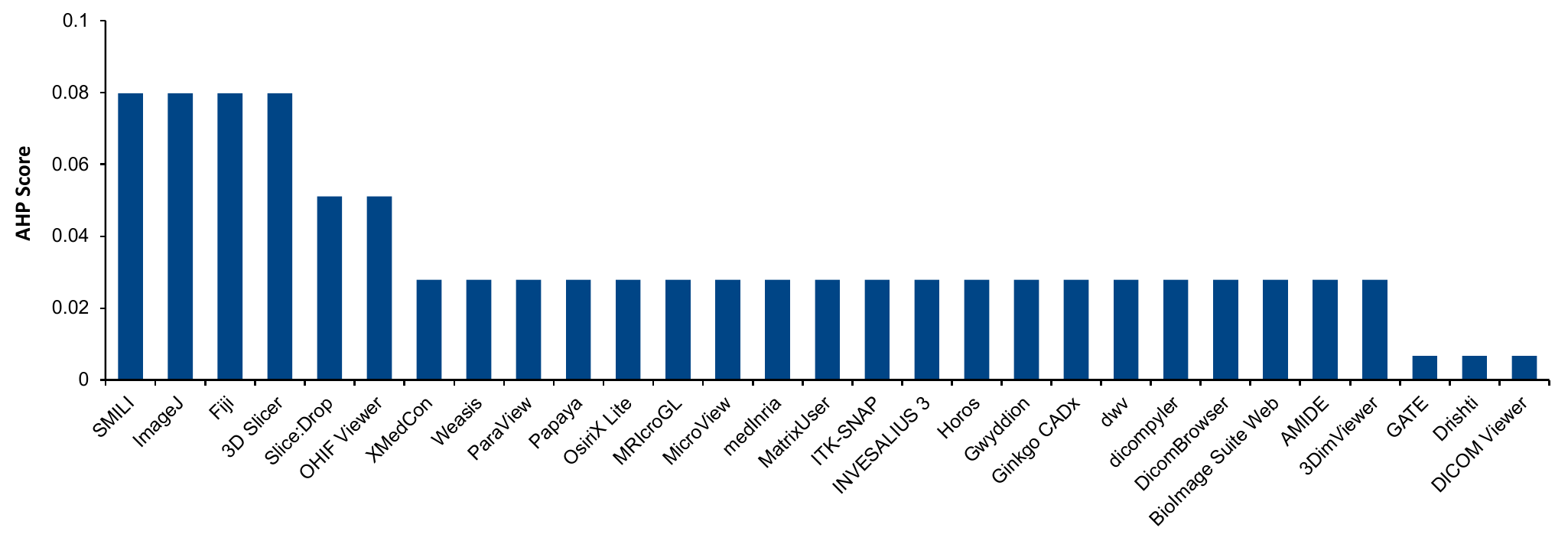}
\caption{AHP surface reliability scores}
\label{fg_reliability_scores}
\end{figure}

Figure~\ref{fg_reliability_scores} shows the AHP results.  As shown in
Section~\ref{sec_result_installability}, most of the software products did not
``break'' during installation, or did not need installation; \textit{dwv} and
\textit{GATE} broke in the building stage, and the processes were not
recoverable; we could not install the dependency for \textit{DICOM Viewer}. Of
the seven software packages with a getting started tutorial and operation steps
in the tutorial, most showed no error when we followed the steps. However,
\textit{GATE} could not open macro files and became unresponsive several times,
without any descriptive error message. When assessing robustness
(Section~\ref{sec_result_robustness}), we found that \textit{Drishti} crashed
when loading damaged image files, without showing any descriptive error message.
We did not find any problems with the on-line version of
\textit{dwv}.

\subsection{Surface Robustness} \label{sec_result_robustness}

\begin{figure}[!ht]
\includegraphics[scale=0.47]{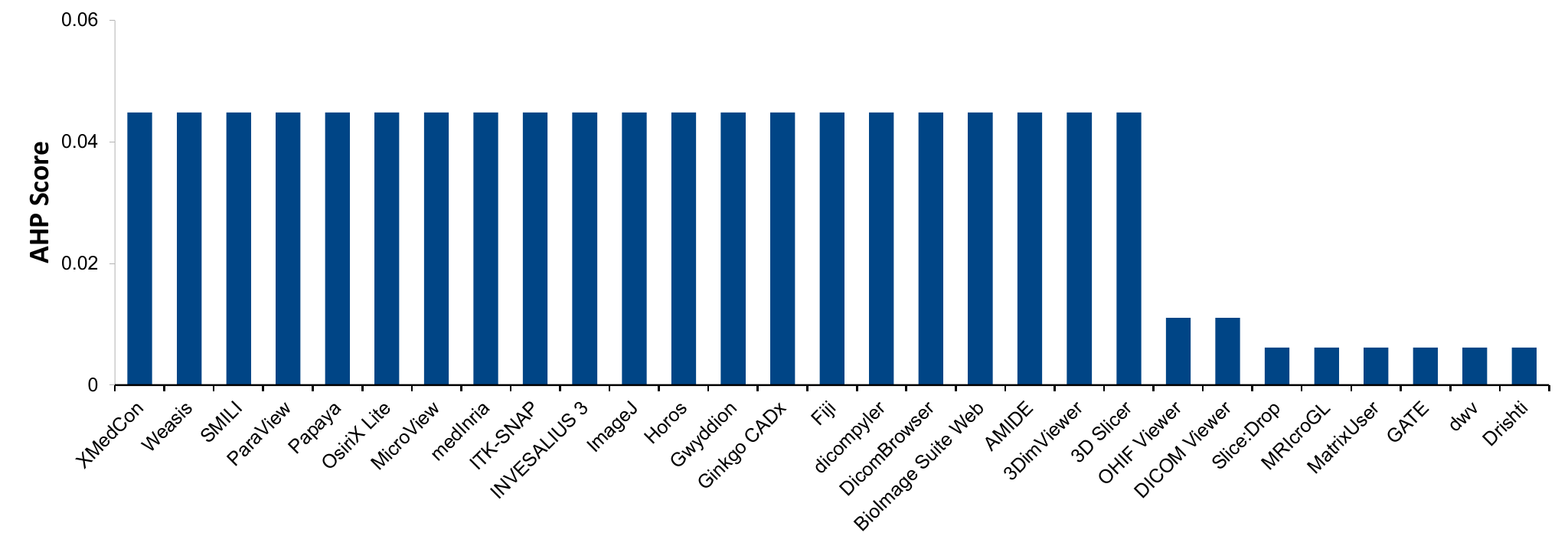}
\caption{AHP surface robustness scores}
\label{fg_robustness_scores}
\end{figure}

Figure \ref{fg_robustness_scores} presents the scores for surface robustness.
The packages with higher scores elegantly handled unexpected/unanticipated
inputs, typically showing a clear error message. We may have underestimated the
score of \textit{OHIF Viewer}, since we needed further customization to load
data.

Digital Imaging and Communications in Medicine (DICOM) ``defines the formats for
medical images that can be exchanged with the data and quality necessary for
clinical use'' \citep{MITA2021}. According to their documentation, all 29
software packages should support the DICOM standard. To test robustness, we
prepared two types of image files: correct and incorrect formats (with the
incorrect format created by relabelled a text file to have the ``.dcm''
extension).  All software packages loaded the correct format image, except for
\textit{GATE}, which failed for unknown reasons.  For the broken format,
\textit{MatrixUser}, \textit{dwv}, and \textit{Slice:Drop} ignored the incorrect
format of the file and loaded it regardless. They did not show any error message
and displayed a blank image. \textit{MRIcroGL} behaved similarly except that it
showed a meaningless image. \textit{Drishti} successfully detected the broken
format of the file, but the software crashed as a result.

\subsection{Surface Usability} \label{sec_result_usability}

\begin{figure}[!ht]
\includegraphics[scale=0.47]{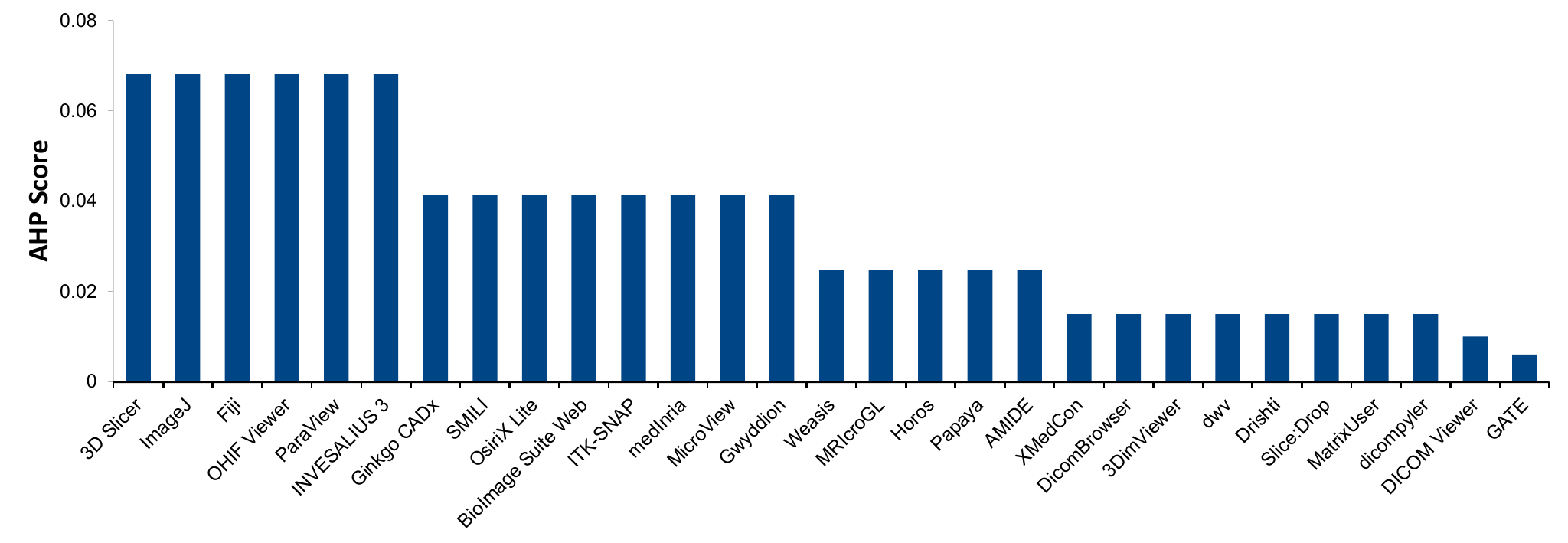}
\caption{AHP surface usability scores}
\label{fg_usability_scores}
\end{figure}

Figure~\ref{fg_usability_scores} shows the AHP scores for surface usability. The
software with higher scores usually provided both comprehensive documented
guidance and a good user experience. \textit{INVESALIUS 3} provided an excellent
example of a detailed and precise user manual. \textit{GATE} also provided
numerous documents, but unfortunately we had difficulty understanding and using
them. We found getting started tutorials for only 11 projects, but a user manual
for 22 projects. \textit{MRIcroGL} was the only project that explicitly
documented expected user characteristics.

\subsection{Maintainability} \label{sec_score_maintainability}

\begin{figure}[!ht]
\includegraphics[scale=0.47]{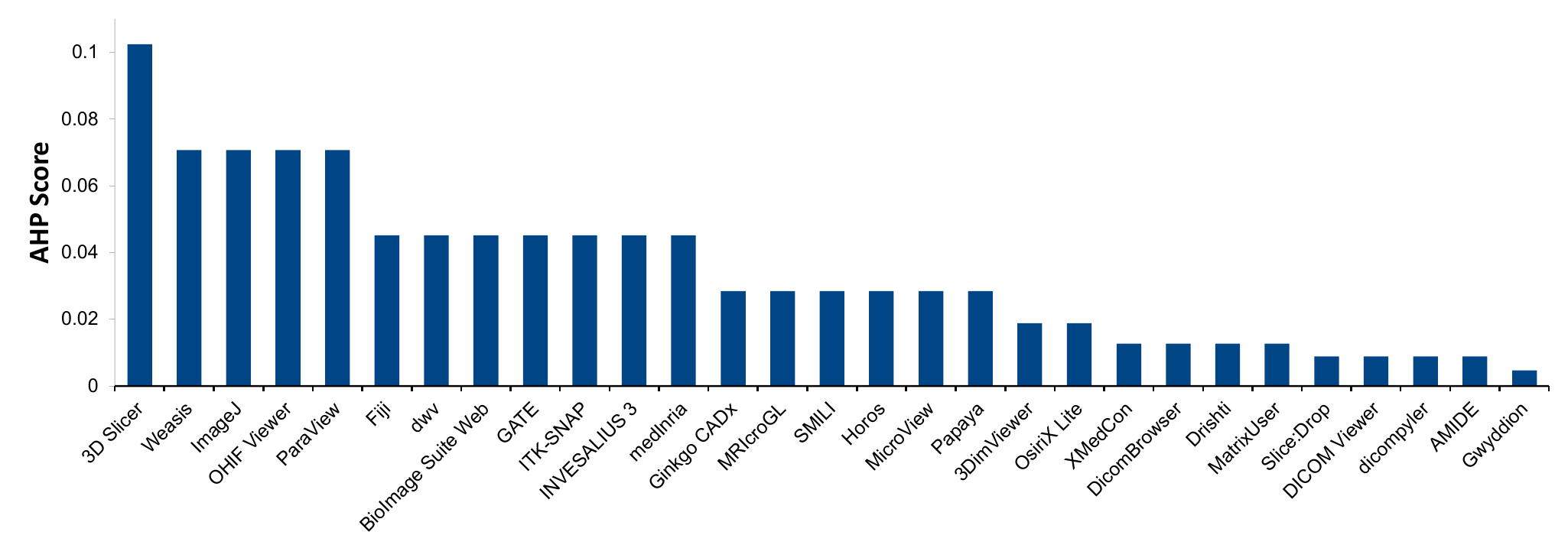}
\caption{AHP maintainability scores}
\label{fg_maintainability_scores}
\end{figure}

Figure~\ref{fg_maintainability_scores} shows the ranking results for
maintainability. We gave \textit{3D Slicer} the highest score because we found
it to have the most comprehensive artifacts. For example, as far as we could
find, only a few of the 29 projects had a product, developer's manual, or API
documentation, and only \textit{3D Slicer}, \textit{ImageJ}, \textit{Fiji}
included all three documents. Moreover, \textit{3D Slicer} has a much higher
percentage of closed issues (92\%) compared to \textit{ImageJ} (52\%) and
\textit{Fiji} (64\%). Table~\ref{tab_maintainability_docs} shows which
projects had these documents, in descending order of their maintainability
scores. 

\begin{table}[!ht]
\centering
\begin{tabular}{lccc}
\toprule
\multicolumn{1}{c}{Software} & Prod.\ Roadmap & Dev.\ Manual & API Doc. \\ 
\midrule
3D Slicer & \checkmark & \checkmark & \checkmark \\
ImageJ & \checkmark & \checkmark & \checkmark \\
Weasis &  & \checkmark &  \\
OHIF Viewer &  & \checkmark & \checkmark \\
Fiji & \checkmark & \checkmark & \checkmark \\
ParaView & \checkmark &  &  \\
SMILI &  &  & \checkmark \\
medInria &  & \checkmark &  \\
INVESALIUS 3 & \checkmark &  &  \\
dwv &  &  & \checkmark \\
BioImage Suite Web &  & \checkmark &  \\
Gwyddion &  & \checkmark & \checkmark \\ 
\bottomrule
\end{tabular}
\caption{Software with the maintainability documents (listed in descending order of 
maintainability score)}
\label{tab_maintainability_docs}
\end{table}

Twenty-seven of the 29 projects used git as the version control tool, with 24 of these
using GitHub. \textit{AMIDE} used Mercurial and \textit{Gwyddion} used
Subversion. \textit{XMedCon}, \textit{AMIDE}, and \textit{Gwyddion} used
SourceForge. \textit{DicomBrowser} and \textit{3DimViewer} used BitBucket. 

\subsection{Reusability} \label{sec_result_reusability}

\begin{figure}[!ht]
\includegraphics[scale=0.47]{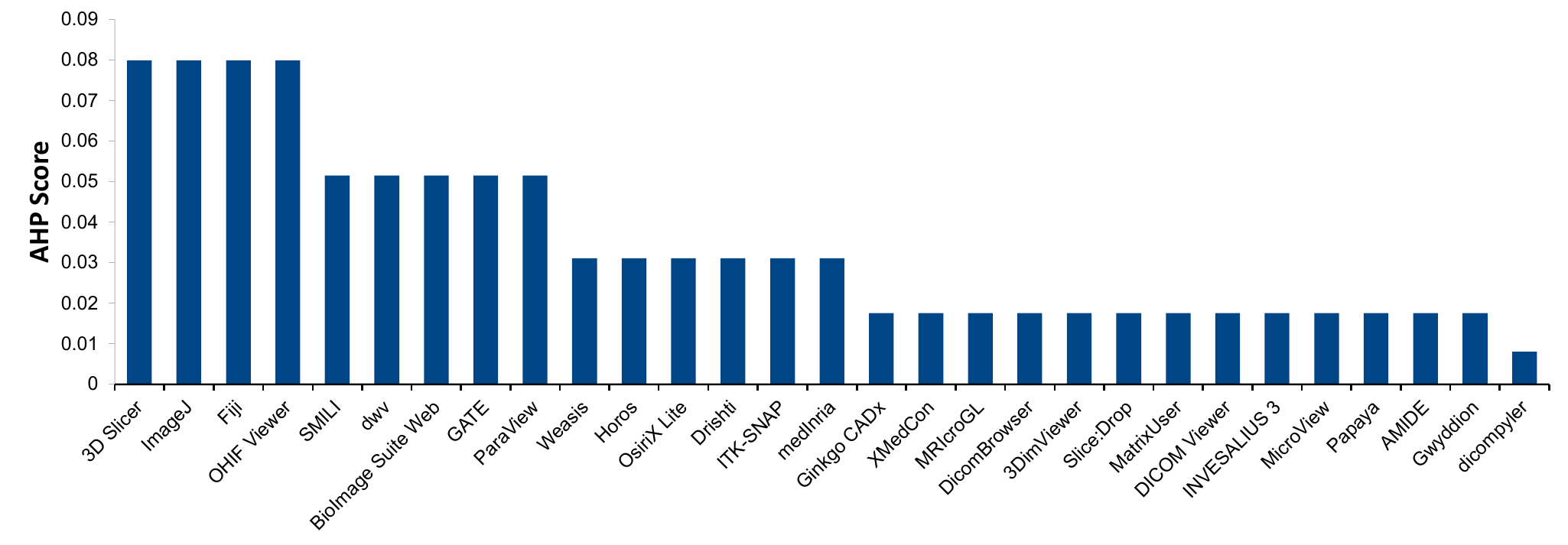}
\caption{AHP reusability scores}
\label{fg_reusability_scores}
\end{figure}

Figure~\ref{fg_reusability_scores} shows the AHP results for reusability. As
described in Section~\ref{sec_grading_software}, we gave higher scores to the
projects with API documentation. As shown in
Table~\ref{tab_maintainability_docs}, seven projects had API documents. We also
assumed that projects with more code files and less LOC per code file are more
reusable. Table \ref{tab_loc_per_file} shows the number of text-based files by
project, which we used to approximate the number of code files. The table also
lists the total number of lines (including comments and blanks), LOC, and
average LOC per file. We arranged the items in descending order of their
reusability scores.

\begin{table}[!ht]
\centering
\begin{tabular}{lllll}
\toprule
\multirow{2}{*}{Software} & \multirow{2}{*}{Text Files} & \multirow{2}{*}{Total Lines} & \multirow{2}{*}{LOC} & \multirow{2}{*}{LOC/file} \\
 &  &  &  &  \\ 
\midrule
OHIF Viewer & 1162 & 86306 & 63951 & 55 \\
3D Slicer & 3386 & 709143 & 501451 & 148 \\
Gwyddion & 2060 & 787966 & 643427 & 312 \\
ParaView & 5556 & 1276863 & 886326 & 160 \\
OsiriX Lite & 2270 & 873025 & 544304 & 240 \\
Horos & 2346 & 912496 & 561617 & 239 \\
medInria & 1678 & 214607 & 148924 & 89 \\
Weasis & 1027 & 156551 & 123272 & 120 \\
BioImage Suite Web & 931 & 203810 & 139699 & 150 \\
GATE & 1720 & 311703 & 207122 & 120 \\
Ginkgo CADx & 974 & 361207 & 257144 & 264 \\
SMILI & 275 & 90146 & 62626 & 228 \\
Fiji & 136 & 13764 & 10833 & 80 \\
Drishti & 757 & 345225 & 268168 & 354 \\
ITK-SNAP & 677 & 139880 & 88530 & 131 \\
3DimViewer & 730 & 240627 & 178065 & 244 \\
DICOM Viewer & 302 & 34701 & 30761 & 102 \\
ImageJ & 40 & 10740 & 9681 & 242 \\
dwv & 188 & 71099 & 47815 & 254 \\
MatrixUser & 216 & 31336 & 23121 & 107 \\
INVESALIUS 3 & 156 & 59328 & 48605 & 312 \\
AMIDE & 183 & 139658 & 102827 & 562 \\
Papaya & 110 & 95594 & 71831 & 653 \\
MicroView & 137 & 36173 & 27470 & 201 \\
XMedCon & 202 & 129991 & 96767 & 479 \\
MRIcroGL & 97 & 50445 & 8493 & 88 \\
Slice:Drop & 77 & 25720 & 19020 & 247 \\
DicomBrowser & 54 & 7375 & 5505 & 102 \\
dicompyler & 48 & 19201 & 15941 & 332 \\ 
\bottomrule
\end{tabular}
\caption{Number of files and lines (sorted in descending order of reusability
scores)}
\label{tab_loc_per_file}
\end{table}

\subsection{Surface Understandability} \label{sec_result_understandability}

\begin{figure}[!ht]
\includegraphics[scale=0.47]{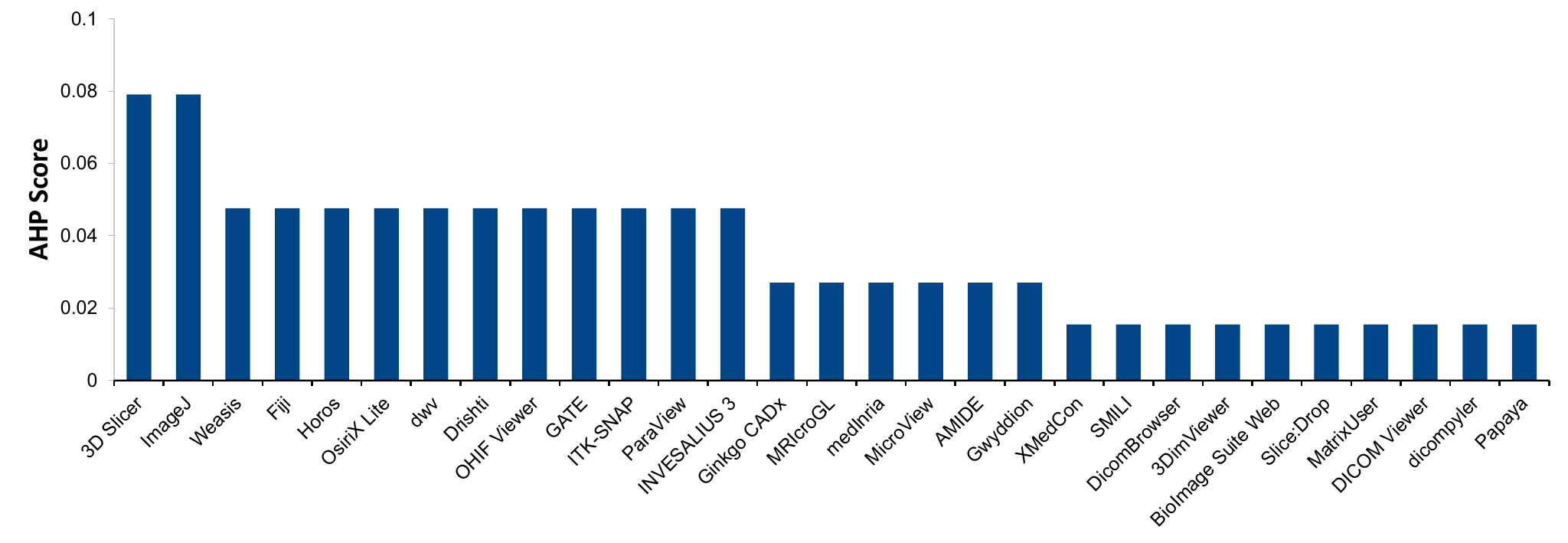}
\caption{AHP surface understandability scores}
\label{fg_surface_understandability_scores}
\end{figure}

Figure~\ref{fg_surface_understandability_scores} shows the scores for surface
understandability. All projects had a consistent coding style with parameters in
the same order for all functions; modularized code; and, clear comments, indicating
what is done, not how. However, we only found explicit identification of a
coding standard for 3 out of the 29: \textit{3D Slicer}, \textit{Weasis}, and
\textit{ImageJ}. We also found hard-coded constants (rather than symbolic
constants) in \textit{medInria}, \textit{dicompyler}, \textit{MicroView}, and
\textit{Papaya}. We did not find any reference to the algorithms used in
projects \textit{XMedCon}, \textit{DicomBrowser}, \textit{3DimViewer},
\textit{BioImage Suite Web}, \textit{Slice:Drop}, \textit{MatrixUser},
\textit{DICOM Viewer}, \textit{dicompyler}, and \textit{Papaya}. 

\subsection{Visibility/Transparency} \label{sec_result_visibility_transparency}

\begin{figure}[!ht]
\includegraphics[scale=0.47]{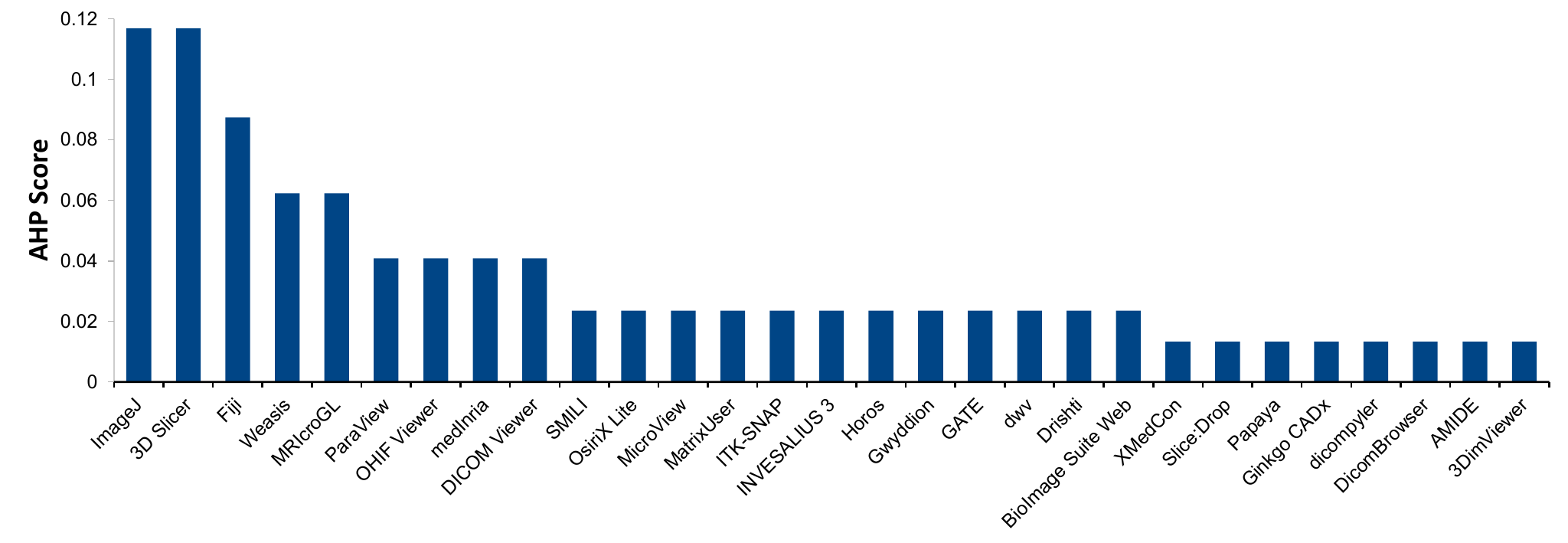}
\caption{AHP visibility/transparency scores}
\label{fg_visibility_transparency_scores}
\end{figure}

Figure~\ref{fg_visibility_transparency_scores} shows the AHP scores for
visibility/transparency. Generally speaking, the teams that actively documented
their development process and plans scored higher.
Table~\ref{tab_Visibility/Transparency_docs} shows the projects that had
documents for the development process, project status, development environment,
and release notes, in descending order of their visibility/transparency
scores.

\begin{table}[!ht]
\centering
\begin{tabular}{lllll}
\toprule
Software & Dev.\ Process & Proj.\ Status & Dev.\ Env. & Rls.\ Notes \\ 
\midrule
3D Slicer & \checkmark & \checkmark & \checkmark & \checkmark \\
ImageJ & \checkmark & \checkmark & \checkmark & \checkmark \\
Fiji & \checkmark & \checkmark & \checkmark &  \\
MRIcroGL &  &  &  & \checkmark \\
Weasis &  &  & \checkmark & \checkmark \\
ParaView &  & \checkmark &  &  \\
OHIF Viewer &  &  & \checkmark & \checkmark \\
DICOM Viewer &  &  & \checkmark & \checkmark \\
medInria &  &  & \checkmark & \checkmark \\
SMILI &  &  &  & \checkmark \\
Drishti &  &  &  & \checkmark \\
INVESALIUS 3 &  &  &  & \checkmark \\
OsiriX Lite &  &  &  & \checkmark \\
GATE &  &  &  & \checkmark \\
MicroView &  &  &  & \checkmark \\
MatrixUser &  &  &  & \checkmark \\
BioImage Suite Web &  &  & \checkmark &  \\
ITK-SNAP &  &  &  & \checkmark \\
Horos &  &  &  & \checkmark \\
dwv &  &  &  & \checkmark \\
Gwyddion &  &  &  & \checkmark \\ 
\bottomrule
\end{tabular}
\caption{Software with visibility/transparency related documents (listed in
descending order of visibility/transparency score)}
\label{tab_Visibility/Transparency_docs}
\end{table}

\subsection{Overall Scores} \label{Sec_OverallQ}

\begin{figure}[!ht]
\includegraphics[scale=0.47]{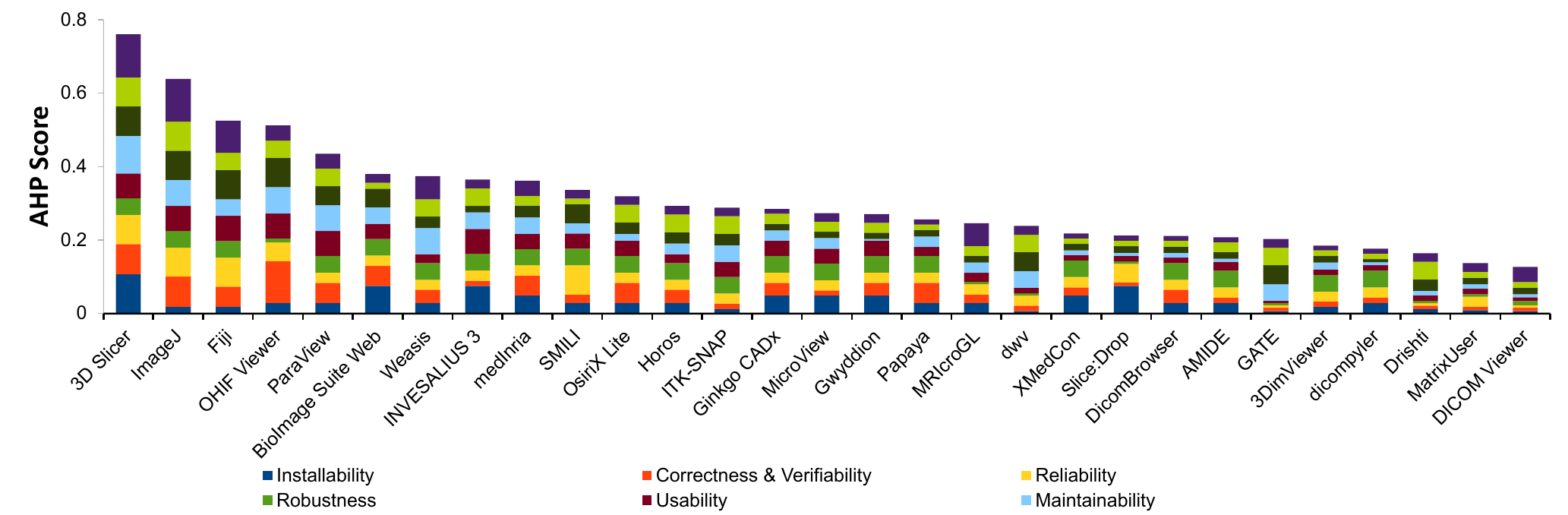}
\caption{Overall AHP scores with an equal weighting for all 9 software qualities}

\label{fg_overall_scores}
\end{figure}

As described in Section~\ref{sec_AHP}, for our AHP measurements, we have nine
criteria (qualities) and 29 alternatives (software packages). In the absence of
a specific real world context, we assumed all nine qualities are equally
important. Figure~\ref{fg_overall_scores} shows the overall scores in descending
order. Since we produced the scores from the AHP process, the total sum of the
29 scores is precisely 1.0.

The top four software products \textit{3D Slicer}, \textit{ImageJ},
\textit{Fiji}, and \textit{OHIF Viewer} have higher scores in most criteria.
\textit{3D Slicer} has a score in the top two for all qualities; \textit{ImageJ}
ranks near the top for all qualities, except for correctness \& verifiability.
\textit{OHIF Viewer} and \textit{Fiji} have similar overall scores, with
\textit{Fiji} doing better in installability and \textit{OHIF Viewer} doing
better in correctness \& verifiability.  Given the installation problems, we may
have underestimated the scores on reliability and robustness for \textit{DICOM
Viewer}, but we compared it equally for the other seven qualities.

\section{Comparison to Community Ranking} \label{Sec_VsCommunityRanking}

To address~\rqref{RQ_CompareHQ2Popular} about how our ranking compares to the
popularity of projects as judged by the scientific community, we make two
comparisons:
\begin{itemize}
\item A comparison of our ranking (from Section~\ref{ch_results}) with the
community ratings on GitHub, as shown by GitHub stars, number of forks, and
number of people watching the projects; and,
\item A comparison of top-rated software from our methodology with the top
recommendations from our domain experts (as mentioned in
Section~\ref{sec_software_selection}).
\end{itemize}

Table~\ref{tab_ranking_vs_GitHub} shows our ranking of the 29 MI projects, and
their GitHub metrics, if applicable. As mentioned in
Section~\ref{sec_score_maintainability}, 24 projects used GitHub. Since GitHub
repositories have different creation dates, we collect the number of months each
stayed on GitHub, and calculate the average number of new stars, people
watching, and forks per 12 months. Section~\ref{sec_grading_software} describes
the method of getting the creation date.  The items in
Table~\ref{tab_ranking_vs_GitHub} are listed in descending order of the average
number of new stars per year.  The non-GitHub items are listed in the order of
our ranking.  We collected all GitHub statistics in July 2021.  

Generally speaking, most of the top-ranking MI software projects also received
greater attention and popularity on GitHub. Between our ranking and the GitHub
stars-per-year ranking, four of the top five software projects appear in both
lists. Our top five packages are scattered among the first eight positions on the
GitHub list. However, as discussed below there are discrepancies between the two
lists.

In some cases projects are popular in the community, but were assigned a low
rank by our methodology.  This is the case for \textit{dwv}. The reason for the
low ranking is that, as mentioned in Section~\ref{sec_result_installability}, we
failed to build it locally, and used the test version on its websites for the
measurements. We followed the instructions and tried to run the command ``yarn
run test'' locally, which did not work. In addition, the test version did not
detect a broken DICOM file and displayed a blank image as described in
Section~\ref{sec_result_robustness}. We might underestimate the scores for
\textit{dwv} due to uncommon technical issues. 

We also ranked \textit{DICOM Viewer} much lower than its popularity. As
mentioned in Section~\ref{sec_result_installability}, it depended on the
NextCloud platform that we could not successfully install. Thus, we might
underestimate the scores of its surface reliability and surface robustness. 

Further reason for discrepancies between our ranking and the community's ranking
is that we weighted all qualities equally. This is not likely how users
implicitly rank the different qualities. As a result, some projects with high
community popularity may have scored lower with our method because of a
relatively higher (compared to the scientific community's implicit ranking)
weighting of the poor scores for some qualities. A further explanation for
discrepancies between our measures and the star measures may also be due to
inaccuracy with using stars to approximate popularity.  Stars are not an ideal
measure because stars represent the community's feeling in the past more than
they measure current preferences \citep{Szulik2017}.  The issue with stars is
that they tend only to be added, not removed.  A final reason for
inconsistencies between our ranking and the community's ranking is that, as for
consumer products, more factors influence popularity than just quality.

\begingroup
\renewcommand{\arraystretch}{0.85}
\begin{table}[!ht]
\centering
\begin{tabular}{llllll}
\toprule
Software & Comm.\ Rank & Our Rank & Stars/yr & Watches/yr & Forks/yr \\ 
\midrule
3D Slicer & 1 & 1 & 284 & 19 & 128 \\
OHIF Viewer & 2 & 4 & 277 & 19 & 224 \\
dwv & 3 & 19 & 124 & 12 & 51 \\
ImageJ & 4 & 2 & 84 & 9 & 30 \\
ParaView & 5 & 5 & 67 & 7 & 28 \\
Horos & 6 & 12 & 49 & 9 & 18 \\
Papaya & 7 & 17 & 45 & 5 & 20 \\
Fiji & 8 & 3 & 44 & 5 & 21 \\
DICOM Viewer & 9 & 29 & 43 & 6 & 9 \\
INVESALIUS 3 & 10 & 8 & 40 & 4 & 17 \\
Weasis & 11 & 7 & 36 & 5 & 19 \\
dicompyler & 12 & 26 & 35 & 5 & 14 \\
OsiriX Lite & 13 & 11 & 34 & 9 & 24 \\
MRIcroGL & 14 & 18 & 24 & 3 & 3 \\
GATE & 15 & 24 & 19 & 6 & 26 \\
Ginkgo CADx & 16 & 14 & 19 & 4 & 6 \\
BioImage Suite Web & 17 & 6 & 18 & 5 & 7 \\
Drishti & 18 & 27 & 16 & 4 & 4 \\
Slice:Drop & 19 & 21 & 10 & 2 & 5 \\
ITK-SNAP & 20 & 13 & 9 & 1 & 4 \\
medInria & 21 & 9 & 7 & 3 & 6 \\
SMILI & 22 & 10 & 3 & 1 & 2 \\
MatrixUser & 23 & 28 & 2 & 0 & 0 \\
MicroView & 24 & 15 & 1 & 1 & 1 \\
Gwyddion & 25 & 16 & n/a & n/a & n/a \\
XMedCon & 26 & 20 & n/a & n/a & n/a \\
DicomBrowser & 27 & 22 & n/a & n/a & n/a \\
AMIDE & 28 & 23 & n/a & n/a & n/a \\
3DimViewer & 29 & 25 & n/a & n/a & n/a \\ 
\bottomrule
\end{tabular}
\caption{Software ranking by our methodology versus the community (Comm.)\
ranking using GitHub metrics (Sorted in descending order of community
popularity, as estimated by the number of new stars per year)}
\label{tab_ranking_vs_GitHub}
\end{table}
\endgroup

As shown in Section~\ref{sec_software_selection}, our domain experts recommended
a list of top software with 12 software products.  All the top 4 entries from
the Domain Expert's list are among the top 12 ranked by our methodology. Three
of the top four on both lists are the same: \textit{3D Slicer}, \textit{ImageJ},
and \textit{Fiji}. \textit{3D Slicer} is top project by both rankings (and by
the GitHub stars measure as well).  The Domain Expert ranked \textit{Horos} as
their second choice, while we ranked it twelfth.  Our third ranked project,
\textit{OHIF Viewer} was not listed by the Domain Expert.  Neither were the
software packages that we ranked from fifth to eleventh (\textit{ParaView},
\textit{Weasis}, \textit{medInria}, \textit{BioImage Suite Web}, \textit{OsiriX
Lite}, \textit{INVESALIUS}, and \textit{Gwyddion}).  The software mentioned by
the Domain Expert that we did not rank were the six recommended packages that
did not have visualization as the primary function (as discussed in
Section~\ref{sec_software_selection}).  The differences between the list
recommended by our methodology and the Domain Expert are not surprising.  As
mentioned above, our methodology weights all qualities equally, but that may not
be the case for the Domain Expert's impressions.  Moreover, although the Domain
Expert has significant experience with MI software, they have not used all 29
packages that were measured.

Although our ranking and the estimate of the community's ranking are not perfect
measures, they do suggest a correlation between best practices and popularity.
We do not know which comes first, the use of best practices or popularity, but
we do know that the top ranked packages tend to incorporate best practices. The
next sections will explore how the practices of the MI community compare to the
broader research software community. We will also investigate the practices from
the top projects that others within the MI community, and within the broader
research software community, can potentially adopt.

\section{Comparison Between MI and Research Software for Artifacts}
\label{Sec_CompareArtifacts}

As part of filling in the measurement template (from
Section~\ref{sec_grading_software}), we summarized the artifacts observed in
each MI package. Table~\ref{artifactspresent} groups the artifacts by frequency
into categories of common (20 to 29 ($>$67\%) packages), uncommon (10 to 19
(33-67\%) packages), and rare (1 to 9 ($<$33\%) packages). \citet{Dong2021-Data}
summarizes the full measurements.  Tables~\ref{tab_maintainability_docs}
and~\ref{tab_Visibility/Transparency_docs} show the details on which projects
use which types of artifacts for documents related to maintainability and
visibility, respectively.

\begin{table}[ht!]
    \begin{center}
    \begin{tabular}{ p{4.6 cm} p{5.6 cm} p{5 cm}}
    \toprule
    Common & Uncommon & Rare \\
    \midrule
    README (29) & Build scripts (18) & Getting Started (9)\\
    Version control (29) & Tutorials (18) & Developer's manual (8)\\
    License (28) & Installation guide (16) & Contributing (8)\\
    Issue tracker (28) & Test cases (15) & API documentation (7)\\
    User manual (22) & Authors (14) & Dependency list (7)\\
    Release info. (22) & Frequently Asked Questions (FAQ) (14) & Troubleshooting guide (6)\\
     & Acknowledgements (12) & Product roadmap (5)\\
     & Changelog (12) & Design documentation (5)\\
     & Citation (11) & Code style guide (3)\\
     & & Code of conduct (1)\\
     & & Requirements (1)\\
    \bottomrule
    \end{tabular}
    \caption{Artifacts Present in MI Packages, Classified by Frequency (The number 
    in brackets is the number of occurrences)}
    \label{artifactspresent}
    \end{center}
\end{table}

We answer~\rqref{RQ_CompareArtifacts} by comparing the artifacts that we
observed in MI repositories to those observed and recommended for research
software in general. Our comparison may point out areas where some MI software
packages fall short of current best practices. This is not intended to be a
criticism of any existing packages, especially since in practice not every
project needs to achieve the highest possible quality. However, rather than
delve into the nuances of which software can justify compromising which
practices we will write our comparison under the ideal assumption that every
project has sufficient resources to match best practices.
    
Table~\ref{Tbl_Guidelines} (based on data from \citep{SmithAndMichalski2022})
shows that MI artifacts generally match the recommendations found in nine
current research software development guidelines:
\begin{itemize}
\item United States Geological Survey Software Planning Checklist
\citep{USGS2019},
\item DLR (German Aerospace Centre) Software Engineering Guidelines
\citep{TobiasEtAl2018}, 
\item Scottish Covid-19 Response Consortium Software Checklist
\citep{BrettEtAl2021},
\item Good Enough Practices in Scientific Computing \citep{WilsonEtAl2016},
\item xSDK (Extreme-scale Scientific Software Development Kit) Community Package
Policies \citep{SmithAndRoscoe2018},
\item Trilinos Developers Guide \citep{HerouxEtAl2008},
\item EURISE (European Research Infrastructure Software Engineers') Network
Technical Reference \citep{ThielEtAl2020},
\item CLARIAH (Common Lab Research Infrastructure for the Arts and Humanities)
Guidelines for Software Quality \citep{vanGompelEtAl2016}, and
\item A Set of Common Software Quality Assurance Baseline Criteria for Research
Projects \citep{OrvizEtAl2017}.
\end{itemize}

In Table~\ref{Tbl_Guidelines} each row corresponds to an artifact.  For a given
row, a checkmark in one of the columns means that the corresponding guideline
recommends this artifact.  The last column shows whether the artifact appears in
the measured set of MI software, either not at all (blank), commonly (C),
uncommonly (U) or rarely (R).  We did our best to interpret the meaning of each
artifact consistently between guidelines and specific MI software, but the
terminology and the contents of artifacts are not standardized.  The challenge
even exists for the ubiquitous README file.  As illustrated by
\citet{PranaEtAl2018}, the content of README files shows significant variation
between projects.  Although some content is reasonably consistent, with 97\% of
README files contain at least one section describing the `What' of the
repository and 89\% offering some `How' content, other categories are more
variable.  For instance, information on `Contribution', `Why', and `Who',
appear in 28\%, 26\% and 53\% of the analyzed files, respectively
\citep{PranaEtAl2018}.  

The frequency of checkmarks in Table~\ref{Tbl_Guidelines} indicates the
popularity of recommending a given artifact, but it does not imply that the most
commonly recommended artifacts are the most important artifacts. Just because a
guideline does not explicitly recommend an artifact, does not mean the guideline
authors do not value that artifact.  They may have excluded it because it is out
of the scope of their recommendations, or outside their experience.  For
instance, an artifact related to uninstall is only explicitly mentioned by
\citet{vanGompelEtAl2016}, but other guideline authors would likely see its
value.  They may simply feel that uninstall is implied by install, or they may
have never asked themselves whether they need separate uninstall instructions.

\begin{table}[H]
\begin{center}
\begin{tabular}{ p{2.5cm}p{1cm}p{1cm}p{1cm}p{1cm}p{1cm}p{1cm}p{1cm}p{1.2cm}p{1cm}p{0.8cm} }
\toprule
~ \ & \citet{USGS2019} & \citet{TobiasEtAl2018} & \citet{BrettEtAl2021} &
\citet{WilsonEtAl2016} & \citet{SmithAndRoscoe2018} & \citet{HerouxEtAl2008} &
\citet{ThielEtAl2020} & \citet{vanGompelEtAl2016} & \citet{OrvizEtAl2017} & MI\\
\midrule
LICENSE & \checkmark & \checkmark & \checkmark & \checkmark & \checkmark & &
\checkmark & \checkmark & \checkmark & C\\
README &  & \checkmark & \checkmark & \checkmark & \checkmark & & \checkmark &
\checkmark & \checkmark & C\\
CONTRIBUTING &  & \checkmark & \checkmark & \checkmark & \checkmark & &
\checkmark & \checkmark & \checkmark & R\\
CITATION &  &  &  & \checkmark & & & & \checkmark & \checkmark & U\\
CHANGELOG &  & \checkmark &  & \checkmark & \checkmark & & \checkmark &  &  & U\\
INSTALL &  &  &  &  & \checkmark & & \checkmark & \checkmark & \checkmark & U\\
\midrule
Uninstall &  &  &  &  & & & & \checkmark & &  \\
Dependency List &  &  & \checkmark & & \checkmark & & & \checkmark &  & R\\
Authors &  &  &  &  &  &  & \checkmark & \checkmark & \checkmark & U\\
Code of Conduct &  &  &  &  & & & \checkmark & & & R\\
Acknowledgements &  &  &  &  &  &  & \checkmark & \checkmark & \checkmark & U\\
Code Style Guide &  & \checkmark &  &  & & & \checkmark & \checkmark & \checkmark & R\\
Release Info. &  & \checkmark &  &  & & \checkmark & \checkmark & & & C\\
Prod.\ Roadmap &  &  &  &  & & \checkmark & \checkmark & \checkmark & & R\\
\midrule
Getting started &  &  &  &  & \checkmark & & \checkmark & \checkmark & \checkmark & R\\
User manual &  &  & \checkmark &  & & & \checkmark & & & C\\
Tutorials &  &  &  &  & & & \checkmark & & & U\\
FAQ &  &  &  &  & & & \checkmark & \checkmark & \checkmark & U\\
\midrule
Issue Track &  & \checkmark & \checkmark & & \checkmark & \checkmark &
\checkmark & & \checkmark & C\\
Version Control &  & \checkmark & \checkmark & \checkmark & \checkmark &
\checkmark & \checkmark & \checkmark & \checkmark & C\\ 
Build Scripts &  & \checkmark &  & \checkmark & \checkmark & \checkmark &
\checkmark & & \checkmark & U\\
\midrule
Requirements &  & \checkmark &  &  & & \checkmark &  &  & \checkmark & R\\
Design Doc.\ &  & \checkmark  & \checkmark &  & \checkmark & & \checkmark &
\checkmark& \checkmark & R\\
API Doc. &  &  &  &  & \checkmark & & \checkmark & \checkmark & \checkmark & R\\
Test Plan &  & \checkmark &  &  & & \checkmark & & & &  \\
Test Cases & \checkmark & \checkmark & \checkmark &  & \checkmark & \checkmark &
\checkmark & \checkmark & \checkmark & U\\
\bottomrule
\end{tabular}
\caption{Comparison of Recommended Artifacts in Software Development Guidelines
to Artifacts in MI Projects (C for Common, U for Uncommon and R for Rare)}
\label{Tbl_Guidelines}
\end{center}
\end{table}

Two of the items that appear in Table~\ref{artifactspresent} do not appear in
the software development guidelines shown in Table~\ref{Tbl_Guidelines}:
Troubleshooting guide and Developer's manual.  Although the guidelines do not
name these two artifacts, the information contained within them overlaps with
the recommended artifacts.  A Troubleshooting guideline contains information
that would typically be found in a User manual.  A Developer's guide overlaps
with information from the README, INSTALL, Uninstall, Dependency List, Release
Information, API documentation and Design documentation.  In our current
analysis, we have identified artifacts by the names given by the software
guidelines and MI examples.  In the future, a more in-depth analysis would look
at the knowledge fragments that are captured in the artifacts, rather than
focusing on the names of the files that collect these fragments together.

Although the MI community shows examples of 88\% (23 of 26) of the practices we
found in research software guidelines (Table~\ref{Tbl_Guidelines}), we did not
observe three recommended artifacts: i) Uninstall, ii) Test plans, and iii)
Requirements.  Uninstall is likely an omission caused by the focus on installing
software. Given the storage capacity of current hardware, developers and users
are not generally concerned with uninstall.  Moreover, as mentioned above,
uninstall is not particularly emphasized in existing recommendations.  

Test plans describe the scope, approach, resources, and the schedule of planned
test activities \citep{VanVliet2000}.  The plan should cover details such as the
makeup of the testing team, automated testing tools and technology to employ,
the testing process, system tests, integration tests and unit tests. We did not
observe test plans for MI software, but that doesn't mean plans weren't created;
it means that the plans are not under version control. Test plans would have to
at least be implicitly created, since we observed test cases with reasonable
frequency for MI software (test cases are categorized as uncommon).

The other apparently neglected document is the requirements specification, which
records the functionalities, expected performance, goals, context, design
constraints, external interfaces and other quality attributes of the software
\citep{IEEE1998}.  If developers write requirements they typically are based on
a template, which provide documentation structure, guidelines, and rules.
Although there is no universally accepted template, examples include
\citet{ESA1991}, \citet{IEEE1998}, \citet{NASA1989}, and
\citet{RobertsonAndRobertson1999Vol}.

MI software is like other research software in its neglect of requirements
documentation.  Although requirements documentation is recommended by some
\citep{TobiasEtAl2018, HerouxEtAl2008, SmithAndKoothoor2016}, in practice
research software developers often do not produce a proper requirements
specification \citep{HeatonAndCarver2015}. \citet{SandersAndKelly2008}
interviewed 16 scientists from 10 disciplines and found that none of the
scientists created requirements specifications, unless regulations in their
field mandated such a document. \citet{Nguyen-HoanEtAl2010} showed requirements
are the least commonly produced type of documentation for research software in
general. When looking at the pain points for research software developers,
\citet{WieseEtAl2019} found that software requirements and management is the
software engineering discipline that most hurts scientific developers,
accounting for 23\% of the technical problems reported by study participants.
The lack of support for requirements is likely due to the perception that
up-front requirements are impossible for research software
\citep{CarverEtAl2007, SegalAndMorris2008}, but if we drop the insistence on
``up-front'' requirements, allowing instead for the requirements to be written
iteratively and incrementally, requirements are feasible \citep{Smith2016}.
\citet{SmithEtAl2007} provides a requirements template tailored to research
software. 

Table~\ref{Tbl_Guidelines} shows several recommended artifacts that are rarely
observed in practice.  A theme among these rare artifacts is that, except for
the user-focused getting started manual, they are developer-focused.  The
dependency list, which is a list of software library dependencies, was rarely
observed, but this information is still likely present, just embedded in build
scripts. The other developer-focused and rare artifacts are as follows:

\begin{itemize}

\item \textbf{A Contributing file} provides new contributors with the information
that they need to start adding/modifying the repository's files.
\citet{Abdalla2016} provides a simple template for creating an open-source
contributor guideline. 

\item \textbf{A Developer Code of Conduct} explicitly states the expectations
for developers on how they should treat one another \citep{TouraniEtAl2017}. The
code outlines rules for communication and establishes enforcement mechanisms for
violations.  As \citet{TouraniEtAl2017} states, the developer code documents
the spirit of a community so that anyone can comfortably contribute regardless
of ethnicity, gender, or sexual orientation. Three popular codes of conduct are
\citep{TouraniEtAl2017}:
\href{https://www.contributor-covenant.org/version/2/1/code_of_conduct/}
{Contributor Covenant}, \href{https://ubuntu.com/community/code-of-conduct}
{Ubuntu Code of Conduct}, and \href{https://www.djangoproject.com/conduct/}
{Django Code of Conduct}. A code of conduct can improve inclusivity, which
brings the benefit of a wider pool of contributors.  For example, a code of
conduct can improve the participation of women \citep{SinghEtAl2021}. A standard
of ethical behaviour can be captured in the code, for projects that are looking
to abide by a code of ethics, such as the IEEE Code of Ethics \citep{IEEE1999},
or the Professional Engineers of Ontario code of ethics \citep[p.\
23--24]{PEO2021}.

\item \textbf{Code Style Guidelines} present standards for writing code. Style
guides codify such elements as formatting, commenting, naming identifiers, best
practices and dangers to avoid \citep{Carty2020}. For instance, most coding
style guides will specify using ALLCAPS when naming symbolic constants.
Understandability improves under standardization, since developers spend
more time on the content of the code, and less time distracted by its style.
Three sample style guides are:
\href{https://google.github.io/styleguide/javaguide.html} {Google Java Style
Guide},
\href{http://cnl.sogang.ac.kr/cnlab/lectures/programming/python/PEP8_Style_Guide.pdf}
{PEP8 Style Guide for Python}, and
\href{https://google.github.io/styleguide/cppguide.html} {Google \CC Style
Guide}.  Linting tools, like \href{https://pypi.org/project/flake8/}{flake8} for
Python, can be used to enforced coding styles, like the PEP8 standard.

\item \textbf{A Product Roadmap} explains the vision and direction for a product
offering \citep{MunchEtAl2019}.  Although they have different forms, all
roadmaps cover the following: \begin{inparaenum}[i)]
	\item Where are we now?,
	\item Where are we going?, and
	\item How can we get there? \end{inparaenum} \citep{PhaalEtAl2005}. As
stated by \citet{Pichler2012}, a product roadmap provides the following
benefits: continuity of purpose, facilitation of collaboration, and assistance
with prioritization. Creating a roadmap involves the following steps:
\begin{inparaenum}[i)]
	\item define and outline a strategic mission and product vision, 
	\item scan the environment, 
	\item revise and distill the product vision to write the product roadmap, and 
	\item estimate the product life cycle and evaluate the mix of planned
development efforts \end{inparaenum} \citep{VahaniittyEtAl2002}.  

\item \textbf{Design Documentation} explicitly states the design goals and
priorities, records the likely changes, decomposes the complex system into
separate modules, and specifies the relationship between the modules. Design
documentation shows the syntax of the interface for the modules, and in some
cases also documents the semantics.  Some potential elements of design
documentation include the following:

\begin{itemize}
	\item Representation of the system design and class design using Unified
	Modelling Language class diagrams. This approach is suited to
	object-oriented design and designs that use patterns \citep{Gamma1995}.
	\item Rigorous documentation of the system design following the template for
	a Module Guide (MG) \citep{ParnasEtAl1984}.  An MG organizes the modules in
	a hierarchy by their secrets.
	\item An explanation of the design using data flow diagrams to show typical use cases
	for input transformation.
	\item A table or graph showing the traceability between the requirements and
	the modules (or classes)
	\item The syntax of the modules (or classes) by providing lists of the state
	variables, exported constants and all exported access programs for each
	module (or class).  This shows the interface that can used to access each
	module's services.
	\item A formal specification of the semantics of input/output relationships
	and state transitions for each module using a Module Interface Specification
	(MIS) \citep{HoffmanAndStrooper1995}. An MIS is an abstract model that
	formally shows each module's access programs and the associated transitions
	and outputs based on their state, environment, and input variables.
	\citet{ElSheikhEtAl2004, SmithAndYu2009} show the example of an MIS for a
	mesh generator.
\end{itemize}

\item \textbf{API Documentation} shows developers the services or data provided
by the software application (or library) through such resources as its methods
or objects \citep{MengEtAl2018}.  Understandability is improved by API
documentation \citep{MengEtAl2018}. API documentation can be generated using
tools like Doxygen, pydoc, and javadoc.

\end{itemize}

The rare artifacts for MI software are similar to the rare artifacts for Lattice
Boltzmann solvers \citep{Michalski2021}, except LBM software is more likely to
have developer related artifacts, like Contributing, Dependency list, and Design
documentation.

To improve MI software in the future, an increased use of checklists could help.
Developers can use checklists to ensure they follow best practices.  Some
examples include checklists merging branches into master \citep{Brown2015},
checklists for saving and sharing changes to the project \citep{WilsonEtAl2016},
checklists for new and departing team members \citep{HerouxAndBernholdt2018},
checklists for processes related to commits and releases \citep{HerouxEtAl2008}
and checklists for overall software quality \citep{ThielEtAl2020, SSI2022}.  For
instance, for Lattice Boltzmann solver software, ESPResSo has a checklist for
managing releases \citep{Michalski2021}. 

The above discussion shows that, taken together, MI projects fall somewhat short
of recommended best practices for research software.  However, MI software is
not alone in this.  Many, if not most, research projects fall short of best
practices.  A gap exists in research software development practices and software
engineering recommendations \citep{Storer2017, Kelly2007, OwojaiyeEtAl2021_CSE}.
\citet{JohansonAndHasselbring2018} observe that the state-of-the-practice for
research software in industry and academia does not incorporate state-of-the-art
SE tools and methods.  This causes sustainability and reliability problems
\citep{FaulkEtAl2009}. Rather than benefit from capturing and reusing previous
knowledge, projects waste time and energy ``reinventing the wheel''
\citep{deSouzaEtAl2019}.

\section{Comparison of Tool Usage Between MI and Other Research Software}
\label{Sec_CompareTools}

Developers use software tools to support the development, verification,
maintenance, and evolution of software, software processes, and artifacts
\citep[p.\ 501]{GhezziEtAl2003}. MI software uses tools for CI/CD, user support,
version control, documentation, and project management.  To answer
\rqref{RQ_CompareToolsProjMngmnt} we summarize the tool usage in these
categories, and compare this to the usage by the research software community.

Table~\ref{tab_user_support_model} summarizes the user support models by the
number of projects using each model (projects may use more than one support
model). We do not know whether the prevalent use of GitHub issues for user
support is by design, or whether this just naturally happens as users seek
help. The common use of GitHub by MI developers is not surprising, given that
GitHub is the largest code host in the world, with over 128 million public
repositories and over 23 million users (as of roughly February 26, 2020)
\citep{Kashyap2020}.

\begin{table}[!ht]
\centering
\begin{tabular}{lc}
\toprule
\multicolumn{1}{c}{User Support Model} & Num.\ Projects \\
\midrule
GitHub issue & 24 \\
Frequently Asked Questions (FAQ) & 12 \\
Forum & 10 \\
E-mail address & 9 \\
GitLab issue, SourceForge discussions & 2 \\
Troubleshooting & 2 \\
Contact form & 1 \\ 
\bottomrule
\end{tabular}
\caption{\label{tab_user_support_model}User support models by number of projects}
\end{table}

From Section~\ref{sec_score_maintainability}, 27 of the 29 projects used git as
the version control tool, one used Mercurial and one used Subversion.  The
hosting is on GitHub for 24 packages, SourceForge for three and BitBucket for
two.  Although teams may have a process for accepting new contributions, no one
discussed this during their interviews. However, most teams (eight of nine)
mentioned using GitHub and pull requests to manage contributions from the
community. The interviewees generally gave very positive feedback on using
GitHub. Some teams previously used a different approach to version control and
eventually transferred to git and GitHub.  The past approaches included
contributions from e-mail (three teams), contributions from forums (one team)
and e-mailing the git repository back and forth between developers (one team).

The common use of version control for MI software illustrates considerable
improvement from the poor adoption of version control tools that Wilson lamented
in 2006 \citep{Wilson2006}.  The proliferation of version control tools for MI
matches the increase in the broader research software community.  A little over
10 years ago \citet{Nguyen-HoanEtAl2010} estimated that only 50\% of research
software projects use version control, but even at that time
\citet{Nguyen-HoanEtAl2010} noted an increase from previous usage levels. A
survey in 2018 shows 81\% of developers use a version control system
\citep{AlNoamanyAndBorghi2018}. \citet{Smith2018} has similar results, showing
version control usage for alive projects in mesh generation, geographic
information systems and statistical software for psychiatry increasing from
75\%, 89\% and 17\% (respectively) to 100\%, 95\% and 100\% (respectively) over
a four-year period ending in 2018. (For completeness the same study showed a
small decrease in version control usage for seismology software over the same
time period, from 41\% down to 36\%).  A recent survey by \citet{CarverEtAl2022}
shows version control use among practitioners at over 95\%, with 83/87 survey
respondents indicating that they use it. All but one of the software guides
cited in Section~\ref{Sec_CompareArtifacts} includes the advice to use version
control. (The USGS guide \citep{USGS2019} was the only set of recommendations to
not mention version control.) The high usage of version control tools in MI
software matches the trend in research software in general.

As mentioned in Section~\ref{sec_result_correctness_verifiability}, we
identified five projects using CI/CD tools (about 17\% of the assessed
projects). We found which projects used CI/CD by examining the documentation and
source code of all projects. The count of CI/CD usage may actually be higher,
since traces of CI/CD usage may not always appear in a repository.  This was the
case for a study of LBM software, where interviews with developers showed that
more projects used CI/CD than was evident from repository artifacts alone
\citep{Michalski2021}.  The 17\% utilization for MI software contrasts with the
high frequency with which research software development guidelines recommend
continuous integration \citep{BrettEtAl2021, Brown2015, ThielEtAl2020,
Zadka2018, vanGompelEtAl2016}. Although there is currently little data available
on CI/CD utilization for research software, our impression is that CI/CD is not
yet common practice, despite its recommendation.  This is certainly the case for
LBM software, where usage numbers are similar to MI software, with only 12.5\%
of 24 LBM packages showing evidence of CI/CD in their repositories
\citep{Michalski2021}.  The survey of \citet{CarverEtAl2022} suggests higher use
of CI/CD with 54\% (54/100) of respondents indicating that they use it. However,
that survey measures something different from the current one by surveying
practitioners, rather than assessing projects.  Additional information on CI/CD
is given in the recommendations (Section~\ref{Sec_ContinuousIntegration}).

For documentation tools and methods mentioned by the interviewees, the most
popular (mentioned by about 30\% of developers) were forum discussions and
videos.  The second most popular options (mentioned by about 20\% of developers)
were GitHub, wiki pages, workshops, and social media. The least frequently
mentioned options (about 10\% of developers) included writing books, and google
forms.  In contrasting MI software with LBM software, the most significant
documentation tool difference is that LBM software often uses document
generation tools, like doxygen and sphinx \citep{Michalski2021}, while MI does
not appear to use these tools. 

Some interviewees mentioned the project management tools they used. Generally
speaking, the interviewees talked about two types of tools:
\begin{inparaenum}[i)]
\item trackers, including GitHub, issue trackers, bug trackers and Jira; and,
\item documentation tools, including GitHub, Wiki page, Google Doc, and
Confluence.
\end{inparaenum}
Of the specifically named tools in the above lists, interviewees mentioned
GitHub 3 times, and each of the other tools once each.

Based on information provided by \citet{JungEtAl2022}, tool utilization for MI
software has much in common with tool utilization for ocean modelling software.
Both use tools for editing, compiling, code management, testing, building, and
project management.  From the data available, ocean modelling differs from MI
software in its use of Kanban boards for project management.

\section[Comparison to Other Research Software]{Comparison of Principles, Process, and
Methodologies to Research Software in General} \label{Sec_CompareMethodologies}

We answer research question \rqref{RQ_CompareMethodologies} by comparing the
principles, processes, and methodologies used for MI software to what can be
gleaned from the literature on research software in general. In our interviews
with developers the responses about development model were vague, with only two
interviewees following a definite development model. In some cases the
interviewees felt their process was similar to an existing development model.
Three teams (about 38\%) either followed agile, or something similar to agile.
Two teams (25\%) either followed a waterfall process, or something similar.
Three teams (about 38\%) explicitly stated that their process was undefined or
self-directed.

Our observations of an informally defined process, with elements of agile
methods, matches what has been observed for research software in general.
Scientific developers naturally use an agile philosophy \citep{AckroydEtAl2008,
CarverEtAl2007, EasterbrookAndJohns2009, Segal2005, HeatonAndCarver2015}, or an
amethododical process \citep{Kelly2013}, or a knowledge acquisition driven
process \citep{Kelly2015}.  A waterfall-like process can work for research
software \citep{Smith2016}, especially if the developers work iteratively and
incrementally, but externally document their work as if they followed a
rationale design process \citep{parnas1986rational}.

No interviewee introduced any strictly defined project management process. The
most common approach was following the issues, such as bugs and feature
requests. Additionally, the \textit{3D Slicer} team had weekly meetings to
discuss the goals for the project; the \textit{INVESALIUS 3} team relied on the
GitHub process for their project management; the \textit{ITK-SNAP} team had a
fixed six-month release pace; only the interviewee from the \textit{OHIF} team
mentioned that the team has a project manager; the \textit{3D Slicer} team and
\textit{BioImage Suite Web} team do nightly builds and tests. The \textit{OHIF}
developer believes that a better project management process can improve junior
developer efficiency while also improving internal and external communication.

We identified the use of unit testing in less than half of the 29 projects. On
the other hand, the interviewees believed that testing (including usability
tests with users) was the top solution to improve correctness, usability, and
reproducibility.  This level of testing matches what was observed for LBM
software \citep{Michalski2021} and is apparently greater than the level of
testing for ocean modelling software.  \citet{JungEtAl2022} reports that ocean
modellers underemphasize testing.

As the observed artifacts in Table~\ref{artifactspresent} show, none of the 29
projects emphasize documentation. None of them had theory manuals, although we
did identify a road map in the \textit{3D Slicer} project.  We did not find
requirements specifications. Table~\ref{tab_opinion_doc} summarizes
interviewees' opinions on documentation. Interviewees from each of the eight
projects thought that documentation was essential to their projects, and most of
them said that it could save their time to answer questions from users and
developers. Most of them saw the need to improve their documentation, and only
three of them thought that their documentations conveyed information clearly
enough. Nearly half of developers also believed that the lack of time prevented
them from improving documentation.

\begin{table}[!ht]
\centering
\begin{tabular}{ll}
\toprule
Opinion on Documentation & Num.\ Ans. \\ 
\midrule
Documentation is vital to the project & 8 \\
Documentation of the project needs improvements & 7 \\
Referring to documentation saves time to answer questions & 6 \\
Lack of time to maintain good documentation & 4 \\
Documentation of the project conveys information clearly & 3 \\
Coding is more fun than documentation & 2 \\
Users help each other by referring to documentation & 1 \\ 
\bottomrule
\end{tabular}
\caption{Opinions on documentation by the numbers of interviewees with the
answers}
\label{tab_opinion_doc}
\end{table}

As Table~\ref{Tbl_Guidelines} suggests, an emphasis on documentation, especially
for new developers, is echoed in research software guidelines. Multiple
guidelines recommend a document explaining how to contribute to a project, often
named CONTRIBUTING. Guidelines also recommend tutorials, user guides and quick
start examples. \citet{SmithAndRoscoe2018} suggests including instructions
specifically for on-boarding new developers. For open-source software in general
(not just research software), \citet{Fogel2005} recommends providing tutorial
style examples, developer guidelines, demos, and screenshots.

\section{Developer Pain Points} \label{painpoints}

Based on interviews with nine developers (described in
Section~\ref{sec_interview_methods}), we answer three research questions (first
mentioned in Section~\ref{sec_motivation}): \rqref{RQ_PainPoints}) What are the
pain points for developers working on research software projects?;
\rqref{RQ_ComparePainPoints}) How do the pain points of developers from MI
compare to the pain points for research software in general?; and
\rqref{RQ_Concerns}) For MI developers what specific best practices are taken to
address the pain points and software quality concerns? 

Our interviews identified pain points related to a lack of time and funding,
technology hurdles, improving correctness, and improving usability.  In this
section, we go through each pain point and contrast the MI experience with
observations from other domains.  We also cover potential ways to address the
pain points, as promoted by the community.  (Later, in
Section~\ref{ch_recommendations}, we propose additional pain mitigation
strategies based on our experience.)  In addition to pain points, we summarize
MI developer strategies for improving maintainability and reproducibility.
Although the interviewees did not explicitly identify these two qualities as
pain points, we did discuss threats to these qualities and ways to improve them
as part of our interview process \citep{SmithEtAl2021}.  The interviewee's
practices for addressing pain points and improving quality can potentially be
emulated by other MI developers. Moreover, these practices may provide examples
that can be followed by other research software domains.

\citet{PintoEtAl2018} lists some pain points that did not come up in our
conversations with MI developers: interruptions while coding, scope bloat, lack
of user feedback, hard to collaborate on software projects, and aloneness.
\citet{WieseEtAl2019} also mention two research software pain points that did
not explicitly arise in our interviews: reproducibility, and software scope
determination.  To the list of pain points not discussed for MI, our study of
LBM software \citep{SmithEtAl2024} adds lack of software experience for the
developers, technical debt, and documentation. We did not observe any pain
points for MI that were not also observed for LBM. From the pain points
mentioned above, although the topics of reproducibility and technical debt did
not come up in our MI interviews, we covered these two topics as part of the
discussion of software qualities, as summarized at the end of this section.
Although previous studies show pain points that were not mentioned by MI
developers, we cannot conclude that these pain points are not relevant for MI
software development, since we only interviewed nine developers for about an
hour each.

\begin{enumerate}

\item[P\refstepcounter{pnum}\thepnum \label{P_LackDevTime}:] \textbf{Lack of
Development Time:} Many interviewees thought lack of time, along with lack of
funding (discussed next), were their most significant obstacles. Other domains
of research software also experience the lack of time pain point
\citep{PintoEtAl2018, PintoEtAl2016, WieseEtAl2019}. Our study of LBM software
\citep{SmithEtAl2024} also highlighted lack of time as a significant pain point.

Potential and proven solutions suggested by the interviewees include:

\begin{itemize}
\item Shifting from development to maintenance when the team does not have
enough developers for building new features and fixing bugs at the same time;
\item Improving documentation to save time answering users' and developers'
questions;
\item Supporting third-party plugins and extensions; and,
\item Using GitHub Actions for CI/CD (Continuous Integration and Continuous
Delivery.)
\end{itemize}

\item[P\refstepcounter{pnum}\thepnum \label{P_LackFunding}:] \textbf{Lack of
Funding:} Developers felt the pain of having to attract funding to develop and
maintain their software. For instance, the interviewees from \textit{3D Slicer}
and \textit{OHIF} said getting funding for software maintenance is more
challenging than finding funding for research. The interviewee from 
\textit{ITK-SNAP} thought more funding was a way to solve the lack of time
problem, because they could hire more dedicated developers. On the other hand,
the interviewee from \textit{Weasis} did not feel that funding could
solve the same problem, since they would still need time to supervise the project. 

Funding challenges have also been noted by others \citep{GewaltigAndCannon2012,
Goble2014, KaterbowAndFeulner2018, SmithEtAl2024}. Researchers that devote time
to software have the additional challenge that funding agencies do not always
count software when they are judging the academic excellence of the applicant.
\citet{WieseEtAl2019} reported developer pains related to publicity, since
publishing norms have historically made it difficult to get credit for creating
software.  As studied by \citet{HowisonAndBullard2016}, research software
(specifically biology software, but the trend likely applies to other research
software domains) is infrequently cited. \citet{PintoEtAl2018} also mentions the
lack of formal reward system for research software.

An interviewee proposed an idea for increasing funding: Licensing the software
to commercial companies to integrate it into their products.
    
\item[P\refstepcounter{pnum}\thepnum \label{P_TechnologyHurdles}:]
\textbf{Technology Hurdles:} The technology hurdles mentioned by MI developers
include: hard to keep up with changes in OS and libraries, difficult to transfer
to new technologies, hard to support multiple OSes, and hard to support lower-end
computers. Developers expressed difficulty balancing between four factors:
cross-platform compatibility, convenience to development and maintenance,
performance, and security.

The pain point survey of \citet{WieseEtAl2019} highlights that technology
hurdles are an issue for research software in general.  Some technical-related
problems mentioned by \citet{WieseEtAl2019} include dependency management,
cross-platform compatibility (also mentioned by \citet{PintoEtAl2018}), CI,
hardware issues and operating system issues. From \citep{SmithEtAl2024}
technology pain points for LBM developers include setting up parallelization and
CI. 

The solutions proposed by the MI developers include the following:

\begin{itemize}
\item Adopting a web-based approach with backend servers, to better support
lower-end computers;
\item Using memory-mapped files to consume less computer memory, to better
support lower-end computers; 
\item Using computing power from the computers GPU for web applications;
\item Maintaining better documentations to ease the development and maintenance
processes;
\item Improving performance via more powerful computers, which one interviewee
pointed out has already happened.
\end{itemize}

As the above list shows, developers perceive that web-based applications will
address the technology hurdle.  Table~\ref{tab_native_vs_web} shows the teams'
choices between native application and web application. Most of the 29 teams (24
of 29, or 83\%) chose to develop native applications. For the eight teams we
interviewed, three of them were building web applications, and the
\textit{MRIcroGL} team was considering a web-based solution.

\begin{table}[!ht]
\centering
\begin{tabular}{lll}
\toprule
Software Team & Native Application & Web Application \\ 
\midrule
3D Slicer & \checkmark & \\
INVESALIUS 3 & \checkmark & \\
dwv & & \checkmark \\
BioImage Suite Web & & \checkmark \\
ITK-SNAP & \checkmark & \\
MRIcroGL & \checkmark & \\
Weasis & \checkmark & \\
OHIF & & \checkmark \\ 
\midrule
Total number among the eight teams & 5 & 3 \\
Total number among the 29 teams & 24 & 5 \\ 
\bottomrule
\end{tabular}
\caption{Teams' choices between native application and web application}
\label{tab_native_vs_web}
\end{table}

The advantage for native applications is higher performance, while web
applications have the advantage of cross-platform compatibility and a simpler
build process.  These web advantages mirror the native disadvantages of
difficulty with cross-platform compatibility and a complex build process.  The
lower performance disadvantage of web applications can be improved with a server
backend, but in this case there are disadvantages for privacy protection and
server costs.  These issues are discussed further in the recommendations
(Section~\ref{sec_webapps}).

\item[P\refstepcounter{pnum}\thepnum \label{P_Correctness}:]
\textbf{Ensuring Correctness:} Interviewees identified multiple threats to
correctness.  The most frequently mentioned threat was complexity.  Complexity
enters the software by various means, including the large variety of data formats,
complicated data standards, differing outputs between medical imaging machines,
and the addition of (non-viewing related) functionality.  Other threats to
correctness identified include the following:

\begin{itemize}
\item Lack of real world image data for testing, in part because of patient
privacy concerns (\citet{WieseEtAl2019} mentions that the pain point of privacy
concerns also arises for research software in general);
\item Tests are expensive and time-consuming because of the need for huge datasets;
\item Software releases are difficult to manage;
\item No systematic unit testing; and,
\item No dedicated quality assurance team.
\end{itemize}

As implied by the above threats to correctness, testing was the most often
mentioned strategy for MI developers for ensuring correctness.  Seven teams
mentioned test related activities, including test-driven development, component
tests, integration tests, smoke tests, regression tests, self tests and
automated tests.  With the common emphasis on testing to improve correctness, MI
software is ahead of some other scientific domains.  For scientific software in
general \citet{PintoEtAl2018} mention the problem of insufficient testing and
\citet{HannayEtAl2009} show that more developers think testing is important than
the number that believe they have a sufficient understanding of testing
concepts.  Our study of LBM software suggests that this domain shares the
challenges of insufficient testing and insufficient understanding of testing
concepts \citep{SmithEtAl2024}. Automated testing is a specific challenge for
LBM software since free testing services do not offer adequate facilities for
large amounts of data \citep{SmithEtAl2024}. Although not specifically mentioned
during our interviews, the large data sets for MI likely also cause a challenge
for using free testing services, like GitHub Actions.

Research software in general often struggles with the oracle problem for testing
because for many potential test cases the developer doesn't have a means to
judge the correctness of their calculated solutions \citep{HannayEtAl2009,
KanewalaAndBieman2013, KellyEtAl2011, WieseEtAl2019}.  The MI developers did not
allude to this challenge, likely because for a give image (test case) it is
possible to determine, potentially by using other software, the expected
analysis results.

A frequently cited strategy for building confidence in correctness (mentioned by
3 interviewees) is a two state development process with stable releases and
nightly builds.  Other strategies for ensuring correctness that came up during
the interviews include CI/CD, using de-identified copies of medical images for
debugging, sending beta versions to medical workers who can access the data to
do the tests, and collecting/maintaining a dataset of problematic images.  Some
additional strategies used by MI developers include:

\begin{itemize}
\item Using open datasets.
\item If (part of) the team belongs to a medical school or a hospital, using the
datasets they can access;
\item If the team has access to MRI scanners, self-building sample images for
testing;
\item If the team has connections with MI equipment manufacturers, asking for
their help on data format problems;
\end{itemize}

The feedback from the interviewees makes it clear that increased connections
between the development team and medical professionals/institutions could ease
the pain of ensuring correctness via testing.

\item[P\refstepcounter{pnum}\thepnum \label{P_Usability}:]
\textbf{Usability:}  

The discussion with the developers focused on usability issues for two classes
of users: the end users and other developers.  The threats to usability for end
users include an unintuitive user interface, inadequate feedback from the
interface (such as lack of a progress bar), users being unable to determine the purpose of
the software, not all users knowing if the software includes certain features, not
all users understanding how to use the command line tool, and not all users
understanding that the software is a web application. For developers the threats to
usability include not being able to find clear instructions on how to deploy the
software, and the architecture being difficult for new developers to understand.

At least to some extent the problems for MI software users are due to holes in
their background knowledge.  The survey of \citet{WieseEtAl2019} for research
software in general also mentioned that users do not always have the expertise
required to install or use the software. \citet{SmithEtAl2024} observes a
similar pattern for LBM software, with several LBM developers noting that users
sometimes try to use incorrect method combinations. Furthermore, some LBM users
think that the packages will work out of the box to solve their cases, while in
reality computational fluid dynamics knowledge needs to be applied to correctly
modify the packages for a new endeavour.

To improve the usability of MI software, the most common strategies mentioned by
developers are as follows:

\begin{itemize}
    \item Use documentation (user manuals, mailing lists, forums) (mentioned by
    4 developers)
    \item Usability tests and interviews with end users; and, (mentioned by 3
    developers)
    \item Adjusting the software according to user feedback. (mentioned by 3
    developers)
\end{itemize}

Other suggested and practiced strategies include a graphical user interface,
testing every release with active users, making simple things simple and
complicated things possible, focusing on limited number of functions, icons with
clear visual expressions, designing the software to be intuitive, having a UX
(User eXperience) designer, dialog windows for important notifications,
providing an example for users to follow, downsampling images to consume less
memory, and providing an option to load only part of the data to boost
performance.  The last two points recognize that an important component of
usability is performance, since poor performance frustrates users.

\end{enumerate}

Up to this point, we have covered the pain points that came up in interviews
with MI developers, along with a summary of the techniques that are currently
used to address these pain points.  Although the developers did not explicitly
identify the qualities of maintainability and reproducibility as pain points in
our interviews, as part of our interview questions
(Section~\ref{sec_interview_methods}) they did share their approaches for
improving these qualities, as discussed below.

\begin{enumerate}
\item[Q\refstepcounter{qnum}\theqnum \label{Q_Maintainability}:]
\textbf{Maintainability:} \citet{Nguyen-HoanEtAl2010} rate maintainability as the
third most important software quality for research software in general. The push
for sustainable software \citep{deSouzaEtAl2019} is motivated by the pain that
past developers have had with accumulating too much technical debt
\citep{KruchtenEtAl2012}.  For LBM software, \citet{SmithEtAl2024} identifies
technical debt as one of the developer pain points.

To improve maintainability, the most popular (with five out of nine interviewees
mentioning it) strategy is to use a modular approach, with often repeated
functions in a library.  Other strategies that were mentioned for improving
maintainability include supporting third-party extensions, an easy-to-understand
architecture, a dedicated architect, starting from simple solutions, and
documentation.  The \textit{3D Slicer} team used a well-defined structure for
the software, which they named as an ``event-driven MVC pattern''. Moreover,
\textit{3D Slicer} discovers and loads necessary modules at runtime, according
to the configuration and installed extensions. The \textit{BioImage Suite Web}
team had designed and re-designed their software multiple times in the last 10+
years. They found that their modular approach effectively supports
maintainability \citep{Joshi2011}. 

\item[Q\refstepcounter{qnum}\theqnum \label{Q_Reproducibility}:]
\textbf{Reproducibility:}  Although the MI developers did not mention
reproducibility explicitly as a pain point, they did mention the need to improve
documentation.  Good documentation does not just address the pain points of lack
of developer time (\ppref{P_LackDevTime}), technology hurdles
(\ppref{P_TechnologyHurdles}), usability \ppref{P_Usability}, and
maintainability.  Documentation is also necessary for reproducibility. The
challenges of inadequate documentation are a known problem for research software
\citep{PintoEtAl2018, WieseEtAl2019} and for non-research software
\citep{LethbridgeEtAl2003}. 

In our interviews, we discussed threats to reproducibility and strategies for
improving it.  The threats that were mentioned include closed-source software,
no user interaction tests, no unit tests, the need to change versions of some
common libraries, variability between CPUs, and misinterpretation of how
manufacturers create medical images. 

The most commonly cited (by 6 teams) strategy to improve reproducibility was
testing (regression tests, unit tests, having good tests). The second most
common strategy (mentioned by 5 teams) is making code, data, and documentation
available, possibly by creating open-source libraries.  Other ideas that were
mentioned include running the same tests on all platforms, a dockerized version
of the software to insulate it from the OS environment, using standard
libraries, monitoring the upgrades of the library dependencies, clearly
documenting the version information, bringing along the exact versions of all
the dependencies with the software, providing checksums of the data, and
benchmarking the software against other software that overlaps in functionality.
Specifically one interviewee suggested using \textit{3D Slicer} as the benchmark
to test their reproducibility.

\end{enumerate}

\section{Recommendations} \label{ch_recommendations}

In this section we provide recommendations to address the pain points from
Section~\ref{painpoints} to answer~\rqref{RQ_Recommend}.  Our recommendations
are not lists of criticisms for what should have been done in the past, or what
should be done now; they are suggestions for consideration in the future. We
expand on some of the ideas that came out of our interviews with developers
(Section~\ref{painpoints}), including continuous integration, moving to web
applications, and enriching the test data sets. We also bring in new ideas from
our experience like employing linters, peer review, design for change and
assurance cases.  Our aim is to mention ideas that are at least somewhat beyond
conventional best practices. The ideas listed here have the potential to become
best practices in the medium to long-term. We list the ideas roughly in the
order of increasing implementation effort.

\subsection{Use Continuous Integration} \label{Sec_ContinuousIntegration}

Continuous integration involves frequent pushes to a code repository.  With
every push the software is built and tested \citep[p.\ 13]
{HumbleAndFarley2010}, \citep{ShahinEtAl2017, Fowler2006}.  CI can take
significant time and effort to set up and integrate into a team's workflow, but
the benefits are significant, as follows:

\begin{itemize}
	\item Elimination of headaches associated with a separate integration phase
	\citep{Fowler2006}, \citep[p.\ 20]{HumbleAndFarley2010}. If developers
	postpone integration, integration problems are inevitable.  Continuous
	integration means that problems are immediately obvious and the source of
	the problem can be isolated to the small increment that was just committed.
	\item Detection and removal of bugs \citep{Fowler2006} via
	automated testing.  To improve productivity, defects are best discovered and
	fixed at the point where they are introduced \citep[p.\
	23]{HumbleAndFarley2010}.  Code is not the only source of errors; they are
	also found in the files and scripts related to configuration management
	\citep[p.\ 18]{HumbleAndFarley2010}.
	\item Everyone is always working on a stable base, since the rejection of
	inadequate commits means that the main branch will always be working.  A
	stable base will always pass all tests.  If the CI system uses generators
	and linters, it will also have current documentation and standard compliant
	code.  A stable base improves developer productivity, allowing them to focus
	on coding, testing, and documentation.
\end{itemize}

CI consists of the following elements:

\begin{itemize}
	\item A version control system \citep{Fowler2006}. To be effective, all
	files should be under version control, not just code files.  Anything that
	is needed to build, install and run the software should be under version
	control, including configuration files, build scripts, test harnesses, and
	operating system configuration files \citep[p. 19]{HumbleAndFarley2010}.
	Fortunately for the MI, as shown in Section~\ref{sec_score_maintainability}
	all our measured projects use version control.
	\item A fully automated build system \citep{Fowler2006}.  As \citet[p.\
	5]{HumbleAndFarley2010} point out, deploying software manually is an
	anti-pattern.  For MI software, Table~\ref{artifactspresent} shows 18 of 29
	packages (62\%) were observed to include build scripts.  Projects without a
	build system will need to add one to pursue using CI.
	\item An automated test system \citep{Fowler2006}. Building quality software
	involves creating automated tests at the unit, component, and acceptance
	test level, and executing these tests whenever someone makes a change to the
	code, its configuration, the environment, or the software stack that it runs
	on \citep[p.\ 83]{HumbleAndFarley2010}. As Table~\ref{artifactspresent}
	shows, test cases are in the uncommon category for MI software artifacts,
	which means that some MI projects will need to increase their testing
	automation if they wish to pursue CI.
	\item An automated system for other tasks, such as code checking,
	documentation building and web-site updating.  These other tasks are not
	essential to CI, but they can be incorporated to improve the quality
	of the code and the communication between developers and users. For
	instance, a static analysis (possibly via linters) of the code may find poor
	programming practice or lack of adherence to adopted coding standards.
	\item An integrated build system to pull everything together.  Every time
	there is a check-in (for instance a pull request), the integration server
	automatically checks out the sources onto the integration machine, starts a
	build, runs tests, and informs the committer of the results. 
\end{itemize}

To enable incorporation into a team's workflow, \citet[p.\
60]{HumbleAndFarley2010} explain that the usual approach for CI is to keep the
build and test process short. Since MI files are large, the tests run with every
check-in may need to focus on simple code interface tests, saving large tests
for less frequent execution.  A more sophisticated option to address the
bottleneck for merges is CIVET (Continuous Integration, Verification,
Enhancement, and Testing), which solves this problem by intelligently pinning,
cancelling, and if necessary, restarting jobs as merges occur
\citep{SlaughterEtAl2021}. A more sophisticated process management system can
also enforce rules for pull requests, like checking that a test specification
includes the test's motivation, a test description, and a design description for
all changes \citet{SlaughterEtAl2021}. 

Setting up a CI system has never been easier than it is today.  A dedicated CI
server (either physically or virtually) can be installed with tools such as
\href{https://www.jenkins.io/} {Jenkins}, \href{http://buildbot.net/}
{Buildbot}, \href{https://www.gocd.org/} {Go}, and
\href{http://integrity.github.io/} {Integrity}. However, installation on your
own server is often unnecessary since there are many hosted CI solutions, such
as: \href{https://travis-ci.org/} {Travis CI},
\href{https://github.com/features/actions} {GitHub Actions} and
\href{https://circleci.com/} {CircleCI}.  All that is required to begin using a
hosted CI is to select the service and then edit a few lines of a YAML
configuration file in the project's root directory.

\citet{ShahinEtAl2017} highlights the following challenges for adopting CI: lack
of awareness and transparency, lack of expertise and skills, coordination and
collaboration challenges, more pressure and workload for team members, general
resistance to change, scepticism and distrust on continuous practices. The most
common reason given for not adopting CI is that developers are not familiar
enough with CI \citep{HiltonEtAl2016}.  \citet{ShahinEtAl2017} observes that
these problems can be mitigated via improving testing activities, planning and
documentation, promoting a team mindset, adopting new rules and policies, and
decomposing development into smaller units.

Continuous integration and delivery helps with addressing several pain points.
For instance, CI/CD helps reduce development time (\ppref{P_LackDevTime}) by
removing the need for a time-consuming integration stage and by automating
regression testing.  Automated regression tests also help with ensuring
correctness (\ppref{P_Correctness}) and the quality of reproducibility
(\qref{Q_Reproducibility}).

\subsection{Move To Web Applications} \label{sec_webapps}

Section~\ref{painpoints} describes the pain point of technology hurdles
(\ppref{P_TechnologyHurdles}), which motivates considering the use of web
applications. Here we give further advice to help with deciding whether to adopt
a web application. The decision will be based on whether, on balance, the web
application improves the four factors identified by developers: compatibility,
maintainability, performance, and security. To enable decision-making, a team
will need to prioritize between these factors, based on their objectives and
experience. The suggestions are intended to provide ideas and avenues for
exploration; a web application will not be the right fit for all projects and
all teams.

\begin{itemize}

\item \textbf{Modern technologies may improve frontend performance.} Web
applications with only a frontend usually perform worse than native
applications. However, new technologies may ease this difference. For example,
some JavaScript libraries can help the frontend harness the power of the
computer's GPU and accelerate graphical computing. In addition, there are new
frameworks helping developers with cross-platform compatibility. For example,
the \href{https://flutter.dev/}{Flutter} project enables support for web,
mobile, and desktop OS with one codebase.  Other options include
\href{https://vuejs.org/} {Vue}, \href{https://angular.io/} {Angular} and
\href{https://reactjs.org/} {React}, and \href{https://elm-lang.org/}{Elm}.  

\item \textbf{Backend servers can potentially deliver high performance.} Web
applications with backend servers may perform even better than native
applications. If a team needs to support lower-end computers, it is good to use
back-end servers for heavy computing tasks.  For backend servers where traffic
and latency is not an issue, options include
\href{https://www.django-rest-framework.org/} {Django}, \href{https://laravel.com/} {Laravel} and
\href{https://nodejs.org/en/} {Node.js}.  The advantage of Django is that it provides access to Python
libraries.  For backend servers where traffic and latency is an issue,
\href{https://github.com/gin-gonic/gin} {Gin} is an option.

\item \textbf{Backend servers can have low costs.} Serverless solutions from
major cloud service providers (like Amazon Web Services (AWS) and Google Cloud
Platform) may be worth exploring. Serverless solutions still use a server, but
the server provider only charges the team when they use the server. The solution
is event-driven, and costs the team by the number of requests processed. Thus,
serverless can be very cost-effective for less intensively used functions.

\item \textbf{Web transmission may diminish security.} Transferring sensitive
data on-line can be a problem for projects requiring high security. Regulations
for some MI applications may forbid doing web transmissions. In this case, a web
application with a backend may not be an option.

\end{itemize}

\subsection{Enrich the Testing Datasets} 
\label{sec_recommendations_testing_dataset}

As described in Section~\ref{painpoints}, ensuring correctness
(\ppref{P_Correctness}) via testing can be problematic because of limited access
to real-world medical imaging datasets.  We build on the suggestions we heard
from our interviewees as follows:

\begin{itemize}
\item \textbf{Build and maintain good connections to datasets.} A team can build
connections with professionals working in the medical domain, who may have
access to private datasets and can perform tests for the team. If a team has
such professionals as internal members, the process can be simplified.

\item \textbf{Collect and maintain datasets over time.} A team may face problems
caused by various unique inputs over the years of software development. This
data should be collected and maintained over time to form a good, comprehensive,
dataset for testing.

\item \textbf{Search for open data sources.} In general, there are many open MI
datasets.  For instance, there are
\href{https://nihcc.app.box.com/v/ChestXray-NIHCC}{Chest X-ray Datasets} by
National Institute of Health \citep{WangEtAl2017},
\href{https://www.cancerimagingarchive.net/}{Cancer Imaging Archive}
\citep{PriorEtAl2017}, \href{https://medpix.nlm.nih.gov/home}{MedPix} by
National Library of Medicine \citep{Smirniotopoulos2014}, and datasets for liver
\citep{BilicEtAl2019} and brain \citep{MenzeEtAl2015} tumor segmentation
benchmarks.  A team developing MI software should be able to find more open
datasets according to their needs.

\item \textbf{Create sample data for testing.} If a team can access tools
creating sample data, they may also self-build datasets for testing. For
example, an MI software development team can use an MRI scanner to create images
of objects, animals, and volunteers. The team can build the images based on
specific testing requirements.

\item \textbf{Remove privacy from sensitive data.} For data with sensitive
information, a team can ask the data owner to remove such information or add
noise to protect privacy. One example is using de-identified copies of medical
images for testing.

\item \textbf{Establish community collaboration in the domain.} During our
interviews with developers in the MI domain, we heard many stories of asking for
supports from other professionals or equipment manufacturers. However, we
believe that broader collaboration between development teams can address this
problem better. Some datasets are too sensitive to share, but if the community
has some kind of ``group discussion'', teams can better express their needs, and
professionals can better offer voluntary support for testing. Ultimately, the
community can establish a nonprofit organization as a third party, which
maintains large datasets, tests Open Source Software (OSS) in the domain, and
protects privacy. 

\end{itemize}

\subsection{Employ Linters} \label{Sec_Linters}

A linter is a tool that statically analyzes code to find programming errors,
suspicious constructs, and stylistic inconsistencies \citep{Wikipedia2022_Lint}.
Linters can be used as an ad hoc check for code files, but they really come into
their own when used as part of a CI system, as discussed in
Section~\ref{Sec_ContinuousIntegration}. Almost none of the research software
guidelines that we consulted, summarized in Section~\ref{Sec_CompareArtifacts},
mention linters.  The one exception is \citet{ThielEtAl2020}.  Despite the lack
of mention in the guidelines, we believe that linters have the potential to
improve code quality at a relatively low cost.  

Linters have the following benefits: finding potential bugs, finding memory
leaks, improving performance, standardizing code with respect to formatting,
removing silly errors before code reviews, and catching potential security
issues \citep{SourceLevel2022_Lint}. Most popular programming languages have an
accompanying linter.  For example, Python has the options of PyLint, flake8 and
Black \citep{Zadka2018}.

We recommend the use of linters because they are relatively easy to incorporate
into a developer's workflow, and they address several MI pain points
(Section~\ref{painpoints}).  For instance, linters address the lack of
development time (\ppref{P_LackDevTime}) by increasing the developer's
productive time via guarding against making frustrating, time-consuming, mundane
mistakes.  Moreover, since a linter can include rules that capture the wisdom of
senior programmers, it can help newer developers avoid common mistakes. With
respect to the technology hurdle pain point (\ppref{P_TechnologyHurdles}),
linters can assist with the move toward web applications
(Section~\ref{sec_webapps}).  For instance, ESLint in React is a pluggable
linter that lets the developer know if they have imported something and not used
it, if a function could be short-handed, if there are indentation
inconsistencies, etc. \citep{Whitehouse2018}. By insisting on code
standardization linters can reduce technical debt and thus improve
maintainability (\qref{Q_Maintainability}). Although linters are tools for code
analysis, the idea of statically checking for adherence to basic rules can be
extended to check documentation. \citet{SmithEtAl2018_StatSoft} shows how the
use of tools to enforce documentation standards partially explains the
relatively higher quality of statistical tools that are part of the
Comprehensive R Archive Network (CRAN).

\subsection{Conduct a Mix of Rigorous and Informal Peer Reviews} \label{Sec_PeerReview}

We advocate incorporating peer review into the development process, as
frequently recommended for research software \citep{HerouxEtAl2008, Givler2020,
OrvizEtAl2017, USGS2019}. In most cases a modern, lightweight review, should be
adequate.  Modern code review is informal, tool-based, asynchronous, and focused
on reviewing code changes \citep{SadowskiEtAl2018}. Managing a project via
GitHub pull requests is an example of a modern approach to reviewing code.
Software development organizations have moved to this lightweight style of code
review because of the inefficiencies of rigorous inspections
\citep{RigbyAndBird2013}.  However, for important parts of the code, developers
may benefit from mixing in a more rigorous approach. 

\citet{Fagan1976} began work on rigorous review via code inspection.  Elements
of a typical inspection include reviewing the code against a checklist (checking
the consistency of variable names, look for terminating loops, etc.), performing
specific review tasks (such as summarizing the code's purpose, cross-referencing
the code to the technical manual, creating a data dictionary for a given module,
etc.) Rigorous inspection finds 60-65\% of latent defects on average, and often
tops 85\% in defect removal efficiency \citep{Jones2008}. The success rate of
code inspection is generally higher than most forms of testing, which average
between 30 --- 35\% for defect removal efficiency \citep{EbertAndJones2009,
Jones2008}. For research software, \citep{KellyAndShepard2000} show a task based
inspection approach can be effective. Task based inspection is an ideal fit with
an issue tracking system, like GitHub.  The review tasks can be issues, so that
they can be easily assigned, monitored and recorded. Potential issues include
assigning junior developers to test getting-started tutorial and installation
instructions.

As indicated in Section~\ref{Sec_CompareMethodologies} some MI projects use
modern code review, via issue tracking and the use of GitHub.  Those MI projects
not incorporating modern code review would likely benefit by adopting it.
Although a rigorous code inspection is likely not worth the required resources,
for critical parts of the code, developers may want to adopt a more rigorous
approach. For instance, developers may drop the modern trend of
asynchronous review and instead occasionally use synchronous review to help
uncover errors and disseminate best practices throughout the team.  For
instance, teams could periodically meet, either in-person or virtually, and have
junior members walk through their code.  In-person reviews will likely help
realize the benefits of modern code review noticed by
\citet{BirdAndBacchelli2013}: defect detection, knowledge transfer, increased
team awareness, and creation of alternative solutions to problems.

Due to improving code quality and increasing knowledge transfer, peer review
addresses the same pain points and qualities as linters
(Section~\ref{Sec_Linters}): \ppref{P_LackDevTime}, \ppref{P_TechnologyHurdles},
and \qref{Q_Maintainability}. Peer review can potentially find misunderstandings
in how the code implements the required theory, which will improve the
software's correctness (\ppref{P_Correctness}). The benefits of peer review for
addressing pain points can be increased by extending the review from just code,
to also reviewing all software artifacts, including documentation, build
scripts, test cases and the development process itself.

\subsection{Design For Change} \label{Sec_DesForChange}

In our ``state of the practice'' assessment exercise for LBM software
\citep{SmithEtAl2024}, we noticed that LBM developers implicitly used
modularization based on the principle of design for change to improve
maintainability (\qref{Q_Maintainability}).  We recommend that MI developers use
the same principle for their modularizations.  Although the advice to modularize
research software to handle complexity is common \citep{WilsonEtAl2014,
StewartEtAl2017, Storer2017}, specific guidelines on how to divide the software
into modules is less prevalent.  Not every decomposition is a good design for
supporting change, as shown by \citet{Parnas1972a}.  For instance, a design with
low cohesion and high coupling \citep[p.\ 48]{GhezziEtAl2003} will make change
difficult. Especially in research software, where change is inevitable,
designers need to produce a modularization that supports change.
\citet{JungEtAl2022} points out that ocean modelling software is currently
feeling the pain of not emphasizing modularization in legacy code.

Specific examples of design for change for LBM software \citep{SmithEtAl2024}
include the following:

\begin{itemize}
\item \href{https://github.com/pylbm/pylbm}{pyLBM} has decoupled geometries and
models of their system using abstraction and modularization of the source code,
to make it easy to add new features.  The pyLBM design allows for independent
changes to the geometry and the model.  pyLBM also redeveloped data structures
to ease future change. 
\item \href{https://github.com/CFD-GO/TCLB}{TCLB} \citep{rokicki2016adjoint} is
designed to allow for the addition of some LBM features, but changes to major
aspects of the system would be difficult. For example, ``implementing a new
model will be an easy contribution'', but changes to the ``Cartesian mesh … will
be a nightmare'' \citep{SmithEtAl2024}. The design of TCLB highlights that not
every conceivable change needs to be supported, only the likely changes.  
\end{itemize}

As the LBM examples above illustrate, developers can accomplish design for
change by first identifying likely changes, either implicitly or explicitly, and
second by hiding each likely change behind a well-defined module interface. This
approach mirrors the recommendations from \citet{Parnas1972a}.
Section~\ref{Sec_CompareArtifacts} lists ideas for how to document the design,
including the likely changes, so that they are more visible to others.

\subsection{Assurance Case} \label{AssuranceCases}

To ensure correctness (\ppref{P_Correctness} and to achieve the quality of
reproducibility (\qref{Q_Reproducibility}), we recommend considering the use of
assurance cases.  \citet{RinehartEtAl2015} defines an assurance case as ``[a]
reasoned and compelling argument, supported by a body of evidence, that a
system, service, or organization will operate as intended for a defined
application in a defined environment.''  An assurance cases provide an organized
and explicit argument that the software and its documentation achieves desired
qualities, such as correctness and reproducibility.  Although assurance cases
have been successfully employed for safety critical systems
\citep{RinehartEtAl2015}, this technique is relatively new for research software
\citep{SmithEtAl2020_AC, Smith2018}.

One way to present an assurance case is through the Goal Structuring Notation
(GSN) \citep{Spriggs2012}, which make arguments clear, easy to read and, hence,
easy to challenge. GSN starts with a Top Goal (Claim), like ``Program X delivers
correct outputs when used for its intended use/purpose in its intended
environment.''  We then decompose this top goal into Sub-Goals, which themselves
may be further decomposed.  The purpose of the decomposition is to take the
abstract higher level goals and bring them down to something concrete that can
be proven.  The decomposition ends with the terminal Sub-Goals that are
supported by Solutions (Evidence). Typical evidence will consist of documents,
expert reviews, test case results, peer review, etc.  Within the GSN framework,
there are also strategy blocks, which describe the rationale for decomposing a
Goal or Sub-Goal into more detailed Sub-Goals. A common tool for creating,
editing, and presenting, a GSN argument is \href{https://astah.net/} {Astah}.  

\begin{figure}[!ht]
\centering
\includegraphics[width=1.0\textwidth]{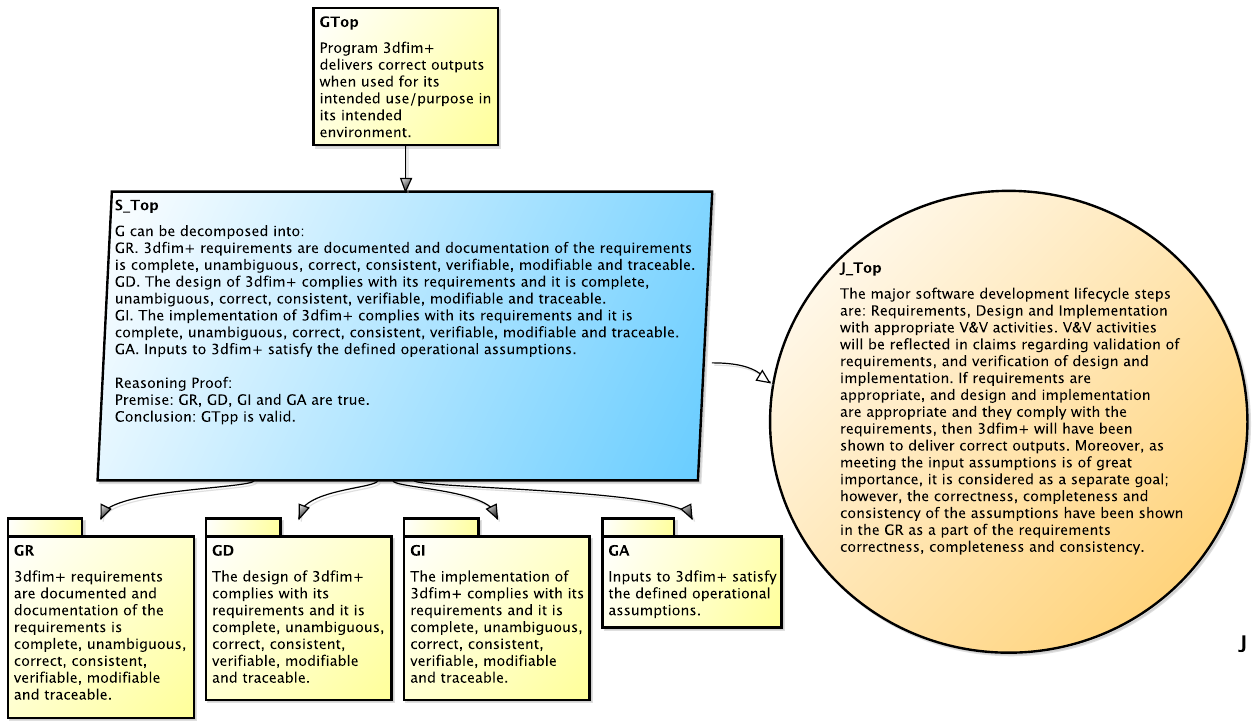}
\caption{Top Goal of the assurance case and its sub-goals}
\label{TopGoal}
\end{figure}

\citet{SmithEtAl2020_AC} shows the example of arguing for the correctness of the
Analysis of Functional NeuroImages (AFNI) package 3dfim+ \citep{Ward2000}.
3dfim+ analyzes the activity of the brain by computing the correlation between
an ideal signal and the measured brain signal for each voxel. The assurance case
for the correctness of 3dfim+ has the top level decomposed into four sub-goals,
as shown in Figure~\ref{TopGoal}.  This example follows the same pattern as used
for medical devices \citep{Wassyng2015}.  The first sub-goal (GR) argues for the
quality of the documentation of the requirements.  The second sub-goal (GD) says
that the design complies with the requirements and the third proposes that the
implementation also complies with the requirements.  The fourth sub-goal (GA)
claims that the inputs to 3dfim+ will satisfy the operational assumptions, since
we need valid input to make an argument for the correctness of the output.

Preparing an assurance case for the pre-existing 3dfim+ software shows the value
of an assurance cases for research software. Although \citet{SmithEtAl2020}
found no errors in the output of the existing software, the rigour of the
proposed approach did lead to finding ambiguities and omissions in the existing
documentation, such as missing information on the coordinate system convention.
In addition, a potential concern for the software itself was identified from the
GA argument: running the software does not produce any warning about the
obligation of the user to provide data that matches the parametric statistical
model employed for the correlation calculations.

\subsection{Generate All Things} \label{Sec_GenAllThings}

To address developer pain points, we propose automatically generating MI code
and its documentation via a ``Generate All Things'' (GAT) approach. A GAT
approach uses models to capture knowledge in domains such as physics, computing,
mathematics, documentation, and certification.  Developers combine and transform
knowledge via explicit ``recipes'', which weave it together to generate the
desired code, documentation, test cases, inspection reports and build scripts. A
recipe can even potentially be written to generate an assurance case
(Section~\ref{AssuranceCases}). Our definition of GAT implies generation of all
software artifacts, not just the code. GAT moves development to a higher level
of abstraction so that domain experts can work without concern for low-level
implementation details. GAT allows developers to optimally generate code and
documentation, reduce the likelihood of errors, eliminate redundancy, and
automate maintenance. With a GAT approach MI developers can experiment with
different algorithm choices, different input formats, different outputs formats,
etc.

Part of GAT is code generation.  In the future, some believe that code
generation will transform coding, documentation, design, and verification
\citep{JohansonAndHasselbring2018, Smith2018}. GAT removes the distraction of
writing software, allowing developers to focus on their science. A GAT approach
removes the maintenance headaches of documentation duplicates and near
duplicates \citep{LucivEtAl2018}, since developers capture knowledge once and
transform it as needed.  Code generation has previously been applied to improve
research software, such as linear algebra software packages like Blitz++
\citep{Veldhuizen1998}, and ATLAS (Automatically Tuned Linear Algebra Software)
\citep{WhaleyEtAl2001}.  Software/hardware generation has been applied for
digital signal processing in Spiral \citep{Pueschel2001}. A generative approach
has also been used for a family of efficient, type-safe Gaussian elimination
algorithms \citet{Carette2008}. \citep{LoggEtAl2012} use code generation when
solving partial differential equations in FEniCS (Finite Element and
Computational Software). \citet{MatkerimEtAl2013} and \citet{OberEtAl2018} use
code generation for High Performance Computing (HPC), using UML (Unified Modelling
Language) for their domain models. \citet{SzymczakEtAl2016} presents initial work on
the GAT approach, \citet{SmithAndCarette2021-BRIC} presents a motivating example,
and \citep{CaretteEtAl2021-Drasil} provides a prototype.

A GAT approach addresses multiple pain points.  For example, a generative
approach can decrease development time (\ppref{P_LackDevTime}) by automation,
once the necessary infrastructure is in place.  The GAT approach addresses the
technology related pain point (\ppref{P_TechnologyHurdles}) because technology
information can be captured in the models and transformed as needed. To ensure
correctness (\ppref{P_Correctness}), a GAT approach should be correct by
construction.  If there are mistakes, GAT has the advantage that they are
propagated throughout the generated artifacts, which greatly increases the
chance that someone will notice the mistake.  Maintainability
(\qref{Q_Maintainability}) is addressed because developers write the recipes
used for generation at a high level making them relatively easy to change.
Usability (\ppref{P_Usability}) is addressed because of the emphasis on
generating up-to-date documentation.  GAT facilitates reproducibility
(\qref{Q_Reproducibility}) because at any time all the code and documentation
can be regenerated.  The generator can include explicit traceability to show the
dependence of the software on specific versions of software libraries.

\section{Threats to Validity} \label{sec_threats_to_validity}

Below we categorize and list the threats to validity that we have identified.
Our categories come from an analysis of software engineering secondary studies
by \citet{AmpatzoglouEtAl2019}, where a secondary study analyzes the data from a
set of primary studies.  \citet{AmpatzoglouEtAl2019} is appropriate because a
common example of a secondary study is a systematic literature review. Our
methodology is a systematic software review --- the primary studies are the
software packages, and our work collects and analyzes these primary studies.  We
identified similar threats to validity in our assessment of the state of the
practice of Lattice Boltzmann Solvers \citep{SmithEtAl2024}.

\subsection{Reliability}

A study is reliable if repetition of the study by different researchers using
the original study's methodology would lead to the same results
\citep{RunesonAndHost2009}. Reliability means that data and analysis are
independent of the specific researcher(s) doing the study.  For the current
study the identified reliability related threats are as follows:

\begin{itemize}
\item One individual does the manual measures for all packages. A different
evaluator might find different results, due to differences in abilities,
experiences, and biases.
\item The manual measurements for the full set of packages took several months.
Over this time the software repositories may have changed and the reviewer's
judgement may have drifted.
\end{itemize}

In \citet{SmithEtAl2016} we reduced concern over the reliability risk associated
with the reviewer's judgement by demonstrating that the measurement process is
reasonably reproducible.  In \citet{SmithEtAl2016} we graded five software
products by two reviewers. Their rankings were almost identical. As long as each
grader uses consistent definitions, the relative comparisons in the AHP results
will be consistent between graders.

\subsection{Construct Validity}

\citet{RunesonAndHost2009} defines construct validity as the adopted
metrics representing what they are intended to measure. Our construct threats are
often related to how we assume our measurements influences the various software
qualities, as summarized in Section~\ref{sec_grading_software}. Specifically,
our construct validity related threats include the following:

\begin{itemize}
\item We make indirect measurement of software qualities since meaningful direct
measures for qualities like maintainability, reusability and verifiability, are
unavailable.  We follow the usual assumption that developers achieve higher
quality by following procedures and adhering to standards \citep[p.\
112]{VanVliet2000}.
\item As mentioned in Section~\ref{sec_result_installability}, we could not
install or build \textit{dwv}, \textit{GATE}, and \textit{DICOM Viewer}. We used
a deployed on-line version for \textit{dwv}, a VM version for \textit{GATE}, but
no alternative for \textit{DICOM Viewer}. We might underestimate their rank due
to these technical issues.
\item Measuring software robustness only involved two pieces of data. This is
likely part of the reason for limited variation in the robustness scores
(Figure~\ref{fg_robustness_scores}). We could add more robustness data by
pushing the software to deal with more unexpected situations, like a broken
Internet connection, but this would require a larger investment of measurement
time. 
\item We may have inaccurately estimated maintainability by assuming a higher
ratio of comments to source code improves maintainability. Moreover, we assumed
that maintainability is improved if a high percentage of issues are closed, but
a project may have a wealth of open issues, and still be maintainable.
\item We assess reusability by the number of code files and LOC per file. This
measure is indicative of modularity, but it does not necessarily mean a good
modularization. The modules may not be general enough to be easily reused, or
the formatting may be poor, or the understandability of the code may be low.
\item The understandability measure relies on 10 random source code files, but
the 10 files will not necessarily be representative. 
\item As discussed in Section~\ref{Sec_OverallQ}, our overall AHP ranking makes
the unrealistic assumption of equal weighting.
\item We approximated popularity by stars and watches
(Section~\ref{Sec_VsCommunityRanking}), but this assumption may not be valid. 
\item As mentioned in Section~\ref{sec_interview_methods}, one interviewee was
too busy to participate in a full interview, so they provided written answers
instead. Since we did not have the chance to explain our questions or ask them
follow-up questions, there is a possibility of misinterpretation of the
questions or answers.
\item In building Table~\ref{Tbl_Guidelines} some judgement was necessary on our
part, since not all guidelines use the same names for artifacts that contain
essentially the same information.
\end{itemize}

\subsection{Internal Validity} \label{Sec_InternalValidity}

Internal validity means that discovered causal relations are trustworthy and
cannot be explained by other factors \citep{RunesonAndHost2009}. In our
methodology the internal validity threats include the following:

\begin{itemize}
\item In our search for software packages
(Section~\ref{sec_software_selection}), we may have missed a relevant package.
\item Our methodology assumes that all relevant software development activities
will leave a trace in the repositories, but this is not necessarily true. For
instance, the possibility exists that CI usage was higher than what we observed
through the artifacts (Section~\ref{Sec_CompareTools}). As another example,
although we saw little evidence of requirements
(Section~\ref{Sec_CompareTools}), maybe teams keep this kind of information
outside their repos, possibly in journal papers or technical reports.
\item We interviewed a relatively small sample of 8 teams.  Their pain points
(Section~\ref{painpoints}) may not be representative of the rest of their
community.
\end{itemize}

\subsection{External Validity}

If the results of a study can be generalized (applied) to other
situations/cases, then the study is externally valid \citep{RunesonAndHost2009}.
We are confident that our search was exhaustive.  We do not believe that we
missed any highly popular examples.  Therefore, the bulk of our validity
concerns are internal (Section~\ref{Sec_InternalValidity}).
However, our hope is that the trends observed, and the lessons learned for MI
software can be applied to other research software.  With that in mind we
identified the following threat to external validity:

\begin{itemize}
\item We cannot generalize our results if the development of MI software is
fundamentally different from other research software.
\end{itemize}

Although there are differences, like the importance of data privacy for MI data,
we found the approach to developing LBM software \citep{SmithEtAl2024} and MI
software to be similar.  Except for the domain specific aspects, we believe that
the trends observed in the current study are externally valid for other research
software.

\section{Future Work} \label{Sec_FutureWork}

The following recommendations for future state of the practice measurement
exercises, for MI or for other domains, could address some threats to validity
mentioned above.  Moreover, some ideas may make the data collection more
efficient.  

\begin{itemize}
    \item We would like to make surface measurements less shallow. For example:
    \begin{itemize}
        \item Surface reliability: our current measurement relies on the
        processes of installation and getting started tutorials. However, not
        all software needs installation or has a getting started tutorial. We
        could devise a list of operation steps (with the help of the Domain
        Expert), perform the same operations with each software, and record any
        errors.
        \item Surface robustness: we used damaged images as inputs for this
        measuring MI software. This process is similar to fuzz testing
        \citep{enwiki:1039424308}, which is one type of fault injection
        \citep{enwiki:1039005082}. We may adopt more fault injection methods,
        and identify tools and libraries to automate this process.
        \item Surface usability: we can design usability tests and test all
        software projects with end-users. The end-users can be volunteers and
        domain experts. Ideas for getting started are available in
        \citet{SmithEtAl2021}.
        \item Surface understandability: our current method does not require
        understanding the source code. As software engineers, perhaps we can
        select a small module of each project, read the source code and
        documentation, try to understand the logic, and score the ease of the
        process.  
        \item Maintainability: we can add a measure modifiability as part of the
        measurement of maintainability.  An experiment could be conducted asking
        participants to make modifications, observing the study subjects during
        the modifications, testing the resulting software and surveying the
        participants \citep{SmithEtAl2021}.
    \end{itemize}
	\item We can further automate the measurements on the grading template. For
	example, with automation scripts and the GitHub API, we may save significant
	time on retrieving the GitHub metrics through a GitHub Metric Collector.
	This Collector can take GitHub repository links as input, automatically
	collect metrics from the GitHub API, and record the results.
	\item We can improve some interview questions. Some examples are:
	\begin{itemize}
	    \item In one question we ask, ``Do you think improving this process can
	    tackle the current problem?''  The problem is that this is a yes-or-no
	    question, which is not informative. We could change the question to ``By
	    improving this process, what current problems can be tackled?''; 
	    \item We can ask for more details about the modular approach, such as
	    ``What principles did you use to divide code into modules? Can you
	    describe an example of using your principles?''.
	\end{itemize}
	\item We can better organize the interview questions. Since we use audio
	conversion tools to transcribe the answers, we should make the transcription
	easier to read. For example, we can order them together for questions about
	the five software qualities and compose a similar structure for each.
	\item We can mark the follow-up interview questions with keywords. For
	example, say ``this is a follow-up question'' every time asking one. Thus,
	we record this sentence in the transcription, and it will be much easier to
	distinguish the follow-up questions from the 20 designed questions.
\end{itemize}

\section{Conclusions} \label{ch_conclusions}

We analyzed the state of the practice for the MI domain with the goal of
understanding current practice, answering our ten research questions
(Section~\ref{sec_motivation}) and providing recommendations for current and
future projects.  Our methods in Section~\ref{ch_methods} form a general process
to evaluate domain-specific software, that we apply to the specific domain of MI
software. We identified 48 MI software candidates, then, with the help of the
Domain Expert selected 29 of them to our final list. 

Section~\ref{ch_results} lists our measurement results for ranking the 29
projects for nine software qualities. Our ranking results appear credible since
they are mostly consistent with the ranking from the scientific community
implied by the GitHub stars-per-year metric. As discussed in
Section~\ref{Sec_VsCommunityRanking}, four of the top five software projects
appear in both our list and in the GitHub popularity list.  Moreover, our top
five packages appear among the first eight positions on the GitHub list.  The
noteworthy discrepancies between the two lists are for the packages that we were
unable to install (\textit{dwv} and \textit{Dicom Viewer}).

Based on our grading scores \textit{3D Slicer}, \textit{ImageJ}, \textit{Fiji}
and \textit{OHIF Viewer} are the top four software performers.  However, the
separation between the top performers and the others is not extreme.  Almost all
packages do well on at least a few qualities, as shown in
Table~\ref{topperformerstable}, which summarizes the packages ranked first and
second for each quality. Almost 70\% (20 of 29) of the software packages appear
in the top two for at least two qualities.  The only packages that do not appear
in Table~\ref{topperformerstable}, or only appear once, are \textit{Papaya},
\textit{MatrixUser}, \textit{MRIcroGL}, \textit{XMedCon}, \textit{dicompyler},
\textit{DicomBrowser}, \textit{AMIDE}, \textit{3DimViewer}, and
\textit{Drishti}. The shortness of this list suggests parity with respect to
adoption of best practices for MI software overall.

\begin{table}[ht!]
	\begin{center}
	\renewcommand{\arraystretch}{1.8}
	\begin{tabular}{ p{3cm}p{13cm} }
		\toprule

		Quality & Ranked 1st or 2nd\\

		\midrule

		Installability & 3D Slicer, BioImage Suite Web, Slice:Drop, INVESALIUS\\

		\pbox{3.0cm}{Correctness \\ and Verifiability} & OHIF Viewer, 3D
		Slicer, ImageJ\\

		Reliability & SMILI, ImageJ, Fiji, 3D Slicer, Slice:Drop, OHIF
		Viewer\\

		Robustness & XMedCon, Weasis, SMILI, ParaView, OsiriX Lite,
		MicroView, medInria, ITK-SNAP, INVESALIUS, ImageJ, Horos, Gwyddion,
		Fiji, dicompyler, DicomBrowser, BioImage Suite Web, AMIDE, 3DimViewer,
		3D Slicer, OHIF Viewer, DICOM Viewer\\

		Usability & 3D Slicer, ImageJ, Fiji, OHIF Viewer, ParaView,
		INVESALIUS, Ginkgo CADx, SMILI, OsiriX Lite, BioImage Suite Web,
		ITK-SNAP, medInria, MicroView, Gwyddion\\

		Maintainability & 3D Slicer, Weasis, ImageJ, OHIF Viewer, ParaView\\

		Reusability & 3D Slicer, ImageJ, Fiji, OHIF Viewer, SMILI, dwv, BioImage
		Suite Web, GATE, ParaView\\

		\pbox{3.0cm}{Understandability} & 3D Slicer, ImageJ, Weasis,
		Fiji, Horos, OsiriX Lite, dwv, Drishti, OHIF Viewer, GATE, ITK-SNAP,
		ParaView, INVESALIUS\\

		\pbox{3.0cm}{Visibility and \\Transparency} & ImageJ, 3D Slicer, Fiji\\

		Overall Quality & 3D Slicer, ImageJ\\

		\bottomrule		
	\end{tabular}
	\caption{Top performers for each quality (sorted by order of quality
	measurement)} \label{topperformerstable}
	\end{center}
\end{table} 

For insight into devising future methods and tools, we interviewed nine
developers (from eight teams) to learn about their pain points
(Section~\ref{painpoints}).  We also discussed qualities of potential concern.
The identified pain points and qualities of concern include: 

\begin{description}
\item [\ppref{P_LackDevTime}] Lack of development time, %P1
\item [\ppref{P_LackFunding}] Lack of funding, %P2
\item [\ppref{P_TechnologyHurdles}] Technology hurdles, %P3
\item [\ppref{P_Correctness}] Ensuring correctness, %P4
\item [\ppref{P_Usability}] Usability, %P5
\item [\qref{Q_Maintainability}] Quality of maintainability, and %Q1
\item [\qref{Q_Reproducibility}] Quality of reproducibility. %Q2
\end{description}  

Despite the pain points, overall MI software is in a healthy state for software
development practices.  In our survey of the selected projects we observed 88\%
of the documentation artifacts recommended by research software development
guidelines (Section~\ref{Sec_CompareArtifacts}).  With respect to tools, MI is
keeping pace with other research software with 100\% of the projects using
version control (with 93\% specifically using git)
(Section~\ref{Sec_CompareTools}).  We observed that the MI developers tend to
follow the typical research software trend of using a quasi-agile software
development process (Section~\ref{Sec_CompareMethodologies}).

Although the state of the practice for MI software is healthy, we did notice
areas where practice seems to lag behind the research software development
guidelines.  For instance, the guidelines recommend three artifacts that were
not observed: uninstall instructions, test plans, and requirements
documentation. We observed the following recommended artifacts, but only rarely:
contributing file, developer code of conduct, code style guidelines, product
roadmap, design documentation, and API documentation
(Section~\ref{Sec_CompareArtifacts}). Although software development tool use
seems healthy, we found the use of CI/CD behind typical usage rates (17\% of the
projects used CI/CD) (Section~\ref{Sec_CompareTools}).  With respect to the
development process, developer identified areas for improvement included testing
(only 50\% of projects were identified to have unit testing) and documentation
(only three out of nine developers felt their documentation was clear enough)
(Section~\ref{Sec_CompareMethodologies}).

Our interviewees proposed strategies to improve the state of the practice, to
address the identified pain points, and to improve software quality.  To their
list (Section~\ref{painpoints}) we added some of our own recommended strategies
(Section~\ref{ch_recommendations}).  Below we summarize the proposed strategies,
with traceability to where we discuss the strategy, and to the relevant pain
points.

\begin{enumerate}

\item Increase documentation to address \ppref{P_LackDevTime},
\ppref{P_TechnologyHurdles}, \ppref{P_Usability}, \qref{Q_Maintainability},
\qref{Q_Reproducibility} (Section~\ref{painpoints})

\item Increase testing by enriching datasets to address \ppref{P_Correctness},
\qref{Q_Reproducibility}
(Section~\ref{painpoints},~\ref{sec_recommendations_testing_dataset})

\item Increase modularity to address \qref{Q_Maintainability} (Section~\ref{painpoints})

\item Use continuous integration to address \ppref{P_LackDevTime},
\ppref{P_Correctness}, \qref{Q_Reproducibility}
(Section~\ref{painpoints},~\ref{Sec_ContinuousIntegration})

\item Move to web applications to address \ppref{P_TechnologyHurdles}
(Section~\ref{painpoints},~\ref{sec_webapps})

\item Employ linters to address \ppref{P_LackDevTime},
\ppref{P_TechnologyHurdles}, \qref{Q_Maintainability}
(Section~\ref{Sec_Linters})

\item Peer reviews to address \ppref{P_LackDevTime},
\ppref{P_TechnologyHurdles}, \ppref{P_Correctness}, \qref{Q_Maintainability}
(Section~\ref{Sec_PeerReview})

\item Design for change to address \qref{Q_Maintainability} (Section~\ref{Sec_DesForChange})

\item Assurance case to address \ppref{P_Correctness}, \qref{Q_Reproducibility}
(Section~\ref{AssuranceCases})

\item Generate all things to address \ppref{P_LackDevTime},
\ppref{P_TechnologyHurdles}, \ppref{P_Correctness}, \ppref{P_Usability},
\qref{Q_Maintainability} and \qref{Q_Reproducibility}.
(Section~\ref{Sec_GenAllThings})

\end{enumerate}

\section*{Acknowledgements}

We would like to thank Peter Michalski and Oluwaseun Owojaiye for fruitful
discussions on topics relevant to this paper.  We would also like to thank Jason
Balaci for advice on web applications.

\section*{Conflict of Interest}

On behalf of all authors, the corresponding author states that there is no
conflict of interest.

\bibliographystyle{ACM-Reference-Format}
\bibliography{MedImageSoft_SOP}


\begin{thebibliography}{203}

%%% the Web address.

\ifx \showCODEN    \undefined \def \showCODEN     #1{\unskip}     \fi
\ifx \showDOI      \undefined \def \showDOI       #1{#1}\fi
\ifx \showISBNx    \undefined \def \showISBNx     #1{\unskip}     \fi
\ifx \showISBNxiii \undefined \def \showISBNxiii  #1{\unskip}     \fi
\ifx \showISSN     \undefined \def \showISSN      #1{\unskip}     \fi
\ifx \showLCCN     \undefined \def \showLCCN      #1{\unskip}     \fi
\ifx \shownote     \undefined \def \shownote      #1{#1}          \fi
\ifx \showarticletitle \undefined \def \showarticletitle #1{#1}   \fi
\ifx \showURL      \undefined \def \showURL       {\relax}        \fi
% invisible to TeX
\providecommand\bibfield[2]{#2}
\providecommand\bibinfo[2]{#2}
\providecommand\natexlab[1]{#1}
\providecommand\showeprint[2][]{arXiv:#2}

\bibitem[Abdalla(2016)]%
        {Abdalla2016}
\bibfield{author}{\bibinfo{person}{Safia Abdalla}.} \bibinfo{year}{2016}\natexlab{}.
\newblock \bibinfo{title}{A template for creating open source contributor guidelines}.
\newblock \bibinfo{howpublished}{\url{https://opensource.com/life/16/3/contributor-guidelines-template-and-tips}}.
\newblock


\bibitem[Ackroyd et~al\mbox{.}(2008)]%
        {AckroydEtAl2008}
\bibfield{author}{\bibinfo{person}{Karen~S. Ackroyd}, \bibinfo{person}{Steve~H. Kinder}, \bibinfo{person}{Geoff~R. Mant}, \bibinfo{person}{Mike~C. Miller}, \bibinfo{person}{Christine~A. Ramsdale}, {and} \bibinfo{person}{Paul~C. Stephenson}.} \bibinfo{year}{2008}\natexlab{}.
\newblock \showarticletitle{Scientific Software Development at a Research Facility}.
\newblock \bibinfo{journal}{\emph{IEEE Software}} \bibinfo{volume}{25}, \bibinfo{number}{4} (\bibinfo{date}{July/August} \bibinfo{year}{2008}), \bibinfo{pages}{44--51}.
\newblock


\bibitem[Administration(2021)]%
        {FDA2021}
\bibfield{author}{\bibinfo{person}{U.S. Food \&~Drug Administration}.} \bibinfo{year}{2021}\natexlab{}.
\newblock \bibinfo{title}{Medical Imaging}.
\newblock \bibinfo{howpublished}{\url{https://www.fda.gov/radiation-emitting-products/radiation-emitting-products-and-procedures/medical-imaging}}.
\newblock
\newblock
\shownote{[Online; accessed 25-July-2021]}.


\bibitem[Afsar(2021)]%
        {Afsar2021}
\bibfield{author}{\bibinfo{person}{Aysel Afsar}.} \bibinfo{year}{2021}\natexlab{}.
\newblock \bibinfo{title}{DICOM Viewer}.
\newblock \bibinfo{howpublished}{\url{https://github.com/ayselafsar/dicomviewer}}.
\newblock
\newblock
\shownote{[Online; accessed 27-May-2021]}.


\bibitem[Ahrens et~al\mbox{.}(2005)]%
        {Ahrens2005}
\bibfield{author}{\bibinfo{person}{J. Ahrens}, \bibinfo{person}{Berk Geveci}, {and} \bibinfo{person}{Charles Law}.} \bibinfo{year}{2005}\natexlab{}.
\newblock \showarticletitle{ParaView: An End-User Tool for Large Data Visualization}.
\newblock \bibinfo{journal}{\emph{Visualization Handbook}} (\bibinfo{date}{01} \bibinfo{year}{2005}).
\newblock


\bibitem[AlNoamany and Borghi(2018)]%
        {AlNoamanyAndBorghi2018}
\bibfield{author}{\bibinfo{person}{Yasmin AlNoamany} {and} \bibinfo{person}{John~A. Borghi}.} \bibinfo{year}{2018}\natexlab{}.
\newblock \showarticletitle{Towards computational reproducibility: researcher perspectives on the use and sharing of software}.
\newblock \bibinfo{journal}{\emph{PeerJ Computer Science}} \bibinfo{volume}{4}, \bibinfo{number}{e163} (\bibinfo{date}{September} \bibinfo{year}{2018}), \bibinfo{pages}{1--25}.
\newblock


\bibitem[Amorim et~al\mbox{.}(2015)]%
        {Amorim2015}
\bibfield{author}{\bibinfo{person}{Paulo Amorim}, \bibinfo{person}{Thiago Franco~de Moraes}, \bibinfo{person}{Helio Pedrini}, {and} \bibinfo{person}{Jorge Silva}.} \bibinfo{year}{2015}\natexlab{}.
\newblock \showarticletitle{InVesalius: An Interactive Rendering Framework for Health Care Support}. \bibinfo{pages}{10}.
\newblock
\showISBNx{978-3-319-27856-8}
\urldef\tempurl%
\url{https://doi.org/10.1007/978-3-319-27857-5_5}
\showDOI{\tempurl}


\bibitem[Ampatzoglou et~al\mbox{.}(2019)]%
        {AmpatzoglouEtAl2019}
\bibfield{author}{\bibinfo{person}{Apostolos Ampatzoglou}, \bibinfo{person}{Stamatia Bibi}, \bibinfo{person}{Paris Avgeriou}, \bibinfo{person}{Marijn Verbeek}, {and} \bibinfo{person}{Alexander Chatzigeorgiou}.} \bibinfo{year}{2019}\natexlab{}.
\newblock \showarticletitle{Identifying, Categorizing and Mitigating Threats to Validity in Software Engineering Secondary Studies}.
\newblock \bibinfo{journal}{\emph{Information and Software Technology}}  \bibinfo{volume}{106} (\bibinfo{date}{02} \bibinfo{year}{2019}).
\newblock
\urldef\tempurl%
\url{https://doi.org/10.1016/j.infsof.2018.10.006}
\showDOI{\tempurl}


\bibitem[Angenent et~al\mbox{.}(2006)]%
        {Angenent2006}
\bibfield{author}{\bibinfo{person}{S. Angenent}, \bibinfo{person}{Eric Pichon}, {and} \bibinfo{person}{Allen Tannenbaum}.} \bibinfo{year}{2006}\natexlab{}.
\newblock \showarticletitle{Mathematical methods in medical image processing}.
\newblock \bibinfo{journal}{\emph{Bulletin (new series) of the American Mathematical Society}}  \bibinfo{volume}{43} (\bibinfo{date}{07} \bibinfo{year}{2006}), \bibinfo{pages}{365--396}.
\newblock
\urldef\tempurl%
\url{https://doi.org/10.1090/S0273-0979-06-01104-9}
\showDOI{\tempurl}


\bibitem[Archie and Marcus(2012)]%
        {Archie2012}
\bibfield{author}{\bibinfo{person}{Kevin Archie} {and} \bibinfo{person}{Daniel Marcus}.} \bibinfo{year}{2012}\natexlab{}.
\newblock \showarticletitle{DicomBrowser: Software for Viewing and Modifying DICOM Metadata}.
\newblock \bibinfo{journal}{\emph{Journal of digital imaging : the official journal of the Society for Computer Applications in Radiology}}  \bibinfo{volume}{25} (\bibinfo{date}{02} \bibinfo{year}{2012}), \bibinfo{pages}{635--45}.
\newblock
\urldef\tempurl%
\url{https://doi.org/10.1007/s10278-012-9462-x}
\showDOI{\tempurl}


\bibitem[Association(2021)]%
        {MITA2021}
\bibfield{author}{\bibinfo{person}{Medical Imaging~Technology Association}.} \bibinfo{year}{2021}\natexlab{}.
\newblock \bibinfo{title}{About DICOM: Overview}.
\newblock \bibinfo{howpublished}{\url{https://www.dicomstandard.org/about-home}}.
\newblock
\newblock
\shownote{[Online; accessed 11-August-2021]}.


\bibitem[{B. H. Menze} et~al\mbox{.}(2015)]%
        {MenzeEtAl2015}
\bibfield{author}{\bibinfo{person}{{B. H. Menze}}, \bibinfo{person}{{A. Jakab}}, \bibinfo{person}{{S. Bauer}}, \bibinfo{person}{{J. Kalpathy-Cramer}}, \bibinfo{person}{{K. Farahani}}, \bibinfo{person}{{J. Kirby}}, \bibinfo{person}{{Y. Burren}}, \bibinfo{person}{{N. Porz}}, \bibinfo{person}{{J. Slotboom}}, \bibinfo{person}{{R. Wiest}}, \bibinfo{person}{{L. Lanczi}}, \bibinfo{person}{{E. Gerstner}}, \bibinfo{person}{{M. -A. Weber}}, \bibinfo{person}{{T. Arbel}}, \bibinfo{person}{{B. B. Avants}}, \bibinfo{person}{{N. Ayache}}, \bibinfo{person}{{P. Buendia}}, \bibinfo{person}{{D. L. Collins}}, \bibinfo{person}{{N. Cordier}}, \bibinfo{person}{{J. J. Corso}}, \bibinfo{person}{{A. Criminisi}}, \bibinfo{person}{{T. Das}}, \bibinfo{person}{{H. Delingette}}, \bibinfo{person}{{{\c C}. Demiralp}}, \bibinfo{person}{{C. R. Durst}}, \bibinfo{person}{{M. Dojat}}, \bibinfo{person}{{S. Doyle}}, \bibinfo{person}{{J. Festa}}, \bibinfo{person}{{F. Forbes}}, \bibinfo{person}{{E. Geremia}}, \bibinfo{person}{{B. Glocker}}, \bibinfo{person}{{P. Golland}}, \bibinfo{person}{{X. Guo}}, \bibinfo{person}{{A. Hamamci}}, \bibinfo{person}{{K. M. Iftekharuddin}}, \bibinfo{person}{{R. Jena}}, \bibinfo{person}{{N. M. John}}, \bibinfo{person}{{E. Konukoglu}}, \bibinfo{person}{{D. Lashkari}}, \bibinfo{person}{{J. A. Mariz}}, \bibinfo{person}{{R. Meier}}, \bibinfo{person}{{S. Pereira}}, \bibinfo{person}{{D. Precup}}, \bibinfo{person}{{S. J. Price}}, \bibinfo{person}{{T. R. Raviv}}, \bibinfo{person}{{S. M. S. Reza}}, \bibinfo{person}{{M. Ryan}}, \bibinfo{person}{{D. Sarikaya}}, \bibinfo{person}{{L. Schwartz}}, \bibinfo{person}{{H. -C. Shin}}, \bibinfo{person}{{J. Shotton}}, \bibinfo{person}{{C. A. Silva}}, \bibinfo{person}{{N. Sousa}}, \bibinfo{person}{{N. K. Subbanna}}, \bibinfo{person}{{G. Szekely}}, \bibinfo{person}{{T. J. Taylor}}, \bibinfo{person}{{O. M. Thomas}}, \bibinfo{person}{{N. J. Tustison}}, \bibinfo{person}{{G. Unal}}, \bibinfo{person}{{F. Vasseur}}, \bibinfo{person}{{M. Wintermark}}, \bibinfo{person}{{D. H. Ye}}, \bibinfo{person}{{L. Zhao}}, \bibinfo{person}{{B. Zhao}}, \bibinfo{person}{{D. Zikic}}, \bibinfo{person}{{M. Prastawa}}, \bibinfo{person}{{M. Reyes}}, {and} \bibinfo{person}{{K. Van Leemput}}.} \bibinfo{year}{2015}\natexlab{}.
\newblock \showarticletitle{The Multimodal Brain Tumor Image Segmentation Benchmark ({BRATS})}.
\newblock \bibinfo{journal}{\emph{IEEE Transactions on Medical Imaging}} \bibinfo{volume}{34}, \bibinfo{number}{10} (\bibinfo{date}{Oct.} \bibinfo{year}{2015}), \bibinfo{pages}{1993--2024}.
\newblock


\bibitem[Bankman(2000)]%
        {Bankman2000}
\bibfield{author}{\bibinfo{person}{Isaac~N. Bankman}.} \bibinfo{year}{2000}\natexlab{}.
\newblock \showarticletitle{Preface}.
\newblock In \bibinfo{booktitle}{\emph{Handbook of Medical Imaging}}, \bibfield{editor}{\bibinfo{person}{Isaac~N. Bankman}} (Ed.). \bibinfo{publisher}{Academic Press}, \bibinfo{address}{San Diego}, \bibinfo{pages}{xi -- xii}.
\newblock
\showISBNx{978-0-12-077790-7}
\urldef\tempurl%
\url{https://doi.org/10.1016/B978-012077790-7/50001-1}
\showDOI{\tempurl}


\bibitem[{Benureau} and {Rougier}(2017)]%
        {BenureauAndRougier2017}
\bibfield{author}{\bibinfo{person}{F. {Benureau}} {and} \bibinfo{person}{N. {Rougier}}.} \bibinfo{year}{2017}\natexlab{}.
\newblock \showarticletitle{{Re-run, Repeat, Reproduce, Reuse, Replicate: Transforming Code into Scientific Contributions}}.
\newblock \bibinfo{journal}{\emph{ArXiv e-prints}} (\bibinfo{date}{Aug.} \bibinfo{year}{2017}).
\newblock
\showeprint[arxiv]{1708.08205}~[cs.GL]


\bibitem[Bilic et~al\mbox{.}(2019)]%
        {BilicEtAl2019}
\bibfield{author}{\bibinfo{person}{Patrick Bilic}, \bibinfo{person}{Patrick~Ferdinand Christ}, \bibinfo{person}{Eugene Vorontsov}, \bibinfo{person}{Grzegorz Chlebus}, \bibinfo{person}{Hao Chen}, \bibinfo{person}{Qi Dou}, \bibinfo{person}{Chi{-}Wing Fu}, \bibinfo{person}{Xiao Han}, \bibinfo{person}{Pheng{-}Ann Heng}, \bibinfo{person}{J{\"{u}}rgen Hesser}, \bibinfo{person}{Samuel Kadoury}, \bibinfo{person}{Tomasz~K. Konopczynski}, \bibinfo{person}{Miao Le}, \bibinfo{person}{Chunming Li}, \bibinfo{person}{Xiaomeng Li}, \bibinfo{person}{Jana Lipkov{\'{a}}}, \bibinfo{person}{John~S. Lowengrub}, \bibinfo{person}{Hans Meine}, \bibinfo{person}{Jan~Hendrik Moltz}, \bibinfo{person}{Chris Pal}, \bibinfo{person}{Marie Piraud}, \bibinfo{person}{Xiaojuan Qi}, \bibinfo{person}{Jin Qi}, \bibinfo{person}{Markus Rempfler}, \bibinfo{person}{Karsten Roth}, \bibinfo{person}{Andrea Schenk}, \bibinfo{person}{Anjany Sekuboyina}, \bibinfo{person}{Ping Zhou}, \bibinfo{person}{Christian H{\"{u}}lsemeyer}, \bibinfo{person}{Marcel Beetz}, \bibinfo{person}{Florian Ettlinger}, \bibinfo{person}{Felix Gr{\"{u}}n}, \bibinfo{person}{Georgios Kaissis}, \bibinfo{person}{Fabian Loh{\"{o}}fer}, \bibinfo{person}{Rickmer Braren}, \bibinfo{person}{Julian Holch}, \bibinfo{person}{Felix Hofmann}, \bibinfo{person}{Wieland~H. Sommer}, \bibinfo{person}{Volker Heinemann}, \bibinfo{person}{Colin Jacobs}, \bibinfo{person}{Gabriel Efrain~Humpire Mamani}, \bibinfo{person}{Bram van Ginneken}, \bibinfo{person}{Gabriel Chartrand}, \bibinfo{person}{An Tang}, \bibinfo{person}{Michal Drozdzal}, \bibinfo{person}{Avi Ben{-}Cohen}, \bibinfo{person}{Eyal Klang}, \bibinfo{person}{Michal~Marianne Amitai}, \bibinfo{person}{Eli Konen}, \bibinfo{person}{Hayit Greenspan}, \bibinfo{person}{Johan Moreau}, \bibinfo{person}{Alexandre Hostettler}, \bibinfo{person}{Luc Soler}, \bibinfo{person}{Refael Vivanti}, \bibinfo{person}{Adi Szeskin}, \bibinfo{person}{Naama Lev{-}Cohain}, \bibinfo{person}{Jacob Sosna}, \bibinfo{person}{Leo Joskowicz}, {and} \bibinfo{person}{Bjoern~H. Menze}.} \bibinfo{year}{2019}\natexlab{}.
\newblock \showarticletitle{The Liver Tumor Segmentation Benchmark (LiTS)}.
\newblock \bibinfo{journal}{\emph{CoRR}}  \bibinfo{volume}{abs/1901.04056} (\bibinfo{year}{2019}).
\newblock
\showeprint[arXiv]{1901.04056}
\urldef\tempurl%
\url{http://arxiv.org/abs/1901.04056}
\showURL{%
\tempurl}


\bibitem[Bird and Bacchelli(2013)]%
        {BirdAndBacchelli2013}
\bibfield{author}{\bibinfo{person}{Christian Bird} {and} \bibinfo{person}{Alberto Bacchelli}.} \bibinfo{year}{2013}\natexlab{}.
\newblock \showarticletitle{Expectations, Outcomes, and Challenges of Modern Code Review}. In \bibinfo{booktitle}{\emph{Proceedings of the International Conference on Software Engineering} (\bibinfo{edition}{proceedings of the international conference on software engineering} ed.)}. \bibinfo{publisher}{IEEE}.
\newblock
\urldef\tempurl%
\url{https://www.microsoft.com/en-us/research/publication/expectations-outcomes-and-challenges-of-modern-code-review/}
\showURL{%
\tempurl}


\bibitem[Bj{\"o}rn(2017)]%
        {Bjorn2017}
\bibfield{author}{\bibinfo{person}{Kari Bj{\"o}rn}.} \bibinfo{year}{2017}\natexlab{}.
\newblock \showarticletitle{Evaluation of Open Source Medical Imaging Software: A Case Study on Health Technology Student Learning Experience}.
\newblock \bibinfo{journal}{\emph{Procedia Computer Science}}  \bibinfo{volume}{121} (\bibinfo{date}{01} \bibinfo{year}{2017}), \bibinfo{pages}{724--731}.
\newblock
\urldef\tempurl%
\url{https://doi.org/10.1016/j.procs.2017.11.094}
\showDOI{\tempurl}


\bibitem[Boehm(2007)]%
        {boehm2007software}
\bibfield{author}{\bibinfo{person}{Barry~W Boehm}.} \bibinfo{year}{2007}\natexlab{}.
\newblock \bibinfo{booktitle}{\emph{Software engineering: Barry W. Boehm's lifetime contributions to software development, management, and research}}. Vol.~\bibinfo{volume}{69}.
\newblock \bibinfo{publisher}{John Wiley \& Sons}.
\newblock


\bibitem[Boyter(2021)]%
        {Boyter2021}
\bibfield{author}{\bibinfo{person}{Ben Boyter}.} \bibinfo{year}{2021}\natexlab{}.
\newblock \bibinfo{title}{Sloc Cloc and Code}.
\newblock \bibinfo{howpublished}{\url{https://github.com/boyter/scc}}.
\newblock
\newblock
\shownote{[Online; accessed 27-May-2021]}.


\bibitem[Brett et~al\mbox{.}(2021)]%
        {BrettEtAl2021}
\bibfield{author}{\bibinfo{person}{Alys Brett}, \bibinfo{person}{James Cook}, \bibinfo{person}{Peter Fox}, \bibinfo{person}{Ian Hinder}, \bibinfo{person}{John Nonweiler}, \bibinfo{person}{Richard Reeve}, {and} \bibinfo{person}{Robert Turner}.} \bibinfo{year}{2021}\natexlab{}.
\newblock \bibinfo{title}{Scottish Covid-19 Response Consortium}.
\newblock \bibinfo{howpublished}{\url{https://github.com/ScottishCovidResponse/modelling-software-checklist/blob/main/software-checklist.md}}.
\newblock


\bibitem[Brown(2015)]%
        {Brown2015}
\bibfield{author}{\bibinfo{person}{Titus Brown}.} \bibinfo{year}{2015}\natexlab{}.
\newblock \bibinfo{title}{Notes from``How to grow a sustainable software development process (for scientific software)''}.
\newblock \bibinfo{howpublished}{\url{http://ivory.idyll.org/blog/2015-growing-sustainable-software-development-process.html}}.
\newblock


\bibitem[Br{\"u}hschwein et~al\mbox{.}(2019)]%
        {Bruhschwein2019}
\bibfield{author}{\bibinfo{person}{Andreas Br{\"u}hschwein}, \bibinfo{person}{Julius Klever}, \bibinfo{person}{Anne-Sophie Hoffmann}, \bibinfo{person}{Denise Huber}, \bibinfo{person}{Elisabeth Kaufmann}, \bibinfo{person}{Sven Reese}, {and} \bibinfo{person}{Andrea Meyer-Lindenberg}.} \bibinfo{year}{2019}\natexlab{}.
\newblock \showarticletitle{Free DICOM-Viewers for Veterinary Medicine: Survey and Comparison of Functionality and User-Friendliness of Medical Imaging PACS-DICOM-Viewer Freeware for Specific Use in Veterinary Medicine Practices}.
\newblock \bibinfo{journal}{\emph{Journal of Digital Imaging}} (\bibinfo{date}{03} \bibinfo{year}{2019}).
\newblock
\urldef\tempurl%
\url{https://doi.org/10.1007/s10278-019-00194-3}
\showDOI{\tempurl}


\bibitem[Carette and Kiselyov(2011)]%
        {Carette2008}
\bibfield{author}{\bibinfo{person}{Jacques Carette} {and} \bibinfo{person}{Oleg Kiselyov}.} \bibinfo{year}{2011}\natexlab{}.
\newblock \showarticletitle{Multi-stage programming with {F}unctors and {M}onads: {E}liminating abstraction overhead from generic code}.
\newblock \bibinfo{journal}{\emph{Sci. Comput. Program.}} \bibinfo{volume}{76}, \bibinfo{number}{5} (\bibinfo{year}{2011}), \bibinfo{pages}{349--375}.
\newblock


\bibitem[Carette et~al\mbox{.}(2021)]%
        {CaretteEtAl2021-Drasil}
\bibfield{author}{\bibinfo{person}{Jacques Carette}, \bibinfo{person}{Spencer Smith}, \bibinfo{person}{Jason Balaci}, \bibinfo{person}{Anthony Hunt}, \bibinfo{person}{Ting-Yu Wu}, \bibinfo{person}{Samuel Crawford}, \bibinfo{person}{Dong Chen}, \bibinfo{person}{Dan Szymczak}, \bibinfo{person}{Brooks MacLachlan}, \bibinfo{person}{Dan Scime}, {and} \bibinfo{person}{Maryyam Niazi}.} \bibinfo{year}{2021}\natexlab{}.
\newblock \bibinfo{booktitle}{\emph{{Drasil}}}.
\newblock
\urldef\tempurl%
\url{https://github.com/JacquesCarette/Drasil/tree/v0.1-alpha}
\showURL{%
\tempurl}


\bibitem[Carty(2020)]%
        {Carty2020}
\bibfield{author}{\bibinfo{person}{David Carty}.} \bibinfo{year}{2020}\natexlab{}.
\newblock \bibinfo{title}{Follow Google's lead with programming style guides}.
\newblock \bibinfo{howpublished}{\url{https://www.techtarget.com/searchsoftwarequality/feature/Follow-Googles-lead-with-programming-style-guides}}.
\newblock


\bibitem[Carver et~al\mbox{.}(2007)]%
        {CarverEtAl2007}
\bibfield{author}{\bibinfo{person}{Jeffrey~C. Carver}, \bibinfo{person}{Richard~P. Kendall}, \bibinfo{person}{Susan~E. Squires}, {and} \bibinfo{person}{Douglass~E. Post}.} \bibinfo{year}{2007}\natexlab{}.
\newblock \showarticletitle{Software Development Environments for Scientific and Engineering Software: A Series of Case Studies}. In \bibinfo{booktitle}{\emph{ICSE '07: Proceedings of the 29th International Conference on Software Engineering}}. \bibinfo{publisher}{IEEE Computer Society}, \bibinfo{address}{Washington, DC, USA}, \bibinfo{pages}{550--559}.
\newblock
\showISBNx{0-7695-2828-7}
\urldef\tempurl%
\url{https://doi.org/10.1109/ICSE.2007.77}
\showDOI{\tempurl}


\bibitem[Carver et~al\mbox{.}(2022)]%
        {CarverEtAl2022}
\bibfield{author}{\bibinfo{person}{Jeffrey~C. Carver}, \bibinfo{person}{Nicholas Weber}, \bibinfo{person}{Karthik Ram}, \bibinfo{person}{Sandra Gesing}, {and} \bibinfo{person}{Daniel~S. Katz}.} \bibinfo{year}{2022}\natexlab{}.
\newblock \showarticletitle{A Survey of the State of the Practice for Research Software in the United States}.
\newblock \bibinfo{journal}{\emph{PeerJ Computer Science}} \bibinfo{number}{8:e963} (\bibinfo{year}{2022}).
\newblock
\urldef\tempurl%
\url{https://doi.org/10.7717/peerj-cs.963}
\showDOI{\tempurl}


\bibitem[Chandra et~al\mbox{.}(2018)]%
        {Chandra2018}
\bibfield{author}{\bibinfo{person}{Shekhar Chandra}, \bibinfo{person}{Jason Dowling}, \bibinfo{person}{Craig Engstrom}, \bibinfo{person}{Ying Xia}, \bibinfo{person}{Anthony Paproki}, \bibinfo{person}{Ales Neubert}, \bibinfo{person}{David Rivest-H{\'e}nault}, \bibinfo{person}{Olivier Salvado}, \bibinfo{person}{Stuart Crozier}, {and} \bibinfo{person}{Jurgen Fripp}.} \bibinfo{year}{2018}\natexlab{}.
\newblock \showarticletitle{A lightweight rapid application development framework for biomedical image analysis}.
\newblock \bibinfo{journal}{\emph{Computer Methods and Programs in Biomedicine}}  \bibinfo{volume}{164} (\bibinfo{date}{07} \bibinfo{year}{2018}).
\newblock
\urldef\tempurl%
\url{https://doi.org/10.1016/j.cmpb.2018.07.011}
\showDOI{\tempurl}


\bibitem[Choplin et~al\mbox{.}(1992)]%
        {Choplin1992}
\bibfield{author}{\bibinfo{person}{Robert Choplin}, \bibinfo{person}{J Boehme}, {and} \bibinfo{person}{C Maynard}.} \bibinfo{year}{1992}\natexlab{}.
\newblock \showarticletitle{Picture archiving and communication systems: an overview}.
\newblock \bibinfo{journal}{\emph{Radiographics : a review publication of the Radiological Society of North America, Inc}}  \bibinfo{volume}{12} (\bibinfo{date}{02} \bibinfo{year}{1992}), \bibinfo{pages}{127--9}.
\newblock
\urldef\tempurl%
\url{https://doi.org/10.1148/radiographics.12.1.1734458}
\showDOI{\tempurl}


\bibitem[Corbly(2014)]%
        {Corbly2014}
\bibfield{author}{\bibinfo{person}{James~Edward Corbly}.} \bibinfo{year}{2014}\natexlab{}.
\newblock \showarticletitle{The Free Software Alternative: Freeware, Open Source Software, and Libraries}.
\newblock \bibinfo{journal}{\emph{Information Technology and Libraries}} \bibinfo{volume}{33}, \bibinfo{number}{3} (\bibinfo{date}{Sep.} \bibinfo{year}{2014}), \bibinfo{pages}{65--75}.
\newblock
\urldef\tempurl%
\url{https://doi.org/10.6017/ital.v33i3.5105}
\showDOI{\tempurl}


\bibitem[de~Souza et~al\mbox{.}(2019)]%
        {deSouzaEtAl2019}
\bibfield{author}{\bibinfo{person}{Mario~Rosado de Souza}, \bibinfo{person}{Robert Haines}, \bibinfo{person}{Markel Vigo}, {and} \bibinfo{person}{Caroline Jay}.} \bibinfo{year}{2019}\natexlab{}.
\newblock \showarticletitle{What Makes Research Software Sustainable? An Interview Study With Research Software Engineers}.
\newblock \bibinfo{journal}{\emph{CoRR}}  \bibinfo{volume}{abs/1903.06039} (\bibinfo{year}{2019}).
\newblock
\showeprint[arxiv]{1903.06039}
\urldef\tempurl%
\url{http://arxiv.org/abs/1903.06039}
\showURL{%
\tempurl}


\bibitem[Dong(2021a)]%
        {Dong2021}
\bibfield{author}{\bibinfo{person}{Ao Dong}.} \bibinfo{year}{2021}\natexlab{a}.
\newblock \emph{\bibinfo{title}{Assessing the State of the Practice for Medical Imaging Software}}.
\newblock \bibinfo{thesistype}{Master's\ thesis}. \bibinfo{school}{McMaster University}, \bibinfo{address}{Hamilton, ON, Canada}.
\newblock


\bibitem[Dong(2021b)]%
        {Dong2021-Data}
\bibfield{author}{\bibinfo{person}{Ao Dong}.} \bibinfo{year}{2021}\natexlab{b}.
\newblock \bibinfo{title}{Software Quality Grades for MI Software}.
\newblock \bibinfo{howpublished}{Mendeley Data, V1, doi: 10.17632/k3pcdvdzj2.1}.
\newblock
\urldef\tempurl%
\url{https://doi.org/10.17632/k3pcdvdzj2.1}
\showDOI{\tempurl}


\bibitem[Easterbrook and Johns(2009)]%
        {EasterbrookAndJohns2009}
\bibfield{author}{\bibinfo{person}{Steve~M. Easterbrook} {and} \bibinfo{person}{Timothy~C. Johns}.} \bibinfo{year}{2009}\natexlab{}.
\newblock \showarticletitle{Engineering the Software for Understanding Climate Change}.
\newblock \bibinfo{journal}{\emph{Comuting in Science \& Engineering}} \bibinfo{volume}{11}, \bibinfo{number}{6} (\bibinfo{date}{November/December} \bibinfo{year}{2009}), \bibinfo{pages}{65--74}.
\newblock
\showISSN{0740-7475}
\urldef\tempurl%
\url{https://doi.org/10.1109/MCSE.2009.193}
\showDOI{\tempurl}


\bibitem[{Ebert} and {Jones}(2009)]%
        {EbertAndJones2009}
\bibfield{author}{\bibinfo{person}{C. {Ebert}} {and} \bibinfo{person}{C. {Jones}}.} \bibinfo{year}{2009}\natexlab{}.
\newblock \showarticletitle{Embedded Software: Facts, Figures, and Future}.
\newblock \bibinfo{journal}{\emph{Computer}} \bibinfo{volume}{42}, \bibinfo{number}{4} (\bibinfo{date}{April} \bibinfo{year}{2009}), \bibinfo{pages}{42--52}.
\newblock
\showISSN{0018-9162}
\urldef\tempurl%
\url{https://doi.org/10.1109/MC.2009.118}
\showDOI{\tempurl}


\bibitem[ElSheikh et~al\mbox{.}(2004)]%
        {ElSheikhEtAl2004}
\bibfield{author}{\bibinfo{person}{Ahmed~H. ElSheikh}, \bibinfo{person}{W.~Spencer Smith}, {and} \bibinfo{person}{Samir~E. Chidiac}.} \bibinfo{year}{2004}\natexlab{}.
\newblock \showarticletitle{Semi-formal design of reliable mesh generation systems}.
\newblock \bibinfo{journal}{\emph{Advances in Engineering Software}} \bibinfo{volume}{35}, \bibinfo{number}{12} (\bibinfo{year}{2004}), \bibinfo{pages}{827--841}.
\newblock


\bibitem[Emms(2019)]%
        {Emms2019}
\bibfield{author}{\bibinfo{person}{Steve Emms}.} \bibinfo{year}{2019}\natexlab{}.
\newblock \bibinfo{title}{16 Best Free Linux Medical Imaging Software}.
\newblock \bibinfo{howpublished}{\url{https://www.linuxlinks.com/medicalimaging/}}.
\newblock
\newblock
\shownote{[Online; accessed 02-February-2020]}.


\bibitem[ESA(1991)]%
        {ESA1991}
\bibfield{author}{\bibinfo{person}{ESA}.} \bibinfo{year}{February 1991}\natexlab{}.
\newblock \bibinfo{booktitle}{\emph{{ESA} Software Engineering Standards, {PSS-05-0} Issue 2}}.
\newblock \bibinfo{type}{{T}echnical {R}eport}. \bibinfo{institution}{European Space Agency}.
\newblock


\bibitem[Fagan(1976)]%
        {Fagan1976}
\bibfield{author}{\bibinfo{person}{M.~E. Fagan}.} \bibinfo{year}{1976}\natexlab{}.
\newblock \showarticletitle{Design and code inspections to reduce errors in program development}.
\newblock \bibinfo{journal}{\emph{IBM Systems Journal}} \bibinfo{volume}{15}, \bibinfo{number}{3} (\bibinfo{year}{1976}), \bibinfo{pages}{182--211}.
\newblock
\urldef\tempurl%
\url{https://doi.org/10.1147/sj.153.0182}
\showDOI{\tempurl}


\bibitem[Faulk et~al\mbox{.}(2009)]%
        {FaulkEtAl2009}
\bibfield{author}{\bibinfo{person}{S. Faulk}, \bibinfo{person}{E. Loh}, \bibinfo{person}{M.~L. V.~D. Vanter}, \bibinfo{person}{S. Squires}, {and} \bibinfo{person}{L.~G. Votta}.} \bibinfo{year}{2009}\natexlab{}.
\newblock \showarticletitle{Scientific Computing's Productivity Gridlock: How Software Engineering Can Help}.
\newblock \bibinfo{journal}{\emph{Computing in Science Engineering}} \bibinfo{volume}{11}, \bibinfo{number}{6} (\bibinfo{date}{Nov} \bibinfo{year}{2009}), \bibinfo{pages}{30--39}.
\newblock
\showISSN{1521-9615}
\urldef\tempurl%
\url{https://doi.org/10.1109/MCSE.2009.205}
\showDOI{\tempurl}


\bibitem[Fillard et~al\mbox{.}(2012)]%
        {Fillard2012}
\bibfield{author}{\bibinfo{person}{Pierre Fillard}, \bibinfo{person}{Nicolas Toussaint}, {and} \bibinfo{person}{Xavier Pennec}.} \bibinfo{year}{2012}\natexlab{}.
\newblock \showarticletitle{Medinria: DT-MRI processing and visualization software}.
\newblock  (\bibinfo{date}{04} \bibinfo{year}{2012}).
\newblock


\bibitem[Fogel(2005)]%
        {Fogel2005}
\bibfield{author}{\bibinfo{person}{Karl Fogel}.} \bibinfo{year}{2005}\natexlab{}.
\newblock \bibinfo{booktitle}{\emph{Producing Open Source Software: How to Run a Successful Free Software Project}}.
\newblock \bibinfo{publisher}{O'Reilly Media, Inc.}
\newblock
\showISBNx{0596007590}


\bibitem[Fowler(2006)]%
        {Fowler2006}
\bibfield{author}{\bibinfo{person}{Martin Fowler}.} \bibinfo{year}{2006}\natexlab{}.
\newblock \bibinfo{title}{Continuous Integration}.
\newblock \bibinfo{howpublished}{\url{https://martinfowler.com/articles/continuousIntegration.html}}.
\newblock


\bibitem[Gamma et~al\mbox{.}(1995)]%
        {Gamma1995}
\bibfield{author}{\bibinfo{person}{E. Gamma}, \bibinfo{person}{R. Helm}, \bibinfo{person}{J. Vlissides}, {and} \bibinfo{person}{I~R Johnson}.} \bibinfo{year}{1995}\natexlab{}.
\newblock \bibinfo{booktitle}{\emph{Design Patterns: Elements of Reusable Object-Oriented Software}}.
\newblock \bibinfo{publisher}{Addison-Wesley Professional}.
\newblock


\bibitem[Gewaltig and Cannon(2012)]%
        {GewaltigAndCannon2012}
\bibfield{author}{\bibinfo{person}{Marc-Oliver Gewaltig} {and} \bibinfo{person}{Robert Cannon}.} \bibinfo{year}{2012}\natexlab{}.
\newblock \showarticletitle{Quality and sustainability of software tools in neuroscience}.
\newblock \bibinfo{journal}{\emph{Cornell University Library}} (\bibinfo{date}{May} \bibinfo{year}{2012}), \bibinfo{pages}{20 pp}.
\newblock


\bibitem[Ghezzi et~al\mbox{.}(2003)]%
        {GhezziEtAl2003}
\bibfield{author}{\bibinfo{person}{Carlo Ghezzi}, \bibinfo{person}{Mehdi Jazayeri}, {and} \bibinfo{person}{Dino Mandrioli}.} \bibinfo{year}{2003}\natexlab{}.
\newblock \bibinfo{booktitle}{\emph{Fundamentals of Software Engineering} (\bibinfo{edition}{2nd} ed.)}.
\newblock \bibinfo{publisher}{Prentice Hall}, \bibinfo{address}{Upper Saddle River, NJ, USA}.
\newblock


\bibitem[Gieniusz(2019)]%
        {Gieniusz2019}
\bibfield{author}{\bibinfo{person}{Tomasz Gieniusz}.} \bibinfo{year}{2019}\natexlab{}.
\newblock \bibinfo{title}{GitStats}.
\newblock \bibinfo{howpublished}{\url{https://github.com/tomgi/git_stats}}.
\newblock
\newblock
\shownote{[Online; accessed 27-May-2021]}.


\bibitem[Givler(2020)]%
        {Givler2020}
\bibfield{author}{\bibinfo{person}{Ray Givler}.} \bibinfo{year}{2020}\natexlab{}.
\newblock \bibinfo{title}{A Checklist of Basic Software Engineering Practices for Data Analysts and Data Scientists}.
\newblock \bibinfo{howpublished}{\url{https://www.linkedin.com/pulse/checklist-basic-software-engineering-practices-data-analysts-givler/?articleId=6681691007303630849}}.
\newblock


\bibitem[GNU(2019)]%
        {GNU2019}
\bibfield{author}{\bibinfo{person}{GNU}.} \bibinfo{year}{2019}\natexlab{}.
\newblock \bibinfo{title}{Categories of free and nonfree software}.
\newblock \bibinfo{howpublished}{\url{https://www.gnu.org/philosophy/categories.html}}.
\newblock
\newblock
\shownote{[Online; accessed 20-May-2021]}.


\bibitem[Goble(2014)]%
        {Goble2014}
\bibfield{author}{\bibinfo{person}{Carole Goble}.} \bibinfo{year}{2014}\natexlab{}.
\newblock \showarticletitle{Better Software, Better Research}.
\newblock \bibinfo{journal}{\emph{IEEE Internet Computing}} \bibinfo{volume}{18}, \bibinfo{number}{5} (\bibinfo{year}{2014}), \bibinfo{pages}{4--8}.
\newblock
\urldef\tempurl%
\url{https://doi.org/10.1109/MIC.2014.88}
\showDOI{\tempurl}


\bibitem[Haak et~al\mbox{.}(2015)]%
        {Haak2015}
\bibfield{author}{\bibinfo{person}{Daniel Haak}, \bibinfo{person}{Charles-E Page}, {and} \bibinfo{person}{Thomas Deserno}.} \bibinfo{year}{2015}\natexlab{}.
\newblock \showarticletitle{A Survey of DICOM Viewer Software to Integrate Clinical Research and Medical Imaging}.
\newblock \bibinfo{journal}{\emph{Journal of digital imaging}}  \bibinfo{volume}{29} (\bibinfo{date}{10} \bibinfo{year}{2015}).
\newblock
\urldef\tempurl%
\url{https://doi.org/10.1007/s10278-015-9833-1}
\showDOI{\tempurl}


\bibitem[Haehn(2013)]%
        {Haehn2013}
\bibfield{author}{\bibinfo{person}{Daniel Haehn}.} \bibinfo{year}{2013}\natexlab{}.
\newblock \showarticletitle{Slice:drop: collaborative medical imaging in the browser}. \bibinfo{pages}{1--1}.
\newblock
\urldef\tempurl%
\url{https://doi.org/10.1145/2503541.2503645}
\showDOI{\tempurl}


\bibitem[Hamill(2004)]%
        {Hamill2004}
\bibfield{author}{\bibinfo{person}{Paul Hamill}.} \bibinfo{year}{2004}\natexlab{}.
\newblock \bibinfo{booktitle}{\emph{Unit test frameworks: Tools for high-quality software development}}.
\newblock \bibinfo{publisher}{O'Reilly Media}.
\newblock


\bibitem[Hannay et~al\mbox{.}(2009a)]%
        {Hannay2009}
\bibfield{author}{\bibinfo{person}{Jo~Erskine Hannay}, \bibinfo{person}{Carolyn MacLeod}, \bibinfo{person}{Janice Singer}, \bibinfo{person}{Hans~Petter Langtangen}, \bibinfo{person}{Dietmar Pfahl}, {and} \bibinfo{person}{Greg Wilson}.} \bibinfo{year}{2009}\natexlab{a}.
\newblock \showarticletitle{How do scientists develop and use scientific software?}. In \bibinfo{booktitle}{\emph{2009 ICSE Workshop on Software Engineering for Computational Science and Engineering}}. \bibinfo{pages}{1--8}.
\newblock
\urldef\tempurl%
\url{https://doi.org/10.1109/SECSE.2009.5069155}
\showDOI{\tempurl}


\bibitem[Hannay et~al\mbox{.}(2009b)]%
        {HannayEtAl2009}
\bibfield{author}{\bibinfo{person}{Jo~Erskine Hannay}, \bibinfo{person}{Carolyn MacLeod}, \bibinfo{person}{Janice Singer}, \bibinfo{person}{Hans~Petter Langtangen}, \bibinfo{person}{Dietmar Pfahl}, {and} \bibinfo{person}{Greg Wilson}.} \bibinfo{year}{2009}\natexlab{b}.
\newblock \showarticletitle{How Do Scientists Develop and Use Scientific Software?}. In \bibinfo{booktitle}{\emph{Proceedings of the 2009 ICSE Workshop on Software Engineering for Computational Science and Engineering}} \emph{(\bibinfo{series}{SECSE '09})}. \bibinfo{publisher}{IEEE Computer Society}, \bibinfo{address}{Washington, DC, USA}, \bibinfo{pages}{1--8}.
\newblock
\showISBNx{978-1-4244-3737-5}
\urldef\tempurl%
\url{https://doi.org/10.1109/SECSE.2009.5069155}
\showDOI{\tempurl}


\bibitem[Hasan(2020)]%
        {Hasan2020}
\bibfield{author}{\bibinfo{person}{Mehedi Hasan}.} \bibinfo{year}{2020}\natexlab{}.
\newblock \bibinfo{title}{Top 25 Best Free Medical Imaging Software for Linux System}.
\newblock \bibinfo{howpublished}{\url{https://www.ubuntupit.com/top-25-best-free-medical-imaging-software-for-linux-system/}}.
\newblock
\newblock
\shownote{[Online; accessed 30-January-2020]}.


\bibitem[Heaton and Carver(2015)]%
        {HeatonAndCarver2015}
\bibfield{author}{\bibinfo{person}{Dustin Heaton} {and} \bibinfo{person}{Jeffrey~C. Carver}.} \bibinfo{year}{2015}\natexlab{}.
\newblock \showarticletitle{Claims About the Use of Software Engineering Practices in Science}.
\newblock \bibinfo{journal}{\emph{Inf. Softw. Technol.}} \bibinfo{volume}{67}, \bibinfo{number}{C} (\bibinfo{date}{Nov.} \bibinfo{year}{2015}), \bibinfo{pages}{207--219}.
\newblock
\showISSN{0950-5849}
\urldef\tempurl%
\url{https://doi.org/10.1016/j.infsof.2015.07.011}
\showDOI{\tempurl}


\bibitem[Heroux and Bernholdt(2018)]%
        {HerouxAndBernholdt2018}
\bibfield{author}{\bibinfo{person}{Michael~A. Heroux} {and} \bibinfo{person}{David~E. Bernholdt}.} \bibinfo{year}{2018}\natexlab{}.
\newblock \bibinfo{title}{Better (Small) Scientific Software Teams, tutorial in {A}rgonne Training Program on Extreme-Scale Computing ({ATPESC})}.
\newblock \bibinfo{howpublished}{\url{https://press3.mcs.anl.gov//atpesc/files/2018/08/ATPESC_2018_Track-6_3_8-8_1030am_Bernholdt-Better_Scientific_Software_Teams.pdf}}.
\newblock
\urldef\tempurl%
\url{https://doi.org/articles/journal_contribution/ATPESC_Software_Productivity_03_Better_Small_Scientific_Software_Teams/6941438}
\showDOI{\tempurl}


\bibitem[Heroux et~al\mbox{.}(2008)]%
        {HerouxEtAl2008}
\bibfield{author}{\bibinfo{person}{Michael~A. Heroux}, \bibinfo{person}{James~M. Bieman}, {and} \bibinfo{person}{Robert~T. Heaphy}.} \bibinfo{year}{2008}\natexlab{}.
\newblock \bibinfo{title}{Trilinos Developers Guide Part {II}: {ASC} Softwar Quality Engineering Practices Version 2.0}.
\newblock \bibinfo{howpublished}{\url{https://faculty.csbsju.edu/mheroux/fall2012_csci330/TrilinosDevGuide2.pdf}}.
\newblock


\bibitem[Hilton et~al\mbox{.}(2016)]%
        {HiltonEtAl2016}
\bibfield{author}{\bibinfo{person}{Michael Hilton}, \bibinfo{person}{Timothy Tunnell}, \bibinfo{person}{Kai Huang}, \bibinfo{person}{Darko Marinov}, {and} \bibinfo{person}{Danny Dig}.} \bibinfo{year}{2016}\natexlab{}.
\newblock \showarticletitle{Usage, costs, and benefits of continuous integration in open-source projects}. In \bibinfo{booktitle}{\emph{2016 31st IEEE/ACM International Conference on Automated Software Engineering (ASE)}}. \bibinfo{pages}{426--437}.
\newblock


\bibitem[Hoffman and Strooper(1995)]%
        {HoffmanAndStrooper1995}
\bibfield{author}{\bibinfo{person}{Daniel~M. Hoffman} {and} \bibinfo{person}{Paul~A. Strooper}.} \bibinfo{year}{1995}\natexlab{}.
\newblock \bibinfo{booktitle}{\emph{Software Design, Automated Testing, and Maintenance: A Practical Approach}}.
\newblock \bibinfo{publisher}{International Thomson Computer Press}, \bibinfo{address}{New York, NY, USA}.
\newblock


\bibitem[horosproject.org(2020)]%
        {horosproject2020}
\bibfield{author}{\bibinfo{person}{horosproject.org}.} \bibinfo{year}{2020}\natexlab{}.
\newblock \bibinfo{title}{Horos}.
\newblock \bibinfo{howpublished}{\url{https://github.com/horosproject/horos}}.
\newblock
\newblock
\shownote{[Online; accessed 27-May-2021]}.


\bibitem[Howison and Bullard(2016)]%
        {HowisonAndBullard2016}
\bibfield{author}{\bibinfo{person}{James Howison} {and} \bibinfo{person}{Julia Bullard}.} \bibinfo{year}{2016}\natexlab{}.
\newblock \showarticletitle{Software in the Scientific Literature: Problems with Seeing, Finding, and Using Software Mentioned in the Biology Literature}.
\newblock \bibinfo{journal}{\emph{J. Assoc. Inf. Sci. Technol.}} \bibinfo{volume}{67}, \bibinfo{number}{9} (\bibinfo{date}{sep} \bibinfo{year}{2016}), \bibinfo{pages}{2137--2155}.
\newblock
\showISSN{2330-1635}
\urldef\tempurl%
\url{https://doi.org/10.1002/asi.23538}
\showDOI{\tempurl}


\bibitem[Humble and Farley(2010)]%
        {HumbleAndFarley2010}
\bibfield{author}{\bibinfo{person}{Jez Humble} {and} \bibinfo{person}{David~G. Farley}.} \bibinfo{year}{2010}\natexlab{}.
\newblock \bibinfo{booktitle}{\emph{Continuous Delivery: Reliable Software Releases through Build, Test, and Deployment Automation}}.
\newblock \bibinfo{publisher}{Addison-Wesley}, \bibinfo{address}{Upper Saddle River, NJ}.
\newblock
\showISBNx{978-0-321-60191-9}
\urldef\tempurl%
\url{http://my.safaribooksonline.com/9780321601919}
\showURL{%
\tempurl}


\bibitem[IEEE(1991)]%
        {IEEEStdGlossarySET1990}
\bibfield{author}{\bibinfo{person}{IEEE}.} \bibinfo{year}{1991}\natexlab{}.
\newblock \bibinfo{booktitle}{\emph{IEEE Standard Glossary of Software Engineering Terminology}}.
\newblock \bibinfo{type}{Standard}. \bibinfo{institution}{IEEE}.
\newblock


\bibitem[IEEE(1998)]%
        {IEEE1998}
\bibfield{author}{\bibinfo{person}{IEEE}.} \bibinfo{year}{1998}\natexlab{}.
\newblock \showarticletitle{Recommended Practice for Software Requirements Specifications}.
\newblock \bibinfo{journal}{\emph{IEEE Std 830-1998}} (\bibinfo{date}{Oct.} \bibinfo{year}{1998}), \bibinfo{pages}{1--40}.
\newblock
\urldef\tempurl%
\url{https://doi.org/10.1109/IEEESTD.1998.88286}
\showDOI{\tempurl}


\bibitem[{IEEE-CS/ACM} and Practices(1999)]%
        {IEEE1999}
\bibfield{author}{\bibinfo{person}{Joint Task Force on Software Engineering~Ethics {IEEE-CS/ACM}} {and} \bibinfo{person}{Professional Practices}.} \bibinfo{year}{1999}\natexlab{}.
\newblock \bibinfo{title}{Code of Ethics, {IEEE} Computer Society}.
\newblock \bibinfo{howpublished}{\url{https://www.computer.org/education/code-of-ethics}}.
\newblock


\bibitem[Innovations(2020)]%
        {ParallaxInnovations2020}
\bibfield{author}{\bibinfo{person}{Parallax Innovations}.} \bibinfo{year}{2020}\natexlab{}.
\newblock \bibinfo{title}{Microview}.
\newblock \bibinfo{howpublished}{\url{https://github.com/parallaxinnovations/MicroView/}}.
\newblock
\newblock
\shownote{[Online; accessed 27-May-2021]}.


\bibitem[Institute(2022)]%
        {SSI2022}
\bibfield{author}{\bibinfo{person}{Software~Sustainability Institute}.} \bibinfo{year}{2022}\natexlab{}.
\newblock \bibinfo{title}{Online sustainability evaluation}.
\newblock \bibinfo{howpublished}{\url{https://www.software.ac.uk/resources/online-sustainability-evaluation}}.
\newblock


\bibitem[Ishizaka and Lusti(2006)]%
        {AlessioEtAl2006}
\bibfield{author}{\bibinfo{person}{Alessio Ishizaka} {and} \bibinfo{person}{Markus Lusti}.} \bibinfo{year}{2006}\natexlab{}.
\newblock \showarticletitle{How to derive priorities in AHP: A comparative study}.
\newblock \bibinfo{journal}{\emph{Central European Journal of Operations Research}}  \bibinfo{volume}{14} (\bibinfo{date}{12} \bibinfo{year}{2006}), \bibinfo{pages}{387--400}.
\newblock
\urldef\tempurl%
\url{https://doi.org/10.1007/s10100-006-0012-9}
\showDOI{\tempurl}


\bibitem[ISO(2001)]%
        {iso2001iec}
\bibfield{author}{\bibinfo{person}{ISO}.} \bibinfo{year}{2001}\natexlab{}.
\newblock \showarticletitle{Iec 9126-1: Software engineering-product quality-part 1: Quality model}.
\newblock \bibinfo{journal}{\emph{Geneva, Switzerland: International Organization for Standardization}}  \bibinfo{volume}{21} (\bibinfo{year}{2001}).
\newblock


\bibitem[ISO/IEC(2011)]%
        {ISO/IEC25010}
\bibfield{author}{\bibinfo{person}{ISO/IEC}.} \bibinfo{year}{2011}\natexlab{}.
\newblock \bibinfo{booktitle}{\emph{Systems and software engineering - Systems and software Quality Requirements and Evaluation (SQuaRE) - System and software quality models}}.
\newblock \bibinfo{type}{Standard}. \bibinfo{institution}{International Organization for Standardization}.
\newblock


\bibitem[ISO/TR(2002)]%
        {ISO/TR16982:2002}
\bibfield{author}{\bibinfo{person}{ISO/TR}.} \bibinfo{year}{2002}\natexlab{}.
\newblock \bibinfo{booktitle}{\emph{Ergonomics of human-system interaction --- Usability methods supporting human-centred design}}.
\newblock \bibinfo{type}{Standard}. \bibinfo{institution}{International Organization for Standardization}.
\newblock


\bibitem[ISO/TR(2018)]%
        {ISO9241-11:2018}
\bibfield{author}{\bibinfo{person}{ISO/TR}.} \bibinfo{year}{2018}\natexlab{}.
\newblock \bibinfo{booktitle}{\emph{Ergonomics of human-system interaction --- Part 11: Usability: Definitions and concepts}}.
\newblock \bibinfo{type}{Standard}. \bibinfo{institution}{International Organization for Standardization}.
\newblock


\bibitem[Jan et~al\mbox{.}(2004)]%
        {Jan2004}
\bibfield{author}{\bibinfo{person}{Sama Jan}, \bibinfo{person}{Giovanni Santin}, \bibinfo{person}{Daniel Strul}, \bibinfo{person}{S Staelens}, \bibinfo{person}{K Assi{\'e}}, \bibinfo{person}{Damien Autret}, \bibinfo{person}{St{\'e}phane Avner}, \bibinfo{person}{Remi Barbier}, \bibinfo{person}{Manuel Bardi{\`e}s}, \bibinfo{person}{Peter Bloomfield}, \bibinfo{person}{David Brasse}, \bibinfo{person}{Vincent Breton}, \bibinfo{person}{Peter Bruyndonckx}, \bibinfo{person}{Irene Buvat}, \bibinfo{person}{AF Chatziioannou}, \bibinfo{person}{Yunsung Choi}, \bibinfo{person}{YH Chung}, \bibinfo{person}{Claude Comtat}, \bibinfo{person}{Denise Donnarieix}, {and} \bibinfo{person}{Christian Morel}.} \bibinfo{year}{2004}\natexlab{}.
\newblock \showarticletitle{GATE: a simulation toolkit for PET and SPECT}.
\newblock \bibinfo{journal}{\emph{Physics in medicine and biology}}  \bibinfo{volume}{49} (\bibinfo{date}{11} \bibinfo{year}{2004}), \bibinfo{pages}{4543--61}.
\newblock
\urldef\tempurl%
\url{https://doi.org/10.1088/0031-9155/49/19/007}
\showDOI{\tempurl}


\bibitem[Johanson and Hasselbring(2018)]%
        {JohansonAndHasselbring2018}
\bibfield{author}{\bibinfo{person}{Arne~N. Johanson} {and} \bibinfo{person}{Wilhelm Hasselbring}.} \bibinfo{year}{2018}\natexlab{}.
\newblock \showarticletitle{Software Engineering for Computational Science: Past, Present, Future}.
\newblock \bibinfo{journal}{\emph{Computing in Science \& Engineering}}  \bibinfo{volume}{Accepted} (\bibinfo{year}{2018}), \bibinfo{pages}{1--31}.
\newblock


\bibitem[Jones(2008)]%
        {Jones2008}
\bibfield{author}{\bibinfo{person}{Capers Jones}.} \bibinfo{year}{2008}\natexlab{}.
\newblock \showarticletitle{Measuring Defect Potentials and Defect Removal Efficiency}.
\newblock \bibinfo{journal}{\emph{Crosstalk, The Journal of Defense Software Engineering}} \bibinfo{volume}{21}, \bibinfo{number}{6} (\bibinfo{date}{June} \bibinfo{year}{2008}), \bibinfo{pages}{11--13}.
\newblock


\bibitem[Joshi et~al\mbox{.}(2011)]%
        {Joshi2011}
\bibfield{author}{\bibinfo{person}{Alark Joshi}, \bibinfo{person}{Dustin Scheinost}, \bibinfo{person}{Hirohito Okuda}, \bibinfo{person}{Dominique Belhachemi}, \bibinfo{person}{Isabella Murphy}, \bibinfo{person}{Lawrence Staib}, {and} \bibinfo{person}{Xenophon Papademetris}.} \bibinfo{year}{2011}\natexlab{}.
\newblock \showarticletitle{Unified Framework for Development, Deployment and Robust Testing of Neuroimaging Algorithms}.
\newblock \bibinfo{journal}{\emph{Neuroinformatics}}  \bibinfo{volume}{9} (\bibinfo{date}{03} \bibinfo{year}{2011}), \bibinfo{pages}{69--84}.
\newblock
\urldef\tempurl%
\url{https://doi.org/10.1007/s12021-010-9092-8}
\showDOI{\tempurl}


\bibitem[Jung et~al\mbox{.}(2022)]%
        {JungEtAl2022}
\bibfield{author}{\bibinfo{person}{Reiner Jung}, \bibinfo{person}{Sven Gundlach}, {and} \bibinfo{person}{Wilhelm Hasselbring}.} \bibinfo{year}{2022}\natexlab{}.
\newblock \showarticletitle{Thematic Domain Analysis for Ocean Modeling}.
\newblock \bibinfo{journal}{\emph{Environmental Modelling \& Software}} (\bibinfo{date}{Jan} \bibinfo{year}{2022}), \bibinfo{pages}{105323}.
\newblock
\showISSN{1364-8152}
\urldef\tempurl%
\url{https://doi.org/10.1016/j.envsoft.2022.105323}
\showDOI{\tempurl}


\bibitem[Kalagiakos(2003)]%
        {kalagiakos2003non}
\bibfield{author}{\bibinfo{person}{Panagiotis Kalagiakos}.} \bibinfo{year}{2003}\natexlab{}.
\newblock \showarticletitle{The Non-Technical Factors of Reusability}. In \bibinfo{booktitle}{\emph{Proceedings of the 29th Conference on EUROMICRO}}. IEEE Computer Society, \bibinfo{pages}{124}.
\newblock


\bibitem[Kanewala and Bieman(2013)]%
        {KanewalaAndBieman2013}
\bibfield{author}{\bibinfo{person}{U. Kanewala} {and} \bibinfo{person}{J.~M. Bieman}.} \bibinfo{year}{2013}\natexlab{}.
\newblock \showarticletitle{Techniques for testing scientific programs without an oracle}. In \bibinfo{booktitle}{\emph{Software Engineering for Computational Science and Engineering (SE-CSE), 2013 5th International Workshop on}}. \bibinfo{pages}{48--57}.
\newblock
\urldef\tempurl%
\url{https://doi.org/10.1109/SECSE.2013.6615099}
\showDOI{\tempurl}


\bibitem[Kashyap(2020)]%
        {Kashyap2020}
\bibfield{author}{\bibinfo{person}{Neeraj Kashyap}.} \bibinfo{year}{2020}\natexlab{}.
\newblock \bibinfo{title}{GitHub's Path to 128M Public Repositories}.
\newblock \bibinfo{howpublished}{\url{https://towardsdatascience.com/githubs-path-to-128m-public-repositories-f6f656ab56b1}}.
\newblock


\bibitem[Katerbow and Feulner(2018)]%
        {KaterbowAndFeulner2018}
\bibfield{author}{\bibinfo{person}{Matthias Katerbow} {and} \bibinfo{person}{Georg Feulner}.} \bibinfo{year}{2018}\natexlab{}.
\newblock \bibinfo{title}{{Recommendations on the development, use and provision of Research Software}}.
\newblock
\newblock
\urldef\tempurl%
\url{https://doi.org/10.5281/zenodo.1172988}
\showDOI{\tempurl}


\bibitem[Kelly(2013)]%
        {Kelly2013}
\bibfield{author}{\bibinfo{person}{Diane Kelly}.} \bibinfo{year}{2013}\natexlab{}.
\newblock \showarticletitle{Industrial Scientific Software: A Set of Interviews on Software Development}. In \bibinfo{booktitle}{\emph{Proceedings of the 2013 Conference of the Center for Advanced Studies on Collaborative Research}} (Ontario, Canada) \emph{(\bibinfo{series}{CASCON '13})}. \bibinfo{publisher}{IBM Corp.}, \bibinfo{address}{Riverton, NJ, USA}, \bibinfo{pages}{299--310}.
\newblock
\urldef\tempurl%
\url{http://dl.acm.org/citation.cfm?id=2555523.2555555}
\showURL{%
\tempurl}


\bibitem[Kelly(2015)]%
        {Kelly2015}
\bibfield{author}{\bibinfo{person}{Diane Kelly}.} \bibinfo{year}{2015}\natexlab{}.
\newblock \showarticletitle{Scientific software development viewed as knowledge acquisition: Towards understanding the development of risk-averse scientific software}.
\newblock \bibinfo{journal}{\emph{Journal of Systems and Software}}  \bibinfo{volume}{109} (\bibinfo{year}{2015}), \bibinfo{pages}{50--61}.
\newblock
\urldef\tempurl%
\url{https://doi.org/10.1016/j.jss.2015.07.027}
\showDOI{\tempurl}


\bibitem[Kelly and Shepard(2000)]%
        {KellyAndShepard2000}
\bibfield{author}{\bibinfo{person}{Diane Kelly} {and} \bibinfo{person}{Terry Shepard}.} \bibinfo{year}{2000}\natexlab{}.
\newblock \showarticletitle{Task-directed software inspection technique: an experiment and case study}. In \bibinfo{booktitle}{\emph{CASCON '00: Proceedings of the 2000 conference of the Centre for Advanced Studies on Collaborative research}} (Mississauga, Ontario, Canada). \bibinfo{publisher}{IBM Press}, \bibinfo{pages}{6}.
\newblock
\urldef\tempurl%
\url{http://portal.acm.org/citation.cfm?id=782040#}
\showURL{%
\tempurl}


\bibitem[Kelly(2007)]%
        {Kelly2007}
\bibfield{author}{\bibinfo{person}{Diane~F. Kelly}.} \bibinfo{year}{2007}\natexlab{}.
\newblock \showarticletitle{A Software Chasm: Software Engineering and Scientific Computing}.
\newblock \bibinfo{journal}{\emph{IEEE Software}} \bibinfo{volume}{24}, \bibinfo{number}{6} (\bibinfo{year}{2007}), \bibinfo{pages}{120--119}.
\newblock
\showISSN{0740-7459}
\urldef\tempurl%
\url{https://doi.org/10.1109/MS.2007.155}
\showDOI{\tempurl}


\bibitem[Kelly et~al\mbox{.}(2011)]%
        {KellyEtAl2011}
\bibfield{author}{\bibinfo{person}{Diane~F. Kelly}, \bibinfo{person}{W.~Spencer Smith}, {and} \bibinfo{person}{Nicholas Meng}.} \bibinfo{year}{2011}\natexlab{}.
\newblock \showarticletitle{Software Engineering for Scientists}.
\newblock \bibinfo{journal}{\emph{Computing in Science \& Engineering}} \bibinfo{volume}{13}, \bibinfo{number}{5} (\bibinfo{date}{Oct.} \bibinfo{year}{2011}), \bibinfo{pages}{7--11}.
\newblock


\bibitem[Kikinis et~al\mbox{.}(2014)]%
        {Kikinis2014}
\bibfield{author}{\bibinfo{person}{Ron Kikinis}, \bibinfo{person}{Steve Pieper}, {and} \bibinfo{person}{Kirby Vosburgh}.} \bibinfo{year}{2014}\natexlab{}.
\newblock \bibinfo{booktitle}{\emph{3D Slicer: A Platform for Subject-Specific Image Analysis, Visualization, and Clinical Support}}. Vol.~\bibinfo{volume}{3}.
\newblock \bibinfo{pages}{277--289}.
\newblock
\showISBNx{978-1-4614-7656-6}
\urldef\tempurl%
\url{https://doi.org/10.1007/978-1-4614-7657-3_19}
\showDOI{\tempurl}


\bibitem[Kim et~al\mbox{.}(2011)]%
        {Kim2011}
\bibfield{author}{\bibinfo{person}{Tae-Yun Kim}, \bibinfo{person}{Jaebum Son}, {and} \bibinfo{person}{Kwanggi Kim}.} \bibinfo{year}{2011}\natexlab{}.
\newblock \showarticletitle{The Recent Progress in Quantitative Medical Image Analysis for Computer Aided Diagnosis Systems}.
\newblock \bibinfo{journal}{\emph{Healthcare informatics research}}  \bibinfo{volume}{17} (\bibinfo{date}{09} \bibinfo{year}{2011}), \bibinfo{pages}{143--9}.
\newblock
\urldef\tempurl%
\url{https://doi.org/10.4258/hir.2011.17.3.143}
\showDOI{\tempurl}


\bibitem[Kruchten et~al\mbox{.}(2012)]%
        {KruchtenEtAl2012}
\bibfield{author}{\bibinfo{person}{Philippe Kruchten}, \bibinfo{person}{Robert~L Nord}, {and} \bibinfo{person}{Ipek Ozkaya}.} \bibinfo{year}{2012}\natexlab{}.
\newblock \showarticletitle{Technical debt: From metaphor to theory and practice}.
\newblock \bibinfo{journal}{\emph{IEEE Software}} \bibinfo{volume}{29}, \bibinfo{number}{6} (\bibinfo{year}{2012}), \bibinfo{pages}{18--21}.
\newblock


\bibitem[Lab(2021)]%
        {Rorden2021}
\bibfield{author}{\bibinfo{person}{Chris~Rorden's Lab}.} \bibinfo{year}{2021}\natexlab{}.
\newblock \bibinfo{title}{MRIcroGL}.
\newblock \bibinfo{howpublished}{\url{https://github.com/rordenlab/MRIcroGL }}.
\newblock
\newblock
\shownote{[Online; accessed 27-May-2021]}.


\bibitem[Lenhard et~al\mbox{.}(2013)]%
        {lenhard2013measuring}
\bibfield{author}{\bibinfo{person}{J{\"o}rg Lenhard}, \bibinfo{person}{Simon Harrer}, {and} \bibinfo{person}{Guido Wirtz}.} \bibinfo{year}{2013}\natexlab{}.
\newblock \showarticletitle{Measuring the installability of service orchestrations using the square method}. In \bibinfo{booktitle}{\emph{2013 IEEE 6th International Conference on Service-Oriented Computing and Applications}}. IEEE, \bibinfo{pages}{118--125}.
\newblock


\bibitem[Lethbridge et~al\mbox{.}(2003)]%
        {LethbridgeEtAl2003}
\bibfield{author}{\bibinfo{person}{T.C. Lethbridge}, \bibinfo{person}{J. Singer}, {and} \bibinfo{person}{A. Forward}.} \bibinfo{year}{2003}\natexlab{}.
\newblock \showarticletitle{How software engineers use documentation: the state of the practice}.
\newblock \bibinfo{journal}{\emph{IEEE Software}} \bibinfo{volume}{20}, \bibinfo{number}{6} (\bibinfo{year}{2003}), \bibinfo{pages}{35--39}.
\newblock
\urldef\tempurl%
\url{https://doi.org/10.1109/MS.2003.1241364}
\showDOI{\tempurl}


\bibitem[Limaye(2012)]%
        {Limaye2012}
\bibfield{author}{\bibinfo{person}{Ajay Limaye}.} \bibinfo{year}{2012}\natexlab{}.
\newblock \showarticletitle{Drishti, A Volume Exploration and Presentation Tool}.
\newblock \bibinfo{journal}{\emph{Proc SPIE}}  \bibinfo{volume}{8506}, \bibinfo{pages}{85060X}.
\newblock
\urldef\tempurl%
\url{https://doi.org/10.1117/12.935640}
\showDOI{\tempurl}


\bibitem[Liu et~al\mbox{.}(2016)]%
        {Liu2016}
\bibfield{author}{\bibinfo{person}{Fang Liu}, \bibinfo{person}{Julia Velikina}, \bibinfo{person}{Walter Block}, \bibinfo{person}{Richard Kijowski}, {and} \bibinfo{person}{Alexey Samsonov}.} \bibinfo{year}{2016}\natexlab{}.
\newblock \showarticletitle{Fast Realistic MRI Simulations Based on Generalized Multi-Pool Exchange Tissue Model}.
\newblock \bibinfo{journal}{\emph{IEEE Transactions on Medical Imaging}}  \bibinfo{volume}{PP} (\bibinfo{date}{10} \bibinfo{year}{2016}), \bibinfo{pages}{1--1}.
\newblock
\urldef\tempurl%
\url{https://doi.org/10.1109/TMI.2016.2620961}
\showDOI{\tempurl}


\bibitem[Loening(2017)]%
        {Loening2017}
\bibfield{author}{\bibinfo{person}{Andy Loening}.} \bibinfo{year}{2017}\natexlab{}.
\newblock \bibinfo{title}{AMIDE}.
\newblock \bibinfo{howpublished}{\url{https://sourceforge.net/p/amide/code/ci/default/tree/amide-current/}}.
\newblock
\newblock
\shownote{[Online; accessed 27-May-2021]}.


\bibitem[Logg et~al\mbox{.}(2012)]%
        {LoggEtAl2012}
\bibfield{editor}{\bibinfo{person}{A. Logg}, \bibinfo{person}{K.-A. Mardal}, {and} \bibinfo{person}{G.~N. Wells}} (Eds.). \bibinfo{year}{2012}\natexlab{}.
\newblock \bibinfo{booktitle}{\emph{Automated Solution of Differential Equations by the Finite Element Method}}. \bibinfo{series}{Lecture Notes in Computational Science and Engineering}, Vol.~\bibinfo{volume}{84}.
\newblock \bibinfo{publisher}{Springer}.
\newblock
\urldef\tempurl%
\url{https://doi.org/10.1007/978-3-642-23099-8}
\showDOI{\tempurl}


\bibitem[Luciv et~al\mbox{.}(2018)]%
        {LucivEtAl2018}
\bibfield{author}{\bibinfo{person}{D.~V. Luciv}, \bibinfo{person}{D.~V. Koznov}, \bibinfo{person}{G.~A. Chernishev}, \bibinfo{person}{A.~N. Terekhov}, \bibinfo{person}{K.~Yu. Romanovsky}, {and} \bibinfo{person}{D.~A. Grigoriev}.} \bibinfo{year}{2018}\natexlab{}.
\newblock \showarticletitle{Detecting Near Duplicates in Software Documentation}.
\newblock \bibinfo{journal}{\emph{Programming and Computer Software}} \bibinfo{volume}{44}, \bibinfo{number}{5} (\bibinfo{date}{01 Sep} \bibinfo{year}{2018}), \bibinfo{pages}{335--343}.
\newblock
\showISSN{1608-3261}
\urldef\tempurl%
\url{https://doi.org/10.1134/S0361768818050079}
\showDOI{\tempurl}


\bibitem[Martelli(2021)]%
        {Martelli2021}
\bibfield{author}{\bibinfo{person}{Yves Martelli}.} \bibinfo{year}{2021}\natexlab{}.
\newblock \bibinfo{title}{dwv}.
\newblock \bibinfo{howpublished}{\url{https://github.com/ivmartel/dwv }}.
\newblock
\newblock
\shownote{[Online; accessed 27-May-2021]}.


\bibitem[Matkerim et~al\mbox{.}(2013)]%
        {MatkerimEtAl2013}
\bibfield{author}{\bibinfo{person}{Bazargul Matkerim}, \bibinfo{person}{Darhan Akhmed-Zaki}, {and} \bibinfo{person}{Manuel Barata}.} \bibinfo{year}{2013}\natexlab{}.
\newblock \showarticletitle{Development High Performance Scientific Computing Application Using Model-Driven Architecture}.
\newblock \bibinfo{journal}{\emph{Applied Mathematical Sciences}} \bibinfo{volume}{7}, \bibinfo{number}{100} (\bibinfo{year}{2013}), \bibinfo{pages}{4961--4974}.
\newblock


\bibitem[McCormick et~al\mbox{.}(2014)]%
        {McCormick2014}
\bibfield{author}{\bibinfo{person}{Matthew McCormick}, \bibinfo{person}{Xiaoxiao Liu}, \bibinfo{person}{Julien Jomier}, \bibinfo{person}{Charles Marion}, {and} \bibinfo{person}{Luis Ibanez}.} \bibinfo{year}{2014}\natexlab{}.
\newblock \showarticletitle{ITK: Enabling Reproducible Research and Open Science}.
\newblock \bibinfo{journal}{\emph{Frontiers in neuroinformatics}}  \bibinfo{volume}{8} (\bibinfo{date}{02} \bibinfo{year}{2014}), \bibinfo{pages}{13}.
\newblock
\urldef\tempurl%
\url{https://doi.org/10.3389/fninf.2014.00013}
\showDOI{\tempurl}


\bibitem[Meng et~al\mbox{.}(2018)]%
        {MengEtAl2018}
\bibfield{author}{\bibinfo{person}{Michael Meng}, \bibinfo{person}{Stephanie Steinhardt}, {and} \bibinfo{person}{Andreas Schubert}.} \bibinfo{year}{2018}\natexlab{}.
\newblock \showarticletitle{Application Programming Interface Documentation: What Do Software Developers Want?}
\newblock \bibinfo{journal}{\emph{Journal of Technical Writing and Communication}} \bibinfo{volume}{48}, \bibinfo{number}{3} (\bibinfo{year}{2018}), \bibinfo{pages}{295--330}.
\newblock
\urldef\tempurl%
\url{https://doi.org/10.1177/0047281617721853}
\showDOI{\tempurl}
\showeprint{https://doi.org/10.1177/0047281617721853}


\bibitem[Michalski(2021)]%
        {Michalski2021}
\bibfield{author}{\bibinfo{person}{Peter Michalski}.} \bibinfo{year}{2021}\natexlab{}.
\newblock \emph{\bibinfo{title}{State of The Practice for Lattice Boltzmann Method Software}}.
\newblock \bibinfo{thesistype}{Master's\ thesis}. \bibinfo{school}{McMaster University}, \bibinfo{address}{Hamilton, Ontario, Canada}.
\newblock


\bibitem[Mu(2019)]%
        {Mu2019}
\bibfield{author}{\bibinfo{person}{Hamza Mu}.} \bibinfo{year}{2019}\natexlab{}.
\newblock \bibinfo{title}{20 Free \& open source DICOM viewers for Windows}.
\newblock \bibinfo{howpublished}{\url{https://medevel.com/free-dicom-viewers-for-windows/}}.
\newblock
\newblock
\shownote{[Online; accessed 31-January-2020]}.


\bibitem[M{\"u}nch et~al\mbox{.}(2019)]%
        {MunchEtAl2019}
\bibfield{author}{\bibinfo{person}{J{\"u}rgen M{\"u}nch}, \bibinfo{person}{Stefan Trieflinger}, {and} \bibinfo{person}{Dominic Lang}.} \bibinfo{year}{2019}\natexlab{}.
\newblock \showarticletitle{Product Roadmap -- From Vision to Reality: A Systematic Literature Review}. In \bibinfo{booktitle}{\emph{2019 IEEE International Conference on Engineering, Technology and Innovation (ICE/ITMC)}}. \bibinfo{pages}{1--8}.
\newblock
\urldef\tempurl%
\url{https://doi.org/10.1109/ICE.2019.8792654}
\showDOI{\tempurl}


\bibitem[Musa et~al\mbox{.}(1987)]%
        {musa1987software}
\bibfield{author}{\bibinfo{person}{JD Musa}, \bibinfo{person}{Anthony Iannino}, {and} \bibinfo{person}{Kazuhira Okumoto}.} \bibinfo{year}{1987}\natexlab{}.
\newblock \bibinfo{title}{Software reliability: prediction and application}.
\newblock
\newblock


\bibitem[NASA(1989)]%
        {NASA1989}
\bibfield{author}{\bibinfo{person}{NASA}.} \bibinfo{year}{1989}\natexlab{}.
\newblock \bibinfo{booktitle}{\emph{Software requirements {DID}, {SMAP-DID-P200-SW}, Release 4.3}}.
\newblock \bibinfo{type}{{T}echnical {R}eport}. \bibinfo{institution}{National Aeronautics and Space Agency}.
\newblock


\bibitem[Nevcas and Klapetek(2012)]%
        {Nevcas2012}
\bibfield{author}{\bibinfo{person}{D Nevcas} {and} \bibinfo{person}{P Klapetek}.} \bibinfo{year}{2012}\natexlab{}.
\newblock \showarticletitle{Gwyddion: an open-source software for spm data analysis}.
\newblock \bibinfo{journal}{\emph{Cent Eur J Phys}}  \bibinfo{volume}{10} (\bibinfo{date}{01} \bibinfo{year}{2012}).
\newblock


\bibitem[Nguyen-Hoan et~al\mbox{.}(2010)]%
        {Nguyen-HoanEtAl2010}
\bibfield{author}{\bibinfo{person}{Luke Nguyen-Hoan}, \bibinfo{person}{Shayne Flint}, {and} \bibinfo{person}{Ramesh Sankaranarayana}.} \bibinfo{year}{2010}\natexlab{}.
\newblock \showarticletitle{A Survey of Scientific Software Development}. In \bibinfo{booktitle}{\emph{Proceedings of the 2010 ACM-IEEE International Symposium on Empirical Software Engineering and Measurement}} (Bolzano-Bozen, Italy) \emph{(\bibinfo{series}{ESEM '10})}. \bibinfo{publisher}{ACM}, \bibinfo{address}{New York, NY, USA}, Article \bibinfo{articleno}{12}, \bibinfo{numpages}{10}~pages.
\newblock
\showISBNx{978-1-4503-0039-1}
\urldef\tempurl%
\url{https://doi.org/10.1145/1852786.1852802}
\showDOI{\tempurl}


\bibitem[Nolf et~al\mbox{.}(2003)]%
        {Nolf2003}
\bibfield{author}{\bibinfo{person}{E Nolf}, \bibinfo{person}{Tony Voet}, \bibinfo{person}{Filip Jacobs}, \bibinfo{person}{R Dierckx}, {and} \bibinfo{person}{Ignace Lemahieu}.} \bibinfo{year}{2003}\natexlab{}.
\newblock \showarticletitle{(X)MedCon * An OpenSource Medical Image Conversion Toolkit}.
\newblock \bibinfo{journal}{\emph{European Journal of Nuclear Medicine and Molecular Imaging}}  \bibinfo{volume}{30} (\bibinfo{date}{08} \bibinfo{year}{2003}), \bibinfo{pages}{S246}.
\newblock
\urldef\tempurl%
\url{https://doi.org/10.1007/s00259-003-1284-0}
\showDOI{\tempurl}


\bibitem[Ober et~al\mbox{.}(2018)]%
        {OberEtAl2018}
\bibfield{author}{\bibinfo{person}{Ileana Ober}, \bibinfo{person}{Marc Palyart}, \bibinfo{person}{Jean-Michel Bruel}, {and} \bibinfo{person}{David Lugato}.} \bibinfo{year}{2018}\natexlab{}.
\newblock \showarticletitle{On the use of models for high-performance scientific computing applications: an experience report}.
\newblock \bibinfo{journal}{\emph{Software {\&} Systems Modeling}} \bibinfo{volume}{17}, \bibinfo{number}{1} (\bibinfo{date}{01 Feb} \bibinfo{year}{2018}), \bibinfo{pages}{319--342}.
\newblock
\showISSN{1619-1374}
\urldef\tempurl%
\url{https://doi.org/10.1007/s10270-016-0518-0}
\showDOI{\tempurl}


\bibitem[Orviz et~al\mbox{.}(2017)]%
        {OrvizEtAl2017}
\bibfield{author}{\bibinfo{person}{Pablo Orviz}, \bibinfo{person}{{\'A}lvaro~L{\'o}pez Garc{\'\i}a}, \bibinfo{person}{Doina~Cristina Duma}, \bibinfo{person}{Giacinto Donvito}, \bibinfo{person}{Mario David}, {and} \bibinfo{person}{Jorge Gomes}.} \bibinfo{year}{2017}\natexlab{}.
\newblock \bibinfo{title}{A set of common software quality assurance baseline criteria for research projects}.
\newblock
\newblock
\urldef\tempurl%
\url{https://doi.org/10.20350/digitalCSIC/12543}
\showDOI{\tempurl}


\bibitem[Owojaiye et~al\mbox{.}(2021)]%
        {OwojaiyeEtAl2021_CSE}
\bibfield{author}{\bibinfo{person}{Oluwaseun Owojaiye}, \bibinfo{person}{W.~Spencer Smith}, \bibinfo{person}{Jacques Carette}, \bibinfo{person}{Peter Michalski}, {and} \bibinfo{person}{Ao Dong}.} \bibinfo{year}{2021}\natexlab{}.
\newblock \showarticletitle{State of Sustainability for Research Software (poster)}. In \bibinfo{booktitle}{\emph{{SIAM-CSE} 2021 Conference on Computational Science and Engineering, Minisymposterium: Software Productivity and Sustainability for CSE}}.
\newblock
\urldef\tempurl%
\url{https://doi.org/10.6084/m9.figshare.14039888.v2}
\showDOI{\tempurl}


\bibitem[Panchal and Keyes(2010)]%
        {Panchal2010}
\bibfield{author}{\bibinfo{person}{A. Panchal} {and} \bibinfo{person}{R. Keyes}.} \bibinfo{year}{2010}\natexlab{}.
\newblock \showarticletitle{SU-GG-T-260: Dicompyler: An Open Source Radiation Therapy Research Platform with a Plugin Architecture}.
\newblock \bibinfo{journal}{\emph{Medical Physics - MED PHYS}}  \bibinfo{volume}{37} (\bibinfo{date}{06} \bibinfo{year}{2010}).
\newblock
\urldef\tempurl%
\url{https://doi.org/10.1118/1.3468652}
\showDOI{\tempurl}


\bibitem[Papademetris et~al\mbox{.}(2005)]%
        {Papademetris2005}
\bibfield{author}{\bibinfo{person}{Xenophon Papademetris}, \bibinfo{person}{Marcel Jackowski}, \bibinfo{person}{Nallakkandi Rajeevan}, \bibinfo{person}{Robert Constable}, {and} \bibinfo{person}{Lawrence Staib}.} \bibinfo{year}{2005}\natexlab{}.
\newblock \showarticletitle{BioImage Suite: An integrated medical image analysis suite}.
\newblock   \bibinfo{volume}{1} (\bibinfo{date}{01} \bibinfo{year}{2005}).
\newblock


\bibitem[Parnas et~al\mbox{.}(1984)]%
        {ParnasEtAl1984}
\bibfield{author}{\bibinfo{person}{D.L. Parnas}, \bibinfo{person}{P.C. Clement}, {and} \bibinfo{person}{D.~M. Weiss}.} \bibinfo{year}{1984}\natexlab{}.
\newblock \showarticletitle{The modular structure of complex systems}. In \bibinfo{booktitle}{\emph{International Conference on Software Engineering}}. \bibinfo{pages}{408--419}.
\newblock


\bibitem[Parnas(1972)]%
        {Parnas1972a}
\bibfield{author}{\bibinfo{person}{David~L. Parnas}.} \bibinfo{year}{1972}\natexlab{}.
\newblock \showarticletitle{On the Criteria To Be Used in Decomposing Systems into Modules}.
\newblock \bibinfo{journal}{\emph{Comm. ACM}} \bibinfo{volume}{15}, \bibinfo{number}{2} (\bibinfo{date}{Dec.} \bibinfo{year}{1972}), \bibinfo{pages}{1053--1058}.
\newblock


\bibitem[Parnas and Clements(1986)]%
        {parnas1986rational}
\bibfield{author}{\bibinfo{person}{David~Lorge Parnas} {and} \bibinfo{person}{Paul~C Clements}.} \bibinfo{year}{1986}\natexlab{}.
\newblock \showarticletitle{A rational design process: How and why to fake it}.
\newblock \bibinfo{journal}{\emph{IEEE transactions on software engineering}} \bibinfo{number}{2} (\bibinfo{year}{1986}), \bibinfo{pages}{251--257}.
\newblock


\bibitem[Phaal et~al\mbox{.}(2005)]%
        {PhaalEtAl2005}
\bibfield{author}{\bibinfo{person}{R. Phaal}, \bibinfo{person}{C.J.P. Farrukh}, {and} \bibinfo{person}{D.R. Probert}.} \bibinfo{year}{2005}\natexlab{}.
\newblock \showarticletitle{Developing a technology roadmapping system}. In \bibinfo{booktitle}{\emph{A Unifying Discipline for Melting the Boundaries Technology Management:}}. \bibinfo{pages}{99--111}.
\newblock
\urldef\tempurl%
\url{https://doi.org/10.1109/PICMET.2005.1509680}
\showDOI{\tempurl}


\bibitem[Pichler(2012)]%
        {Pichler2012}
\bibfield{author}{\bibinfo{person}{Roman Pichler}.} \bibinfo{year}{2012}\natexlab{}.
\newblock \bibinfo{title}{Working with an Agile Product Roadmap}.
\newblock \bibinfo{howpublished}{\url{https://www.romanpichler.com/blog/agile-product-roadmap/}}.
\newblock


\bibitem[Pinto et~al\mbox{.}(2016)]%
        {PintoEtAl2016}
\bibfield{author}{\bibinfo{person}{Gustavo Pinto}, \bibinfo{person}{Igor Steinmacher}, {and} \bibinfo{person}{Marco~Aur{\'e}lio Gerosa}.} \bibinfo{year}{2016}\natexlab{}.
\newblock \showarticletitle{More Common Than You Think: An In-depth Study of Casual Contributors}. In \bibinfo{booktitle}{\emph{2016 IEEE 23rd International Conference on Software Analysis, Evolution, and Reengineering (SANER)}}, Vol.~\bibinfo{volume}{1}. \bibinfo{pages}{112--123}.
\newblock
\urldef\tempurl%
\url{https://doi.org/10.1109/SANER.2016.68}
\showDOI{\tempurl}


\bibitem[Pinto et~al\mbox{.}(2018)]%
        {PintoEtAl2018}
\bibfield{author}{\bibinfo{person}{Gustavo Pinto}, \bibinfo{person}{Igor Wiese}, {and} \bibinfo{person}{Luis~Felipe Dias}.} \bibinfo{year}{2018}\natexlab{}.
\newblock \showarticletitle{How Do Scientists Develop and Use Scientific Software? An External Replication}. In \bibinfo{booktitle}{\emph{Proceedings of 25th IEEE International Conference on Software Analysis, Evolution and Reengineering}}. \bibinfo{pages}{582--591}.
\newblock
\urldef\tempurl%
\url{https://doi.org/10.1109/SANER.2018.8330263}
\showDOI{\tempurl}


\bibitem[Prabhu et~al\mbox{.}(2011)]%
        {Prabhu2011}
\bibfield{author}{\bibinfo{person}{Prakash Prabhu}, \bibinfo{person}{Thomas~B. Jablin}, \bibinfo{person}{Arun Raman}, \bibinfo{person}{Yun Zhang}, \bibinfo{person}{Jialu Huang}, \bibinfo{person}{Hanjun Kim}, \bibinfo{person}{Nick~P. Johnson}, \bibinfo{person}{Feng Liu}, \bibinfo{person}{Soumyadeep Ghosh}, \bibinfo{person}{Stephen Beard}, \bibinfo{person}{Taewook Oh}, \bibinfo{person}{Matthew Zoufaly}, \bibinfo{person}{David Walker}, {and} \bibinfo{person}{David~I. August}.} \bibinfo{year}{2011}\natexlab{}.
\newblock \showarticletitle{A Survey of the Practice of Computational Science} \emph{(\bibinfo{series}{SC '11})}. \bibinfo{publisher}{Association for Computing Machinery}, \bibinfo{address}{New York, NY, USA}, Article \bibinfo{articleno}{19}, \bibinfo{numpages}{12}~pages.
\newblock
\showISBNx{9781450311397}
\urldef\tempurl%
\url{https://doi.org/10.1145/2063348.2063374}
\showDOI{\tempurl}


\bibitem[Prana et~al\mbox{.}(2018)]%
        {PranaEtAl2018}
\bibfield{author}{\bibinfo{person}{Gede Artha~Azriadi Prana}, \bibinfo{person}{Christoph Treude}, \bibinfo{person}{Ferdian Thung}, \bibinfo{person}{Thushari Atapattu}, {and} \bibinfo{person}{David Lo}.} \bibinfo{year}{2018}\natexlab{}.
\newblock \bibinfo{title}{Categorizing the Content of GitHub README Files}.
\newblock
\newblock
\showeprint[arxiv]{1802.06997}~[cs.SE]


\bibitem[Prior et~al\mbox{.}(2017)]%
        {PriorEtAl2017}
\bibfield{author}{\bibinfo{person}{F. Prior}, \bibinfo{person}{Kirk Smith}, \bibinfo{person}{Ashish Sharma}, \bibinfo{person}{Justin Kirby}, \bibinfo{person}{Lawrence Tarbox}, \bibinfo{person}{Ken Clark}, \bibinfo{person}{William Bennett}, \bibinfo{person}{Tracy Nolan}, {and} \bibinfo{person}{John Freymann}.} \bibinfo{year}{2017}\natexlab{}.
\newblock \showarticletitle{The public cancer radiology imaging collections of The Cancer Imaging Archive}.
\newblock \bibinfo{journal}{\emph{Scientific Data}}  \bibinfo{volume}{4} (\bibinfo{date}{09} \bibinfo{year}{2017}), \bibinfo{pages}{sdata2017124}.
\newblock
\urldef\tempurl%
\url{https://doi.org/10.1038/sdata.2017.124}
\showDOI{\tempurl}


\bibitem[{Professional Engineers Act}(2021)]%
        {PEO2021}
\bibfield{author}{\bibinfo{person}{{Professional Engineers Act}}.} \bibinfo{year}{2021}\natexlab{}.
\newblock \bibinfo{title}{Professional Engineers Act, RSO 1990, c P. 28}.
\newblock \bibinfo{howpublished}{\url{https://canlii.ca/t/5568z}}.
\newblock


\bibitem[Project(2006)]%
        {LINFO2006}
\bibfield{author}{\bibinfo{person}{The Linux~Information Project}.} \bibinfo{year}{2006}\natexlab{}.
\newblock \bibinfo{title}{Freeware Definition}.
\newblock \bibinfo{howpublished}{\url{http://www.linfo.org/freeware.html}}.
\newblock
\newblock
\shownote{[Online; accessed 20-May-2021]}.


\bibitem[P\"{u}schel et~al\mbox{.}(2001)]%
        {Pueschel2001}
\bibfield{author}{\bibinfo{person}{Markus P\"{u}schel}, \bibinfo{person}{Bryan Singer}, \bibinfo{person}{Manuela Veloso}, {and} \bibinfo{person}{Jos\'{e} M.~F. Moura}.} \bibinfo{year}{2001}\natexlab{}.
\newblock \showarticletitle{Fast Automatic Generation of {DSP} Algorithms}. In \bibinfo{booktitle}{\emph{International Conference on Computational Science (ICCS)}} \emph{(\bibinfo{series}{Lecture Notes In Computer Science}, Vol.~\bibinfo{volume}{2073})}. \bibinfo{publisher}{Springer}, \bibinfo{pages}{97--106}.
\newblock


\bibitem[Research Imaging~Institute(2019)]%
        {UTHSCSA2019}
\bibfield{author}{\bibinfo{person}{UTHSCSA Research Imaging~Institute}.} \bibinfo{year}{2019}\natexlab{}.
\newblock \bibinfo{title}{Papaya}.
\newblock \bibinfo{howpublished}{\url{https://github.com/rii-mango/Papaya}}.
\newblock
\newblock
\shownote{[Online; accessed 27-May-2021]}.


\bibitem[Rigby and Bird(2013)]%
        {RigbyAndBird2013}
\bibfield{author}{\bibinfo{person}{Peter~C. Rigby} {and} \bibinfo{person}{Christian Bird}.} \bibinfo{year}{2013}\natexlab{}.
\newblock \showarticletitle{Convergent Contemporary Software Peer Review Practices}. In \bibinfo{booktitle}{\emph{Proceedings of the 2013 9th Joint Meeting on Foundations of Software Engineering}} (Saint Petersburg, Russia) \emph{(\bibinfo{series}{ESEC/FSE 2013})}. \bibinfo{publisher}{Association for Computing Machinery}, \bibinfo{address}{New York, NY, USA}, \bibinfo{pages}{202--212}.
\newblock
\showISBNx{9781450322379}
\urldef\tempurl%
\url{https://doi.org/10.1145/2491411.2491444}
\showDOI{\tempurl}


\bibitem[Rinehart et~al\mbox{.}(2015)]%
        {RinehartEtAl2015}
\bibfield{author}{\bibinfo{person}{David~J. Rinehart}, \bibinfo{person}{John~C. Knight}, {and} \bibinfo{person}{Jonathan Rowanhill}.} \bibinfo{year}{2015}\natexlab{}.
\newblock \bibinfo{booktitle}{\emph{Current Practices in Constructing and Evaluating Assurance Cases with Applications to Aviation}}.
\newblock \bibinfo{type}{{T}echnical {R}eport} CR-2014-218678. \bibinfo{institution}{National Aeronautics and Space Administration (NASA)}, \bibinfo{address}{Langley Research Centre, Hampton, Virginia}.
\newblock


\bibitem[Robertson and Robertson(1999)]%
        {RobertsonAndRobertson1999Vol}
\bibfield{author}{\bibinfo{person}{Suzanne Robertson} {and} \bibinfo{person}{James Robertson}.} \bibinfo{year}{1999}\natexlab{}.
\newblock \bibinfo{booktitle}{\emph{Mastering the Requirements Process}}.
\newblock \bibinfo{publisher}{ACM Press/Addison-Wesley Publishing Co}, \bibinfo{address}{New York, NY, USA}, Chapter Volere Requirements Specification Template, \bibinfo{pages}{353--391}.
\newblock


\bibitem[Roduit(2021)]%
        {Roduit2021}
\bibfield{author}{\bibinfo{person}{Nicolas Roduit}.} \bibinfo{year}{2021}\natexlab{}.
\newblock \bibinfo{title}{Weasis}.
\newblock \bibinfo{howpublished}{\url{https://github.com/nroduit/nroduit.github.io}}.
\newblock
\newblock
\shownote{[Online; accessed 27-May-2021]}.


\bibitem[Rokicki and Laniewski-Wollk(2016)]%
        {rokicki2016adjoint}
\bibfield{author}{\bibinfo{person}{J. Rokicki} {and} \bibinfo{person}{L. Laniewski-Wollk}.} \bibinfo{year}{2016}\natexlab{}.
\newblock \showarticletitle{Adjoint lattice Boltzmann for topology optimization on multi-GPU architecture}.
\newblock \bibinfo{journal}{\emph{Computers \& Mathematics with Applications}} \bibinfo{volume}{71}, \bibinfo{number}{3} (\bibinfo{year}{2016}), \bibinfo{pages}{833--848}.
\newblock


\bibitem[Rueden et~al\mbox{.}(2017)]%
        {Rueden2017}
\bibfield{author}{\bibinfo{person}{Curtis Rueden}, \bibinfo{person}{Johannes Schindelin}, \bibinfo{person}{Mark Hiner}, \bibinfo{person}{Barry Dezonia}, \bibinfo{person}{Alison Walter}, {and} \bibinfo{person}{Kevin Eliceiri}.} \bibinfo{year}{2017}\natexlab{}.
\newblock \showarticletitle{ImageJ2: ImageJ for the next generation of scientific image data}.
\newblock \bibinfo{journal}{\emph{BMC Bioinformatics}}  \bibinfo{volume}{18} (\bibinfo{date}{11} \bibinfo{year}{2017}).
\newblock
\urldef\tempurl%
\url{https://doi.org/10.1186/s12859-017-1934-z}
\showDOI{\tempurl}


\bibitem[Runeson and H{\"o}st(2009)]%
        {RunesonAndHost2009}
\bibfield{author}{\bibinfo{person}{Per Runeson} {and} \bibinfo{person}{Martin H{\"o}st}.} \bibinfo{year}{2009}\natexlab{}.
\newblock \showarticletitle{Guidelines for conducting and reporting case study research in software engineering}.
\newblock \bibinfo{journal}{\emph{Empirical Software Engineering}} \bibinfo{volume}{14}, \bibinfo{number}{2} (\bibinfo{date}{19 Dec} \bibinfo{year}{2009}), \bibinfo{pages}{131--164}.
\newblock
\showISSN{1573-7616}
\urldef\tempurl%
\url{https://doi.org/10.1007/s10664-008-9102-8}
\showDOI{\tempurl}


\bibitem[Saaty(1990)]%
        {Saaty1990}
\bibfield{author}{\bibinfo{person}{Thomas~L. Saaty}.} \bibinfo{year}{1990}\natexlab{}.
\newblock \showarticletitle{How to make a decision: The analytic hierarchy process}.
\newblock \bibinfo{journal}{\emph{European Journal of Operational Research}} \bibinfo{volume}{48}, \bibinfo{number}{1} (\bibinfo{year}{1990}), \bibinfo{pages}{9--26}.
\newblock
\showISSN{0377-2217}
\urldef\tempurl%
\url{https://doi.org/10.1016/0377-2217(90)90057-I}
\showDOI{\tempurl}
\newblock
\shownote{Desicion making by the analytic hierarchy process: Theory and applications}.


\bibitem[Sadowski et~al\mbox{.}(2018)]%
        {SadowskiEtAl2018}
\bibfield{author}{\bibinfo{person}{Caitlin Sadowski}, \bibinfo{person}{Emma S{\"o}derberg}, \bibinfo{person}{Luke Church}, \bibinfo{person}{Michal Sipko}, {and} \bibinfo{person}{Alberto Bacchelli}.} \bibinfo{year}{2018}\natexlab{}.
\newblock \showarticletitle{Modern code review: a case study at google}. In \bibinfo{booktitle}{\emph{Proceedings of the 40th International Conference on Software Engineering: Software Engineering in Practice}}. \bibinfo{pages}{181--190}.
\newblock


\bibitem[Samala(2014)]%
        {Samala2014}
\bibfield{author}{\bibinfo{person}{Ravi Samala}.} \bibinfo{year}{2014}\natexlab{}.
\newblock \bibinfo{title}{Can anyone suggest free software for medical images segmentation and volume?}
\newblock \bibinfo{howpublished}{\url{https://www.researchgate.net/post/Can_anyone_suggest_free_software_for_medical_images_segmentation_and_volume}}.
\newblock
\newblock
\shownote{[Online; accessed 31-January-2020]}.


\bibitem[Sanders and Kelly(2008)]%
        {SandersAndKelly2008}
\bibfield{author}{\bibinfo{person}{Rebecca Sanders} {and} \bibinfo{person}{Diane Kelly}.} \bibinfo{year}{2008}\natexlab{}.
\newblock \showarticletitle{Dealing with Risk in Scientific Software Development}.
\newblock \bibinfo{journal}{\emph{IEEE Software}}  \bibinfo{volume}{4} (\bibinfo{date}{July/August} \bibinfo{year}{2008}), \bibinfo{pages}{21--28}.
\newblock


\bibitem[SARL(2019)]%
        {PixmeoSARL2019}
\bibfield{author}{\bibinfo{person}{Pixmeo SARL}.} \bibinfo{year}{2019}\natexlab{}.
\newblock \bibinfo{title}{OsiriX Lite}.
\newblock \bibinfo{howpublished}{\url{https://github.com/pixmeo/osirix}}.
\newblock
\newblock
\shownote{[Online; accessed 27-May-2021]}.


\bibitem[Schindelin et~al\mbox{.}(2012)]%
        {Schindelin2012}
\bibfield{author}{\bibinfo{person}{Johannes Schindelin}, \bibinfo{person}{Ignacio Arganda-Carreras}, \bibinfo{person}{Erwin Frise}, \bibinfo{person}{Verena Kaynig}, \bibinfo{person}{Mark Longair}, \bibinfo{person}{Tobias Pietzsch}, \bibinfo{person}{Stephan Preibisch}, \bibinfo{person}{Curtis Rueden}, \bibinfo{person}{Stephan Saalfeld}, \bibinfo{person}{Benjamin Schmid}, \bibinfo{person}{Jean-Yves Tinevez}, \bibinfo{person}{Daniel White}, \bibinfo{person}{Volker Hartenstein}, \bibinfo{person}{Kevin Eliceiri}, \bibinfo{person}{Pavel Tomancak}, {and} \bibinfo{person}{Albert Cardona}.} \bibinfo{year}{2012}\natexlab{}.
\newblock \showarticletitle{Fiji: An Open-Source Platform for Biological-Image Analysis}.
\newblock \bibinfo{journal}{\emph{Nature methods}}  \bibinfo{volume}{9} (\bibinfo{date}{06} \bibinfo{year}{2012}), \bibinfo{pages}{676--82}.
\newblock
\urldef\tempurl%
\url{https://doi.org/10.1038/nmeth.2019}
\showDOI{\tempurl}


\bibitem[Schlauch et~al\mbox{.}(2018)]%
        {TobiasEtAl2018}
\bibfield{author}{\bibinfo{person}{Tobias Schlauch}, \bibinfo{person}{Michael Meinel}, {and} \bibinfo{person}{Carina Haupt}.} \bibinfo{year}{2018}\natexlab{}.
\newblock \bibinfo{title}{DLR Software Engineering Guidelines}.
\newblock
\newblock
\urldef\tempurl%
\url{https://doi.org/10.5281/zenodo.1344612}
\showDOI{\tempurl}


\bibitem[Schroeder et~al\mbox{.}(2006)]%
        {SchroederEtAl2006}
\bibfield{author}{\bibinfo{person}{Will Schroeder}, \bibinfo{person}{Bill Lorensen}, {and} \bibinfo{person}{Ken Martin}.} \bibinfo{year}{2006}\natexlab{}.
\newblock \bibinfo{booktitle}{\emph{The visualization toolkit}}.
\newblock \bibinfo{publisher}{Kitware}.
\newblock


\bibitem[Segal(2005)]%
        {Segal2005}
\bibfield{author}{\bibinfo{person}{Judith Segal}.} \bibinfo{year}{2005}\natexlab{}.
\newblock \showarticletitle{When Software Engineers Met Research Scientists: A Case Study}.
\newblock \bibinfo{journal}{\emph{Empirical Software Engineering}} \bibinfo{volume}{10}, \bibinfo{number}{4} (\bibinfo{date}{Oct.} \bibinfo{year}{2005}), \bibinfo{pages}{517--536}.
\newblock
\showISSN{1382-3256}
\urldef\tempurl%
\url{https://doi.org/10.1007/s10664-005-3865-y}
\showDOI{\tempurl}


\bibitem[Segal and Morris(2008)]%
        {SegalAndMorris2008}
\bibfield{author}{\bibinfo{person}{Judith Segal} {and} \bibinfo{person}{Chris Morris}.} \bibinfo{year}{2008}\natexlab{}.
\newblock \showarticletitle{Developing Scientific Software}.
\newblock \bibinfo{journal}{\emph{IEEE Software}} \bibinfo{volume}{25}, \bibinfo{number}{4} (\bibinfo{date}{July/August} \bibinfo{year}{2008}), \bibinfo{pages}{18--20}.
\newblock


\bibitem[Shahin et~al\mbox{.}(2017)]%
        {ShahinEtAl2017}
\bibfield{author}{\bibinfo{person}{Mojtaba Shahin}, \bibinfo{person}{Muhammad Ali~Babar}, {and} \bibinfo{person}{Liming Zhu}.} \bibinfo{year}{2017}\natexlab{}.
\newblock \showarticletitle{Continuous Integration, Delivery and Deployment: A Systematic Review on Approaches, Tools, Challenges and Practices}.
\newblock \bibinfo{journal}{\emph{IEEE Access}}  \bibinfo{volume}{5} (\bibinfo{year}{2017}), \bibinfo{pages}{3909--3943}.
\newblock
\urldef\tempurl%
\url{https://doi.org/10.1109/ACCESS.2017.2685629}
\showDOI{\tempurl}


\bibitem[Singh et~al\mbox{.}(2021)]%
        {SinghEtAl2021}
\bibfield{author}{\bibinfo{person}{Vandana Singh}, \bibinfo{person}{Brice Bongiovanni}, {and} \bibinfo{person}{William Brandon}.} \bibinfo{year}{2021}\natexlab{}.
\newblock \showarticletitle{Codes of conduct in Open Source Software---for warm and fuzzy feelings or equality in community?}
\newblock \bibinfo{journal}{\emph{Software Quality Journal}} (\bibinfo{year}{2021}).
\newblock
\showISBNx{1573-1367}
\urldef\tempurl%
\url{https://doi.org/10.1007/s11219-020-09543-w}
\showDOI{\tempurl}


\bibitem[Slaughter et~al\mbox{.}(2021)]%
        {SlaughterEtAl2021}
\bibfield{author}{\bibinfo{person}{Andrew Slaughter}, \bibinfo{person}{Cody Permann}, \bibinfo{person}{Jason Miller}, \bibinfo{person}{Brian Alger}, {and} \bibinfo{person}{Stephen Novascone}.} \bibinfo{year}{2021}\natexlab{}.
\newblock \showarticletitle{Continuous Integration, In-Code Documentation, and Automation for Nuclear Quality Assurance Conformance}.
\newblock \bibinfo{journal}{\emph{Nuclear Technology}}  \bibinfo{volume}{207} (\bibinfo{date}{01} \bibinfo{year}{2021}), \bibinfo{pages}{1--8}.
\newblock
\urldef\tempurl%
\url{https://doi.org/10.1080/00295450.2020.1826804}
\showDOI{\tempurl}


\bibitem[Smirniotopoulos(2014)]%
        {Smirniotopoulos2014}
\bibfield{author}{\bibinfo{person}{James Smirniotopoulos}.} \bibinfo{year}{2014}\natexlab{}.
\newblock \bibinfo{title}{MedPix Medical Image Database}.
\newblock
\newblock
\urldef\tempurl%
\url{https://doi.org/10.13140/2.1.3403.3608}
\showDOI{\tempurl}


\bibitem[Smith et~al\mbox{.}(2018a)]%
        {SmithAndRoscoe2018}
\bibfield{author}{\bibinfo{person}{Barry Smith}, \bibinfo{person}{Roscoe Bartlett}, {and} \bibinfo{person}{xSDK Developers}.} \bibinfo{year}{2018}\natexlab{a}.
\newblock \bibinfo{title}{xSDK Community Package Policies}.
\newblock
\newblock
\urldef\tempurl%
\url{https://doi.org/10.6084/m9.figshare.4495136.v6}
\showDOI{\tempurl}


\bibitem[Smith et~al\mbox{.}(2020a)]%
        {SmithEtAl2020}
\bibfield{author}{\bibinfo{person}{Spencer Smith}, \bibinfo{person}{Jacques Carette}, \bibinfo{person}{Olu Owojaiye}, \bibinfo{person}{Peter Michalski}, {and} \bibinfo{person}{Ao Dong}.} \bibinfo{year}{2020}\natexlab{a}.
\newblock \bibinfo{title}{Quality Definitions of Qualities}.  (\bibinfo{year}{2020}).
\newblock
\newblock
\shownote{Manuscript in preparation}.


\bibitem[Smith and Michalski(2022)]%
        {SmithAndMichalski2022}
\bibfield{author}{\bibinfo{person}{Spencer Smith} {and} \bibinfo{person}{Peter Michalski}.} \bibinfo{year}{2022}\natexlab{}.
\newblock \showarticletitle{Digging Deeper Into the State of the Practice for Domain Specific Research Software}. In \bibinfo{booktitle}{\emph{Proceedings of the International Conference on Computational Science, {ICCS}}}. \bibinfo{pages}{1--15}.
\newblock


\bibitem[Smith et~al\mbox{.}(2024)]%
        {SmithEtAl2024}
\bibfield{author}{\bibinfo{person}{Spencer Smith}, \bibinfo{person}{Peter Michalski}, \bibinfo{person}{Jacques Carette}, {and} \bibinfo{person}{Zahra Keshavarz-Motamed}.} \bibinfo{year}{2024}\natexlab{}.
\newblock \showarticletitle{State of the Practice for {L}attice {B}oltzmann {M}ethod Software}.
\newblock \bibinfo{journal}{\emph{Archives of Computational Methods in Engineering}} \bibinfo{volume}{31}, \bibinfo{number}{1} (\bibinfo{date}{Jan} \bibinfo{year}{2024}), \bibinfo{pages}{313--350}.
\newblock
\urldef\tempurl%
\url{https://doi.org/10.1007/s11831-023-09981-2}
\showDOI{\tempurl}


\bibitem[Smith et~al\mbox{.}(2018c)]%
        {smith2018statistical}
\bibfield{author}{\bibinfo{person}{Spencer Smith}, \bibinfo{person}{Yue Sun}, {and} \bibinfo{person}{Jacques Carette}.} \bibinfo{year}{2018}\natexlab{c}.
\newblock \bibinfo{title}{Statistical Software for Psychology: Comparing Development Practices Between CRAN and Other Communities}.
\newblock
\newblock
\showeprint[arxiv]{1802.07362}~[cs.SE]


\bibitem[Smith et~al\mbox{.}(2018e)]%
        {Smith2018Seismology}
\bibfield{author}{\bibinfo{person}{Spencer Smith}, \bibinfo{person}{Zheng Zeng}, {and} \bibinfo{person}{Jacques Carette}.} \bibinfo{year}{2018}\natexlab{e}.
\newblock \showarticletitle{Seismology software: state of the practice}.
\newblock \bibinfo{journal}{\emph{Journal of Seismology}}  \bibinfo{volume}{22} (\bibinfo{date}{05} \bibinfo{year}{2018}).
\newblock
\urldef\tempurl%
\url{https://doi.org/10.1007/s10950-018-9731-3}
\showDOI{\tempurl}


\bibitem[Smith(2016)]%
        {Smith2016}
\bibfield{author}{\bibinfo{person}{W.~Spencer Smith}.} \bibinfo{year}{2016}\natexlab{}.
\newblock \showarticletitle{A Rational Document Driven Design Process for Scientific Computing Software}.
\newblock In \bibinfo{booktitle}{\emph{Software Engineering for Science}}, \bibfield{editor}{\bibinfo{person}{Jeffrey~C. Carver}, \bibinfo{person}{Neil~Chue Hong}, {and} \bibinfo{person}{George Thiruvathukal}} (Eds.). \bibinfo{publisher}{Taylor \& Francis}, Chapter Section {I} -- Examples of the Application of Traditional Software Engineering Practices to Science, \bibinfo{pages}{33--63}.
\newblock


\bibitem[Smith(2018)]%
        {Smith2018}
\bibfield{author}{\bibinfo{person}{W.~Spencer Smith}.} \bibinfo{year}{2018}\natexlab{}.
\newblock \showarticletitle{Beyond Software Carpentry}. In \bibinfo{booktitle}{\emph{2018 International Workshop on Software Engineering for Science (held in conjunction with ICSE'18)}}. \bibinfo{pages}{1--8}.
\newblock


\bibitem[Smith and Carette(2021)]%
        {SmithAndCarette2021-BRIC}
\bibfield{author}{\bibinfo{person}{W.~Spencer Smith} {and} \bibinfo{person}{Jacques Carette}.} \bibinfo{year}{2021}\natexlab{}.
\newblock \showarticletitle{Sustainable Software via Generation}. In \bibinfo{booktitle}{\emph{Proceedings of the 1st Annual Booth Resource and Innovation Cluster (BRIC) Symposium}}. \bibinfo{pages}{21}.
\newblock


\bibitem[Smith et~al\mbox{.}(2021)]%
        {SmithEtAl2021}
\bibfield{author}{\bibinfo{person}{W.~Spencer Smith}, \bibinfo{person}{Jacques Carette}, \bibinfo{person}{Peter Michalski}, \bibinfo{person}{Ao Dong}, {and} \bibinfo{person}{Oluwaseun Owojaiye}.} \bibinfo{year}{2021}\natexlab{}.
\newblock \bibinfo{title}{Methodology for Assessing the State of the Practice for Domain {X}}.
\newblock \bibinfo{howpublished}{\url{https://arxiv.org/abs/2110.11575}}.
\newblock


\bibitem[Smith and Koothoor(2016)]%
        {SmithAndKoothoor2016}
\bibfield{author}{\bibinfo{person}{W.~Spencer Smith} {and} \bibinfo{person}{Nirmitha Koothoor}.} \bibinfo{year}{2016}\natexlab{}.
\newblock \showarticletitle{A Document-Driven Method for Certifying Scientific Computing Software for Use in Nuclear Safety Analysis}.
\newblock \bibinfo{journal}{\emph{Nuclear Engineering and Technology}} \bibinfo{volume}{48}, \bibinfo{number}{2} (\bibinfo{date}{April} \bibinfo{year}{2016}), \bibinfo{pages}{404--418}.
\newblock
\showISSN{1738-5733}
\urldef\tempurl%
\url{https://doi.org/10.1016/j.net.2015.11.008}
\showDOI{\tempurl}


\bibitem[Smith et~al\mbox{.}(2007)]%
        {SmithEtAl2007}
\bibfield{author}{\bibinfo{person}{W.~Spencer Smith}, \bibinfo{person}{Lei Lai}, {and} \bibinfo{person}{Ridha Khedri}.} \bibinfo{year}{2007}\natexlab{}.
\newblock \showarticletitle{Requirements Analysis for Engineering Computation: A Systematic Approach for Improving Software Reliability}.
\newblock \bibinfo{journal}{\emph{Reliable Computing, Special Issue on Reliable Engineering Computation}} \bibinfo{volume}{13}, \bibinfo{number}{1} (\bibinfo{date}{Feb.} \bibinfo{year}{2007}), \bibinfo{pages}{83--107}.
\newblock
\showISSN{1573-1340}
\urldef\tempurl%
\url{https://doi.org/10.1007/s11155-006-9020-7}
\showDOI{\tempurl}


\bibitem[Smith et~al\mbox{.}(2016a)]%
        {SmithEtAl2016}
\bibfield{author}{\bibinfo{person}{W.~Spencer Smith}, \bibinfo{person}{Adam Lazzarato}, {and} \bibinfo{person}{Jacques Carette}.} \bibinfo{year}{2016}\natexlab{a}.
\newblock \showarticletitle{State of Practice for Mesh Generation Software}.
\newblock \bibinfo{journal}{\emph{Advances in Engineering Software}}  \bibinfo{volume}{100} (\bibinfo{date}{Oct.} \bibinfo{year}{2016}), \bibinfo{pages}{53--71}.
\newblock


\bibitem[Smith et~al\mbox{.}(2018b)]%
        {smith2018state}
\bibfield{author}{\bibinfo{person}{W.~Spencer Smith}, \bibinfo{person}{Adam Lazzarato}, {and} \bibinfo{person}{Jacques Carette}.} \bibinfo{year}{2018}\natexlab{b}.
\newblock \bibinfo{title}{State of the Practice for GIS Software}.
\newblock
\newblock
\showeprint[arxiv]{1802.03422}~[cs.SE]


\bibitem[Smith et~al\mbox{.}(2016b)]%
        {smith2016state}
\bibfield{author}{\bibinfo{person}{W~Spencer Smith}, \bibinfo{person}{D~Adam Lazzarato}, {and} \bibinfo{person}{Jacques Carette}.} \bibinfo{year}{2016}\natexlab{b}.
\newblock \showarticletitle{State of the practice for mesh generation and mesh processing software}.
\newblock \bibinfo{journal}{\emph{Advances in Engineering Software}}  \bibinfo{volume}{100} (\bibinfo{year}{2016}), \bibinfo{pages}{53--71}.
\newblock


\bibitem[Smith et~al\mbox{.}(2020b)]%
        {SmithEtAl2020_AC}
\bibfield{author}{\bibinfo{person}{W.~Spencer Smith}, \bibinfo{person}{Mojdeh~Sayari Nejad}, {and} \bibinfo{person}{Alan Wassyng}.} \bibinfo{year}{2020}\natexlab{b}.
\newblock \showarticletitle{Raising the Bar: Assurance Cases for Scientific Computing Software}.
\newblock \bibinfo{journal}{\emph{Computing in Science and Engineering}} \bibinfo{volume}{23}, \bibinfo{number}{1} (\bibinfo{date}{Feb.} \bibinfo{year}{2020}), \bibinfo{pages}{47--57}.
\newblock


\bibitem[Smith et~al\mbox{.}(2018d)]%
        {SmithEtAl2018_StatSoft}
\bibfield{author}{\bibinfo{person}{W.~Spencer Smith}, \bibinfo{person}{Yue Sun}, {and} \bibinfo{person}{Jacques Carette}.} \bibinfo{year}{2018}\natexlab{d}.
\newblock \bibinfo{title}{Statistical Software for Psychology: Comparing Development Practices Between {CRAN} and Other Communities}.
\newblock \bibinfo{howpublished}{\url{https://arxiv.org/abs/1802.07362}}.
\newblock
\newblock
\shownote{33 pp.}.


\bibitem[Smith and Yu(2009)]%
        {SmithAndYu2009}
\bibfield{author}{\bibinfo{person}{W.~Spencer Smith} {and} \bibinfo{person}{Wen Yu}.} \bibinfo{year}{2009}\natexlab{}.
\newblock \showarticletitle{A Document Driven Methodology for Improving the Quality of a Parallel Mesh Generation Toolbox}.
\newblock \bibinfo{journal}{\emph{Advances in Engineering Software}} \bibinfo{volume}{40}, \bibinfo{number}{11} (\bibinfo{date}{Nov.} \bibinfo{year}{2009}), \bibinfo{pages}{1155--1167}.
\newblock
\urldef\tempurl%
\url{https://doi.org/10.1016/j.advengsoft.2009.05.003}
\showDOI{\tempurl}


\bibitem[SourceLevel(2022)]%
        {SourceLevel2022_Lint}
\bibfield{author}{\bibinfo{person}{SourceLevel}.} \bibinfo{year}{2022}\natexlab{}.
\newblock \bibinfo{title}{What is a linter and why your team should use it?}
\newblock \bibinfo{howpublished}{\url{https://sourcelevel.io/blog/what-is-a-linter-and-why-your-team-should-use-it}}.
\newblock


\bibitem[Spriggs(2012)]%
        {Spriggs2012}
\bibfield{author}{\bibinfo{person}{John Spriggs}.} \bibinfo{year}{2012}\natexlab{}.
\newblock \bibinfo{booktitle}{\emph{{GSN} - The Goal Structuring Notation}}.
\newblock \bibinfo{publisher}{Springer-Verlag}, \bibinfo{address}{London}.
\newblock


\bibitem[Stewart et~al\mbox{.}(2017)]%
        {StewartEtAl2017}
\bibfield{author}{\bibinfo{person}{Graeme Stewart} {et~al\mbox{.}}} \bibinfo{year}{2017}\natexlab{}.
\newblock \showarticletitle{{A Roadmap for HEP Software and Computing R\&D for the 2020s}}.
\newblock \bibinfo{journal}{\emph{arXiv}} (\bibinfo{year}{2017}).
\newblock
\showeprint[arxiv]{1712.06982}~[physics.comp-ph]


\bibitem[Storer(2017)]%
        {Storer2017}
\bibfield{author}{\bibinfo{person}{Tim Storer}.} \bibinfo{year}{2017}\natexlab{}.
\newblock \showarticletitle{Bridging the Chasm: A Survey of Software Engineering Practice in Scientific Programming}.
\newblock \bibinfo{journal}{\emph{ACM Comput. Surv.}} \bibinfo{volume}{50}, \bibinfo{number}{4}, Article \bibinfo{articleno}{47} (\bibinfo{date}{Aug.} \bibinfo{year}{2017}), \bibinfo{numpages}{32}~pages.
\newblock
\showISSN{0360-0300}
\urldef\tempurl%
\url{https://doi.org/10.1145/3084225}
\showDOI{\tempurl}


\bibitem[Szulik(2017)]%
        {Szulik2017}
\bibfield{author}{\bibinfo{person}{Keenan Szulik}.} \bibinfo{year}{2017}\natexlab{}.
\newblock \bibinfo{title}{Don't judge a project by its GitHub stars alone}.
\newblock \bibinfo{howpublished}{\url{https://blog.tidelift.com/dont-judge-a-project-by-its-github-stars-alone}}.
\newblock


\bibitem[Szymczak et~al\mbox{.}(2016)]%
        {SzymczakEtAl2016}
\bibfield{author}{\bibinfo{person}{Daniel Szymczak}, \bibinfo{person}{W.~Spencer Smith}, {and} \bibinfo{person}{Jacques Carette}.} \bibinfo{year}{2016}\natexlab{}.
\newblock \showarticletitle{Position Paper: A Knowledge-Based Approach to Scientific Software Development}. In \bibinfo{booktitle}{\emph{Proceedings of SE4Science'16 in conjunction with the International Conference on Software Engineering (ICSE)}}. In conjunction with ICSE 2016, \bibinfo{address}{Austin, Texas, United States}.
\newblock
\newblock
\shownote{4 pp.}.


\bibitem[TESCAN(2020)]%
        {TESCAN2020}
\bibfield{author}{\bibinfo{person}{TESCAN}.} \bibinfo{year}{2020}\natexlab{}.
\newblock \bibinfo{title}{3DimViewer}.
\newblock \bibinfo{howpublished}{\url{https://bitbucket.org/3dimlab/3dimviewer/src/master/}}.
\newblock
\newblock
\shownote{[Online; accessed 27-May-2021]}.


\bibitem[Thiel(2020)]%
        {ThielEtAl2020}
\bibfield{author}{\bibinfo{person}{Carsten Thiel}.} \bibinfo{year}{2020}\natexlab{}.
\newblock \bibinfo{title}{{EURISE} Network Technical Reference}.
\newblock \bibinfo{howpublished}{\url{https://technical-reference.readthedocs.io/en/latest/}}.
\newblock


\bibitem[Tourani et~al\mbox{.}(2017)]%
        {TouraniEtAl2017}
\bibfield{author}{\bibinfo{person}{Parastou Tourani}, \bibinfo{person}{Bram Adams}, {and} \bibinfo{person}{Alexander Serebrenik}.} \bibinfo{year}{2017}\natexlab{}.
\newblock \showarticletitle{Code of conduct in open source projects}. In \bibinfo{booktitle}{\emph{2017 IEEE 24th International Conference on Software Analysis, Evolution and Reengineering (SANER)}}. \bibinfo{pages}{24--33}.
\newblock
\urldef\tempurl%
\url{https://doi.org/10.1109/SANER.2017.7884606}
\showDOI{\tempurl}


\bibitem[{USGS}(2019)]%
        {USGS2019}
\bibfield{author}{\bibinfo{person}{{USGS}}.} \bibinfo{year}{2019}\natexlab{}.
\newblock \bibinfo{title}{{USGS} (United States Geological Survey) Software Plannning Checklist}.
\newblock \bibinfo{howpublished}{\url{https://www.usgs.gov/media/files/usgs-software-planning-checklist}}.
\newblock


\bibitem[V{\"a}h{\"a}niitty et~al\mbox{.}(2002)]%
        {VahaniittyEtAl2002}
\bibfield{author}{\bibinfo{person}{Jarno V{\"a}h{\"a}niitty}, \bibinfo{person}{Casper Lassenius}, {and} \bibinfo{person}{Kristian Rautiainen}.} \bibinfo{year}{2002}\natexlab{}.
\newblock \showarticletitle{An Approach to Product Roadmapping in Small Software Product Businesses}. University of Technologie, Software Business and Engineering Institute, \bibinfo{address}{Helsinki, Finland}.
\newblock


\bibitem[Vaidya and Kumar(2006)]%
        {VaidyaEtAl2006}
\bibfield{author}{\bibinfo{person}{Omkarprasad~S. Vaidya} {and} \bibinfo{person}{Sushil Kumar}.} \bibinfo{year}{2006}\natexlab{}.
\newblock \showarticletitle{Analytic hierarchy process: An overview of applications}.
\newblock \bibinfo{journal}{\emph{European Journal of Operational Research}} \bibinfo{volume}{169}, \bibinfo{number}{1} (\bibinfo{year}{2006}), \bibinfo{pages}{1--29}.
\newblock
\showISSN{0377-2217}
\urldef\tempurl%
\url{https://doi.org/10.1016/j.ejor.2004.04.028}
\showDOI{\tempurl}


\bibitem[van Gompel et~al\mbox{.}(2016)]%
        {vanGompelEtAl2016}
\bibfield{author}{\bibinfo{person}{Maarten van Gompel}, \bibinfo{person}{Jauco Noordzij}, \bibinfo{person}{Reinier de Valk}, {and} \bibinfo{person}{Andrea Scharnhorst}.} \bibinfo{year}{2016}\natexlab{}.
\newblock \bibinfo{title}{Guidelines for Software Quality, {CLARIAH} Task Force 54.100}.
\newblock \bibinfo{howpublished}{\url{https://github.com/CLARIAH/software-quality-guidelines/blob/master/softwareguidelines.pdf}}.
\newblock


\bibitem[van Vliet(2000)]%
        {VanVliet2000}
\bibfield{author}{\bibinfo{person}{Hans van Vliet}.} \bibinfo{year}{2000}\natexlab{}.
\newblock \bibinfo{booktitle}{\emph{Software Engineering (2nd ed.): Principles and Practice}}.
\newblock \bibinfo{publisher}{John Wiley \& Sons, Inc.}, \bibinfo{address}{New York, NY, USA}.
\newblock
\showISBNx{0-471-97508-7}


\bibitem[Veldhuizen(1998)]%
        {Veldhuizen1998}
\bibfield{author}{\bibinfo{person}{Todd.~L. Veldhuizen}.} \bibinfo{year}{1998}\natexlab{}.
\newblock \showarticletitle{Arrays in {Blitz++}}. In \bibinfo{booktitle}{\emph{Proceedings of the 2nd International Scientific Computing in Object-Oriented Parallel Environments ({ISCOPE}'98), Lecture Notes in Computer Science}}. \bibinfo{publisher}{Springer-Verlag}.
\newblock


\bibitem[Wang et~al\mbox{.}(2017)]%
        {WangEtAl2017}
\bibfield{author}{\bibinfo{person}{Xiaosong Wang}, \bibinfo{person}{Yifan Peng}, \bibinfo{person}{Le Lu}, \bibinfo{person}{Zhiyong Lu}, \bibinfo{person}{Mohammadhadi Bagheri}, {and} \bibinfo{person}{Ronald Summers}.} \bibinfo{year}{2017}\natexlab{}.
\newblock \showarticletitle{ChestX-ray8: Hospital-scale Chest X-ray Database and Benchmarks on Weakly-Supervised Classification and Localization of Common Thorax Diseases}.
\newblock \bibinfo{journal}{\emph{arXiv:1705.02315}} (\bibinfo{date}{05} \bibinfo{year}{2017}).
\newblock


\bibitem[Ward(2000)]%
        {Ward2000}
\bibfield{author}{\bibinfo{person}{B.~Douglas Ward}.} \bibinfo{year}{2000}\natexlab{}.
\newblock \bibinfo{booktitle}{\emph{Program 3dfim+}}.
\newblock Biophysics Research Institute, Medical College of Wisconsin.
\newblock
\urldef\tempurl%
\url{https://afni.nimh.nih.gov/afni/doc/manual/3dfim+.pdf}
\showURL{%
\tempurl}


\bibitem[Wassyng et~al\mbox{.}(2015)]%
        {Wassyng2015}
\bibfield{author}{\bibinfo{person}{Alan Wassyng}, \bibinfo{person}{Neeraj~Kumar Singh}, \bibinfo{person}{Mischa Geven}, \bibinfo{person}{Nicholas Proscia}, \bibinfo{person}{Hao Wang}, \bibinfo{person}{Mark Lawford}, {and} \bibinfo{person}{Tom Maibaum}.} \bibinfo{year}{2015}\natexlab{}.
\newblock \showarticletitle{Can Product-Specific Assurance Case Templates Be Used as Medical Device Standards?}
\newblock \bibinfo{journal}{\emph{{IEEE} Design {\&} Test}} \bibinfo{volume}{32}, \bibinfo{number}{5} (\bibinfo{year}{2015}), \bibinfo{pages}{45--55}.
\newblock
\urldef\tempurl%
\url{https://doi.org/10.1109/MDAT.2015.2462720}
\showDOI{\tempurl}


\bibitem[Whaley et~al\mbox{.}(2001)]%
        {WhaleyEtAl2001}
\bibfield{author}{\bibinfo{person}{R.~C. Whaley}, \bibinfo{person}{A. Petitet}, {and} \bibinfo{person}{J.~J. Dongarra}.} \bibinfo{year}{2001}\natexlab{}.
\newblock \showarticletitle{Automated empirical optimization of software and the {ATLAS} project}.
\newblock \bibinfo{journal}{\emph{Parallel Comput.}} \bibinfo{volume}{27}, \bibinfo{number}{1--2} (\bibinfo{year}{2001}), \bibinfo{pages}{3--35}.
\newblock


\bibitem[Whitehouse(2018)]%
        {Whitehouse2018}
\bibfield{author}{\bibinfo{person}{Ross Whitehouse}.} \bibinfo{year}{2018}\natexlab{}.
\newblock \bibinfo{title}{Setting up ESLint in React}.
\newblock \bibinfo{howpublished}{\url{https://medium.com/@RossWhitehouse/setting-up-eslint-in-react-c20015ef35f7}}.
\newblock


\bibitem[{Wiese} et~al\mbox{.}(2019)]%
        {WieseEtAl2019}
\bibfield{author}{\bibinfo{person}{I.~S. {Wiese}}, \bibinfo{person}{I. {Polato}}, {and} \bibinfo{person}{G. {Pinto}}.} \bibinfo{year}{2019}\natexlab{}.
\newblock \showarticletitle{Naming the Pain in Developing Scientific Software}.
\newblock \bibinfo{journal}{\emph{IEEE Software}} (\bibinfo{year}{2019}), \bibinfo{pages}{1--1}.
\newblock
\showISSN{0740-7459}
\urldef\tempurl%
\url{https://doi.org/10.1109/MS.2019.2899838}
\showDOI{\tempurl}


\bibitem[Wikipedia(2022)]%
        {Wikipedia2022_Lint}
\bibfield{author}{\bibinfo{person}{Wikipedia}.} \bibinfo{year}{2022}\natexlab{}.
\newblock \bibinfo{title}{Lint (software)}.
\newblock \bibinfo{howpublished}{\url{https://en.wikipedia.org/wiki/Lint_(software)}}.
\newblock


\bibitem[{Wikipedia contributors}(2021a)]%
        {enwiki:1039005082}
\bibfield{author}{\bibinfo{person}{{Wikipedia contributors}}.} \bibinfo{year}{2021}\natexlab{a}.
\newblock \bibinfo{title}{Fault injection --- {Wikipedia}{,} The Free Encyclopedia}.
\newblock
\newblock
\urldef\tempurl%
\url{https://en.wikipedia.org/w/index.php?title=Fault_injection&oldid=1039005082}
\showURL{%
\tempurl}
\newblock
\shownote{[Online; accessed 28-August-2021]}.


\bibitem[{Wikipedia contributors}(2021b)]%
        {enwiki:1039424308}
\bibfield{author}{\bibinfo{person}{{Wikipedia contributors}}.} \bibinfo{year}{2021}\natexlab{b}.
\newblock \bibinfo{title}{Fuzzing --- {Wikipedia}{,} The Free Encyclopedia}.
\newblock
\newblock
\urldef\tempurl%
\url{https://en.wikipedia.org/w/index.php?title=Fuzzing&oldid=1039424308}
\showURL{%
\tempurl}
\newblock
\shownote{[Online; accessed 28-August-2021]}.


\bibitem[{Wikipedia contributors}(2021c)]%
        {enwiki:1034877594}
\bibfield{author}{\bibinfo{person}{{Wikipedia contributors}}.} \bibinfo{year}{2021}\natexlab{c}.
\newblock \bibinfo{title}{Medical image computing --- {Wikipedia}{,} The Free Encyclopedia}.
\newblock
\newblock
\urldef\tempurl%
\url{https://en.wikipedia.org/w/index.php?title=Medical_image_computing&oldid=1034877594}
\showURL{%
\tempurl}
\newblock
\shownote{[Online; accessed 25-July-2021]}.


\bibitem[{Wikipedia contributors}(2021d)]%
        {enwiki:1034887445}
\bibfield{author}{\bibinfo{person}{{Wikipedia contributors}}.} \bibinfo{year}{2021}\natexlab{d}.
\newblock \bibinfo{title}{Medical imaging --- {Wikipedia}{,} The Free Encyclopedia}.
\newblock
\newblock
\urldef\tempurl%
\url{https://en.wikipedia.org/w/index.php?title=Medical_imaging&oldid=1034887445}
\showURL{%
\tempurl}
\newblock
\shownote{[Online; accessed 25-July-2021]}.


\bibitem[Wilson et~al\mbox{.}(2014)]%
        {WilsonEtAl2014}
\bibfield{author}{\bibinfo{person}{Greg Wilson}, \bibinfo{person}{D.~A. Aruliah}, \bibinfo{person}{C.~Titus Brown}, \bibinfo{person}{Neil~P. Chue~Hong}, \bibinfo{person}{Matt Davis}, \bibinfo{person}{Richard~T. Guy}, \bibinfo{person}{Steven H.~D. Haddock}, \bibinfo{person}{Kathryn~D. Huff}, \bibinfo{person}{Ian~M. Mitchell}, \bibinfo{person}{Mark~D. Plumbley}, \bibinfo{person}{Ben Waugh}, \bibinfo{person}{Ethan~P. White}, {and} \bibinfo{person}{Paul Wilson}.} \bibinfo{year}{2014}\natexlab{}.
\newblock \showarticletitle{Best Practices for Scientific Computing}.
\newblock \bibinfo{journal}{\emph{PLoS Biol}} \bibinfo{volume}{12}, \bibinfo{number}{1} (\bibinfo{date}{Jan.} \bibinfo{year}{2014}), \bibinfo{pages}{e1001745}.
\newblock
\urldef\tempurl%
\url{https://doi.org/10.1371/journal.pbio.1001745}
\showDOI{\tempurl}


\bibitem[Wilson et~al\mbox{.}(2016)]%
        {WilsonEtAl2016}
\bibfield{author}{\bibinfo{person}{Greg Wilson}, \bibinfo{person}{Jennifer Bryan}, \bibinfo{person}{Karen Cranston}, \bibinfo{person}{Justin Kitzes}, \bibinfo{person}{Lex Nederbragt}, {and} \bibinfo{person}{Tracy~K. Teal}.} \bibinfo{year}{2016}\natexlab{}.
\newblock \showarticletitle{Good Enough Practices in Scientific Computing}.
\newblock \bibinfo{journal}{\emph{CoRR}}  \bibinfo{volume}{abs/1609.00037} (\bibinfo{year}{2016}).
\newblock
\urldef\tempurl%
\url{http://arxiv.org/abs/1609.00037}
\showURL{%
\tempurl}


\bibitem[Wilson(2006)]%
        {Wilson2006}
\bibfield{author}{\bibinfo{person}{Gregory~V. Wilson}.} \bibinfo{year}{2006}\natexlab{}.
\newblock \showarticletitle{Where's the Real Bottleneck in Scientific Computing? {S}cientists would do well to pick some tools widely used in the software industry}.
\newblock \bibinfo{journal}{\emph{American Scientist}} \bibinfo{volume}{94}, \bibinfo{number}{1} (\bibinfo{year}{2006}).
\newblock
\urldef\tempurl%
\url{http://www.americanscientist.org/issues/pub/wheres-the-real-bottleneck-in-scientific-computing}
\showURL{%
\tempurl}


\bibitem[Wollny(2020)]%
        {Wollny2020}
\bibfield{author}{\bibinfo{person}{Gert Wollny}.} \bibinfo{year}{2020}\natexlab{}.
\newblock \bibinfo{title}{Ginkgo CADx}.
\newblock \bibinfo{howpublished}{\url{https://github.com/gerddie/ginkgocadx}}.
\newblock
\newblock
\shownote{[Online; accessed 27-May-2021]}.


\bibitem[Yushkevich et~al\mbox{.}(2006)]%
        {Yushkevich2006}
\bibfield{author}{\bibinfo{person}{Paul~A. Yushkevich}, \bibinfo{person}{Joseph Piven}, \bibinfo{person}{Heather Cody~Hazlett}, \bibinfo{person}{Rachel Gimpel~Smith}, \bibinfo{person}{Sean Ho}, \bibinfo{person}{James~C. Gee}, {and} \bibinfo{person}{Guido Gerig}.} \bibinfo{year}{2006}\natexlab{}.
\newblock \showarticletitle{User-Guided {3D} Active Contour Segmentation of Anatomical Structures: Significantly Improved Efficiency and Reliability}.
\newblock \bibinfo{journal}{\emph{Neuroimage}} \bibinfo{volume}{31}, \bibinfo{number}{3} (\bibinfo{year}{2006}), \bibinfo{pages}{1116--1128}.
\newblock


\bibitem[Zadka(2018)]%
        {Zadka2018}
\bibfield{author}{\bibinfo{person}{Moshe Zadka}.} \bibinfo{year}{2018}\natexlab{}.
\newblock \bibinfo{title}{How to open source your Python library}.
\newblock \bibinfo{howpublished}{\url{https://opensource.com/article/18/12/tips-open-sourcing-python-libraries}}.
\newblock


\bibitem[Zhang et~al\mbox{.}(2008)]%
        {Zhang2008}
\bibfield{author}{\bibinfo{person}{Xiaofeng Zhang}, \bibinfo{person}{Nadine Smith}, {and} \bibinfo{person}{Andrew Webb}.} \bibinfo{year}{2008}\natexlab{}.
\newblock \showarticletitle{1 - Medical Imaging}.
\newblock In \bibinfo{booktitle}{\emph{Biomedical Information Technology}}, \bibfield{editor}{\bibinfo{person}{David~Dagan Feng}} (Ed.). \bibinfo{publisher}{Academic Press}, \bibinfo{address}{Burlington}, \bibinfo{pages}{3--27}.
\newblock
\showISBNx{978-0-12-373583-6}
\urldef\tempurl%
\url{https://doi.org/10.1016/B978-012373583-6.50005-0}
\showDOI{\tempurl}


\bibitem[Ziegler et~al\mbox{.}(2020)]%
        {Ziegler2020}
\bibfield{author}{\bibinfo{person}{Erik Ziegler}, \bibinfo{person}{Trinity Urban}, \bibinfo{person}{Danny Brown}, \bibinfo{person}{James Petts}, \bibinfo{person}{Steve~D. Pieper}, \bibinfo{person}{Rob Lewis}, \bibinfo{person}{Chris Hafey}, {and} \bibinfo{person}{Gordon~J. Harris}.} \bibinfo{year}{2020}\natexlab{}.
\newblock \showarticletitle{Open Health Imaging Foundation Viewer: An Extensible Open-Source Framework for Building Web-Based Imaging Applications to Support Cancer Research}.
\newblock \bibinfo{journal}{\emph{JCO Clinical Cancer Informatics}} \bibinfo{number}{4} (\bibinfo{year}{2020}), \bibinfo{pages}{336--345}.
\newblock
\urldef\tempurl%
\url{https://doi.org/10.1200/CCI.19.00131}
\showDOI{\tempurl}
\showeprint{https://doi.org/10.1200/CCI.19.00131}
\newblock
\shownote{PMID: 32324447}.


\end{thebibliography}

\end{document}